\def\nome#1{{ \label{#1} }}
\def\writenote#1{{ }}

\def\COMMENTA#1{{}}

\def\partder#1#2{ {\partial #1 \over \partial #2} }
\def\primato#1{{{#1}^\prime}}
\def\SSs#1{{{\scriptscriptstyle #1}}}
\def\gTRE{\mbox{${^3}\! g$}}

\def\Be{\begin{equation}}
\def\Ba{\begin{array}}
\def\Ee{\end{equation}}
\def\Ea{\end{array}}

\def\Ds{\displaystyle }

\def\Sss{\scriptscriptstyle }

\def\OldRefP#1{{}}                              

\def\half{{1\over2}}

\def\EL{\mbox{$\cal E\! L$}}
\def\deq{{\buildrel \rm \circ \over =}}

\def\BMx{{\mbox{\boldmath $x$}}}
\def\BMy{{\mbox{\boldmath $y$}}}
\def\BMz{{\mbox{\boldmath $z$}}}

\def\Fc{{\cal F}}

\def\Kc{{\cal K}}
\def\Hc{{\cal H}}
\def\Jc{{\cal J}}
\def\Lc{{\cal L}}
\def\Mc{{\cal M}}
\def\Nc{{\cal N}}
\def\Pc{{\cal P}}

\def\Sc{{\cal S}}

\def\Vc{{\cal V}}

\documentstyle[12pt]{article}
\begin{document}

\addtolength{\baselineskip}{.5 \baselineskip}
\parskip = 2ex

\renewcommand{\theequation}{\thesection.\arabic{equation}}


\begin{flushright}
Short title: {\bf Newtonian Gravitation as a Gauge Theory.} \\
P.A.C.S. number:+04.50,04.90,03.50
\end{flushright}
\begin{center}
{ \LARGE \bf STANDARD AND GENERALIZED \\
    NEWTONIAN GRAVITIES AS \\
    ``GAUGE'' THEORIES OF THE \\
     EXTENDED GALILEI GROUP - \\[2 mm]
     I: THE STANDARD THEORY }
\end{center}
\begin{center}
        R.~DE PIETRI${}^1$  \\[1 mm]
{\it Department of  Physics and Astronomy, University of Pittsburgh, } \\
{\it Pittsburgh, PA 15260, USA }\\
\vskip 2 mm
L.~LUSANNA  \\[1 mm]
{\it I.N.F.N., Sezione di Firenze}  \\
{\it Largo E. Fermi 2, } {\it 50127 Arcetri (FI), Italy} \\
\vskip 1mm
and    \\
\vskip 2mm
        M. PAURI\footnote{On leave from: 
{Dipartimento di Fisica - Sezione di Fisica Teorica},
{Universit\`a di Parma, 43100 Parma, Italy,  } 
{and   I.N.F.N., Sezione di Milano, Gruppo Collegato di Parma}}
 \\[1 mm]
{\it Center for Philosophy of Science - University of Pittsburgh} \\
{\it 817D Cathedral of Learning, Pittsburgh, PA 15260, USA }\\
\end{center}
\vskip 2 mm
\begin{center} {\large \bf Abstract} \end{center}
\begin{quote}
Newton's standard theory of gravitation is reformulated
as a {\it gauge} theory of the {\it extended} 
Galilei Group. The Action principle is obtained
by matching the {\it gauge} technique and a suitable
limiting procedure from the ADM-De Witt action of
general relativity coupled to a relativistic
mass-point. 
\end{quote}

\newpage
\setcounter{equation}{0}
\section{Introduction}

The present work is the first of a series in which a
suitable reformulation of the {\it gauge} procedure is
exploited for dealing with classical non-relativistic
systems. In particular, this is the first of two papers in
which the {\it gauge} technique is applied to {\it extended}
Galilei group. The general scope of this treatment is to
reformulate firstly standard Newton`s theory  as a {\it
general manifestly-covariant} Galilean {\it gauge} theory,
and, secondly, to seek possible generalizations of it.  \par
As is well-known, a geometrical four-dimensional 
formulation of Newton's gravitational theory has been
developed already in the thirties by Elie Cartan
\cite{Cartan}. More recent formulations of the classical
theory of gravitation in geometrical terms have been
proposed by Havas \cite{Havas}, Anderson \cite{ANDE},
Trautman \cite{TrautA}, K\"unzle \cite{Kunz} and Kucha\v{r}
\cite{Kuch}.  Analyses of the classical theory as a
non-relativistic limit of  general relativity has been made
by Dautcourt \cite{Daut}, K\"unzle \cite{Kunz}, Ehlers
\cite{EHL}, Malament \cite{MALA} and others. In all these
papers, the Newtonian  theory is reobtained by describing
its inertial-gravitational structure in terms of an affine
connection compatible with the temporal flow $t_{\mu}$ and a
rank-three spatial metric $h^{\mu\nu}$. While the curvature
of the four-dimensional affine connection is different from
zero because of the presence of matter, the Newtonian
flatness of the absolute three-space is guaranteed by the
further  requirement that  Poisson's equation be satisfied,
in the covariant form $R_{\mu\nu} = 4\pi G \rho (z)  t_\mu
t_\nu$, where $R_{\mu\nu}$ is the Ricci tensor of the affine
connection and $\rho(z)$ is the matter density. In this way
the four-dimensional description is {\it dynamical}, while
the three-dimensional one is not.  A further typical feature
of this geometrical formulation is the fact that,  unlike
the case of general relativity in which there is a unique 
compatible affine connection, the curved four-dimensional
affine  structure can be separated out in a flat affine
({\it inertial})  structure and a gravitational ({\it
force}) potential; this splitting, however, cannot be done
in a unique way, unless special boundary  conditions are
externally provided. \par While the four-dimensional point
of view about Newton's theory  shows remarkable geometrical
insights and even provides a better foundation  for
Newtonian kinematics than does the traditional point of view
(see,  in this connection Earman and Friedman \cite{FRIE}),
it does not lend itself to any easy generalization. And
since what we want to obtain in the end is precisely a true
generalization of Newton's theory allowing for a dynamical
three-space, we will adopt here a completely independent
procedure based on a three-dimensional level of description
from the beginning. With this in view, the present paper
should be read also as  a first necessary step towards the
searched generalized formulation. Our approach develops
through the following steps: 

(1) First, we exploit the {\it gauge} methodology originally
applied by Utiyama \cite{Uty} to the Lorentz group within
the field theoretic framework, in order to find all the
inertial-gravitational fields which can be coupled to a 
non-relativistic mass-point. To achieve this result, we
apply a suitably adapted Utiyama  procedure to the Galilei
group. Specifically, we consider a Galilei invariant Action
corresponding to the {\it projective}  canonical realization
which describes a free particle of mass {\it m}. The
requirement of invariance (properly {\it quasi}-invariance,
according to what usually obtains in the  case of groups
with non trivial cohomology structure) of  the Action under
{\it localized} Galilei transformations, leads directly to
the following results:   (a)  the introduction of eleven
compensating {\it gauge} fields (one more than  the order of
the standard Galilei group due to the {\it central
extension}  of it);  (b) the characterization of their
Galilean transformation properties, and ;  (c) the explicit
form of the Action describing the dynamics of the mass-point
interacting with  the {\it gauge} fields playing the role of
external fields.  A geometrical interpretation of these
latter fields is then exhibited by evidentiating their
relation to the  so-called {\it Galilei} and {\it Newtonian}
Structures studied by  K\"unzle \cite{Kunz} and Kucha\v{r}
\cite{Kuch}.

(2) Second, we look for a possible {\it field} Action
capable of describing  the {\it dynamics} of these fields.
In realizing this program we are guided by the following
facts: (a) The non- relativistic limit of the relativistic
Lagrangian for a mass-point in a  pseudo-Riemannian space is
precisely the Galilei matter Lagrangian  we have already
obtained through the {\it gauge} procedure;  (b) the
non-relativistic limit of Einstein equations leads to the
geometric Cartan structure with Newton's equations; on the
other hand: (c) none among the existing four-dimensional 
formulations of Newtonian gravitation is cast in a
variational form. \par The explicit construction of the
fundamental Galilei Action is  performed by matching the
results obtained through the above  {\it gauge} technique
and a suitable non-relativistic limiting procedure  (for
$c^2 \rightarrow \infty$) from the four-dimensional level. 
Precisely, the limiting procedure is applied to the
Einstein-Hilbert-De Witt  action for the gravitational field
plus a matter action corresponding to a single mass point,
under the assumption  of the existence of a global 3+1
splitting of the total Action,  and of a suitable
parametrization of the 4-metric tensor in terms of powers of
$c^2$. Once the expansion in powers of $1/c^2$ is explicitly
calculated, we make the {\it Ansatz} of identifying the
wanted Galilean Action {\sl A} with the $zero^{th}$  order
term of the expansion itself. \par The resulting Action
contains 27 fields, i.e., 16 fields  {\it over and above}
the {\it gauge} fields obtained through the {\it gauge} 
technique. These fields are not coupled to matter and lack
{\it a-priori} any definite transformation  property. Their
role is nonetheless essential (in the spirit of the 
Einstein-Kretschmann debate\cite{Kret}\rlap ,~ one would
say) as {\it auxiliary} fields,  to the effect that they
guarantee a {\it Galilean general-covariance} of the
three-dimensional theory (where of course the {\it absolute}
nature of {\it time} is preserved).  In fact, once
appropriate transformation properties for the {\it
auxiliary} fields are postulated, the fundamental Galilei
Action turns out to be {\it quasi-invariant} under  the {\it
local} Galilei transformations.  \par As expected, a
constraint analysis shows that the theory has {\it no
physical field degrees of freedom}  so that it is
essentially Newton's theory expressed as a {\it gauge}
invariant theory or,  stated in other words, as a theory 
cast in a form valid in arbitrary  ({\it absolute time}
respecting)  {\it Galilean  reference frames}. This
formulation implies, of course, flatness of the three-space
metric $g_{ij}$ (expressed through the vanishing of the
three-dimensional Ricci tensor: $R_{ij} = 0)$ and validity
of the  Poisson equation in a suitable Galilei-covariant
form (see Section 7.1). As far as we know, a
manifestly-covariant formulation of Newton's theory of
gravitation has never been proposed until now. \par In
Section 2, the free mass-point realization of the extended
Galilei group is expounded together some preliminaries and
notations. Section 3 is devoted to the {\it gaugeization} of
the group: {\it gauge} compensating fields and their group
transformation properties are derived by imposing {\it
quasi-invariance} of the action. The equations of motion of
the mass-pont in presence of the  {\it gauge} "external"
fields are discussed in Section 4 together with a proper
characterization of the  relations between various kinds of
"observers" and "forms" of the fields.  Section 5 is 
dedicated to a summary of the known facts about Galilean and
Newtonian geometrical structures and to the correspondence
between these  latter and our {\it gauge} fields. In section
6 the main {\it Ansatz} for the selection of the fundamental
Galilean Action is discussed.  Section 7 is devoted to a
discussion of the constraint analysis of the covariant form
of Newton's theory in {\it arbitrary - absolute time
respecting} reference systems (Section 7.1) and in {\it
Galilean} reference systems (Section 7.2), respectively.

\setcounter{equation}{0}
\section{Preliminaries on the Galilei free mass-point 
realization}

 The Lagrangian and the action for a free non 
relativistic mass-point can be written,
\Be
L_M = \frac{1}{2} m \delta_{ij} \dot{x}^i \dot{x}^j ~, ~~
{\cal A}_M = \int_{t_1}^{t_2} dt~~L_M ~,
\left( { \delta_{ij} = \left\{ \Ba{ll} 1,\;\; & i  =  j \\
                                       0,     & i\neq j
                               \Ea  \right.
       }\right) ~~,
\nome{action}
\Ee
respectively.

The variational principle \mbox{${\delta}{\cal
A}_M=0$}, with variations which vanish at the end
points, gives the Euler-Lagrange equations:
\Be
{\EL}_i \equiv \partder{L_M}{x^i}
              - \frac{d}{dt} \partder{L_M}{\dot{x}^i}
            = - m \delta_{ij} \ddot{x}^j \deq 0 ~,
\nome{euler}
\Ee
where~ $\deq$ ~means that the equality is satisfied on
the extremals.

The infinitesimal Galilei transformations of the
configuration variables will be written as:
\Be
\left\{
\Ba{rl}
    \delta t         &= - \varepsilon \\
    \delta x^i       &= \varepsilon^i + c^{~~i}_{jk} \omega^j x^k
                             - v^i t   \\
    \delta \dot{x}^i &= c^{~~i}_{jk} \omega^j \dot{x}^k
                             -v^i  ~~,\\
\Ea
\right.
\nome{galiglobal}
\Ee
where $\varepsilon, \varepsilon^i, \omega^i, v^i$ are
the infinitesimal parameters of time translation, space
translations, space rotations and pure Galilei
transformations (Galilei boosts), respectively, and the
$c_{jk}^{~~i}$'s are the standard structure-constants
of the SO(3) group.

We remark that, for a given infinitesimal time
transformation $t
\rightarrow  t^* = t+\delta t$, $f(t) \rightarrow f^*(t^*)$, the
symbol $\delta$ means $\delta f(t) \equiv f^*(t^*) -
f(t)$. On the other hand, the {\it equal time}
infinitesimal transformation will be denoted by
${\delta_0} f(t) \equiv f^*(t) - f(t) = \delta f -
\delta t \cdot{df\over dt}$~. The {\it equal-time}
configuration variables transformations corresponding
to the transformation (\ref{galiglobal}) are:
\Be
\left\{
\Ba{rl}
  \delta_0 ~t        &= 0 \\
  \delta_0 x^i       &= \varepsilon^i + c^{~~i}_{jk} \omega^j x^k
                      -t v^i + \varepsilon \dot{x}^i   \\
  \delta_0 \dot{x}^i &= c^{~~i}_{jk} \omega^j \dot{x}^k
                      -v^i   + \varepsilon \ddot{x}^i ~~.\\
\Ea
\right.
\nome{ETgaliglobal}
\Ee
Note that, while $\delta_0$ commutes with time derivative, so that
\Be
  \delta_0 \frac{df(t)}{dt} = \frac{d}{dt} \delta_0 f(t)  ~~,
\Ee
we have instead
\Be
\Ba{rcl}
  \delta \Ds \frac{df(t)}{dt} 
    &=& \Ds  \delta_0 \frac{df(t)}{dt} + \frac{d^2f(t)}{dt^2} 
             \delta t \\[3 mm]
    &=& \Ds \frac{d}{dt} \delta f(t) 
       -\frac{df(t)}{dt}  \frac{d(\delta t)}{dt}  ~~.
\Ea
\Ee
Finally, notice that, since the
Galilei transformations are not, in general, fixed-time transformations,
the variation of the action must be explicitly written in the form:
\Be
\Delta {\Sc}_M = \int_{t_1}^{t_2} dt \left[
                       { \delta L_M + L_M \frac{d}{dt} \delta t}
                         \right]  
               = \int_{t_1}^{t_2} dt \left[
                       { \delta_0 L_M + \frac{d}{dt} (L_M \delta t) }
                         \right]  
~~.
\nome{variaz}
\Ee
Under the transformations (\ref{galiglobal}), we have:
\Be
\Delta {\Sc}_M = \int_{t_1}^{t_2} dt
\left[{ \frac{d}{dt} \left({ - m \delta_{ij} {x}^i v^j}
\right) } \right] ~~,
\nome{varLM}
\Ee
so that the action (\ref{action}) results {\it quasi-invariant} under
them. As a consequence of this {\it quasi-invariance} of the action,
we have the Noether identity:
\Be
   \dot{G} \equiv - \delta_0 x^i \;\EL_i \deq 0 ~~~,
\nome{noether}
\Ee
where the constant of the motion $G$ is given by:
\Be
G = \partder{L_M}{\dot{x}^i} {\delta_0} x^i - \varepsilon L_M
    + m \delta_{ij} x^i v^j ~.
\nome{noetherD}
\Ee
Before proceeding, let us fix some notation in connection with
more general classes of functions that we will have to consider;
precisely: 
1) $f(\BMz , t)$, with $\BMz$ and $t$ independent variables; and
2) $f(\BMx(t),t)$.

The variations in case 1) when $t\rightarrow t^* = t + \delta t$,  
$\BMz\rightarrow \BMz^* = \BMz + \delta \BMz$, will be denoted
by
\Be
\Ba{rcl}
  \delta   f(\BMz,t) &=& \Ds f^*(\BMz^*,t^*) - f(\BMz,t) \\[1 mm]
  \delta_0 f(\BMz,t) &=& \Ds f^*(\BMz  ,t  ) - f(\BMz,t) = \\[3 mm]
                     &=& \Ds  \delta   f(\BMz,t)
                         - \partder{f(\BMz,t)}{z^k} \delta z^k    
                         - \partder{f(\BMz,t)}{t}   \delta t    ~~~.
\Ea
\Ee                            
On the other hand, in case 2), it is convenient to distinguish three
kinds of variations: if 
\Be
\left\{
\Ba{lcl}
t       &\rightarrow& t^* = t + \delta t  \\[1 mm]  
\BMx (t)&\rightarrow& \Ds \BMx^* (t^*) = \BMx (t) + \delta \BMx (t)
                     = \BMx (t) + \delta_0 \BMx (t) 
                      + \frac{d\BMx (t)}{dt} \delta t ~~, 
\Ea
\right.
\Ee
we shall define:
\Be
\Ba{rcl}
\delta   f(\BMx(t),t)&=& \Ds f^*(\BMx^*(t^*),t^*) - f(\BMx(t),t) 
\\[3 mm] 
\delta_0 f(\BMx(t),t)&=& \Ds f^*(\BMx(t)  ,t  ) - f(\BMx(t),t) = 
\\[3 mm]
                     &=& \Ds \delta   f(\BMx(t),t)
                        - \partder{f(\BMx(t),t)}{x^k}  \delta x^k(t)    
                        - \partder{f(\BMx(t),t)}{t}    \delta t 
\\[3 mm]
\delta_{0[t]} f(\BMx(t),t) 
                   &=& \Ds f^*(\BMx^*(t)  ,t  ) - f(\BMx(t),t) = 
\\[3 mm]
                   &=& \Ds  \delta   f(\BMx(t),t)
                       - \left[
                          \partder{f(\BMx(t),t)}{x^k}  
                          \frac{dx^k(t)}{dt}    
                         +\partder{f(\BMx(t),t)}{t} 
                         \right]   \delta t \\
\Ea
\nome{2.12}
\Ee                            

Let us now turn to the Hamiltonian formalism. The canonical momenta
and the Hamiltonian function are ({$\bar{f}(p,q)$ denotes a
function in phase-space}), 
\Be
\Ba{rl}
  p_i    &= \Ds \partder{L_M}{\dot{x}^i} 
          = m\;\delta_{ij} \dot{x}^j \\[3 mm]
 \bar{H} &= p_i \dot{x}^i - L_M 
          = {1 \over 2 m} \;\delta^{ij} p_i p_j ~~,\\
\Ea
\nome{canonical}
\Ee
respectively. Then, the conserved quantity $G$ becomes:
\Be
\Ba{rl}
  \bar{G} &= \varepsilon \bar{H}  + \varepsilon^i p_i
                     + \omega^i \bar{J}_i + v^i \bar{K}_i \\
          &= \varepsilon \bar{H} + (\eta^i-t v^i) p_i
                     + m \delta_{ij} v^i x^j  \deq 0 ~~, \\
\Ea
\nome{eqNoether}
\Ee
where, for future convenience, we have introduced the infinitesimal
transformation descriptors
\Be
\eta^i(x^j) = \varepsilon^i + c^{~~i}_{jk} \omega^j x^k  ~~.
\nome{defeta}
\Ee
From Eq.(\ref{eqNoether}), we obtain the following independent
constants of the motion:
\Be
 \bar{H}~, ~~p_i~, ~~\bar{K}_i = m \delta_{ij}~x^j - t~p_i ~,
            ~~\bar{J}_i = c^{~~k}_{ij} x^j p_k ~.
\nome{constants}
\Ee

The constants of motion (\ref{constants}) provide a
{\it projective} canonical realization of the Lie
algebra of the Galilei group in terms of Poisson
brackets:
\Be
\Ba{rl}
           \{ \Kc_i , \Hc   \}  &=  \Pc_i  \\
           \{ \Kc_i , \Pc_j \}  &=  \delta_{ij} m \\
           \{ \Pc_i , \Jc_j \}  &=  c^{~~k}_{ij}  \Pc_k \\
           \{ \Kc_i , \Jc_j \}  &=  c^{~~k}_{ij}  \Kc_k \\
           \{ \Jc_i , \Jc_j \}  &=  c^{~~k}_{ij}  \Jc_k ~, 
\Ea
\nome{liealgebra}
\Ee
where:
\Be
\Hc = \bar{H}~, ~~\Pc_i = p_i~, ~~\Kc_i = \bar{K}_i~, 
       ~~\Jc_i = \bar{J}_i ~.
\nome{dent}
\Ee
Alternatively, the realization can be considered as a
{\it true} realization \cite{Marmo} of the
centrally-extended Galilei group via the {\it central
charge} $\Mc = m$~.

The generator of the {\it equal-time} Galilei
transformation in phase-space, say $\bar{\delta}_0$, is
given by the expression $\bar{G}$ defined in
eq.(\ref{eqNoether}). We have
\Be
\left\{ {
\Ba{rcl}
   \bar{\delta}_0 t   &=& 0 \\
   \bar{\delta}_0 x^i &=& \Ds \{ x^i , \bar{G} \}
                        = \varepsilon\frac{1}{m} \delta^{ij} p_j
                          + \eta^i  \\[4 mm]
                    &\Rightarrow& \Ds \left.{
                     \bar{\delta}_0 x^i
                     }\right|_{p=\partial L_M / \partial\dot{x}} 
             = \delta_0 x^i = \delta x^i + \varepsilon \dot{x}^i ~~.
\Ea
}\right.
\nome{Equaltime}
\Ee
Notice that, within the Hamiltonian formalism, the
variation ${\delta_0}\dot{x}^i$ can be obtained only by
using the equations of motion. In fact, we have:
\Be
\Ba{rl}
{\delta_0}\dot{x}^i &= \Ds \partder{ \eta^i(x)}{x^k} \dot{x}^k - v^i 
                          - \varepsilon \ddot{x}^i
                    = \Ds \partder{ \eta^i(x)}{x^k} \dot{x}^k - v^i
                        - \varepsilon\frac{1}{m}
                            \delta^{ij} \EL_{j} \\[4 mm]
                  &\deq \Ds \left.{
                         \frac{\delta^{ij}}{m} \bar{\delta} p_j
                         }\right|_{p=\partial L_M / \partial\dot{x}} 
~~~.
\Ea                         
\nome{aaa}
\Ee
Under the transformations (\ref{ETgaliglobal}), 
which are the configuration-space
analogues of the transformations (\ref{Equaltime}), we have:
\Be
\Delta {\Sc}_M = \int_{t_1}^{t_2} dt
\left[{ \frac{d}{dt} \left({ \varepsilon L_M 
- m \delta_{ij} {x}^i v^j}
\right) } \right] ~.
\nome{varLMzero}
\Ee
The resulting Noether's constants of the motion are
clearly the same as those associated with the
transformations (\ref{galiglobal}) and
(\ref{Equaltime}).

This complication can be easily avoided by turning to a
re-parameterization invariant formulation of the free
mass-point system. Using coordinates $t(\lambda)
,x^i(\lambda) $, the Lagrangian and the Action become
\Be
\Ba{rl}
\hat{L}_M (\lambda) &= \Ds \frac{1}{2} m 
     {\delta_{ij} \primato{x}^i(\lambda)
     \primato{x}^j(\lambda) \over \primato{t}(\lambda)} 
     ~~~, \\[4 mm]
\hat{\Sc}_M &= \Ds \int_{{\lambda}_1}^{{\lambda}_2}
     d{\lambda}~~\hat{L}_M(\lambda) ~~~,
\Ea
\nome{actionpar}
\Ee
respectively, where $\primato{f}(\lambda ) \equiv
\frac{d}{d{\lambda}} f(\lambda )$. As before, in this
enlarged space, we can define again an infinitesimal
transformation $\hat{\delta}$ and an ``equal-$\lambda$''
one, say $\hat{\delta}_0$. The associated Euler-Lagrange
equations and canonical momenta are
\Be
\left\{ {
\Ba{rl}
   \widehat{\EL}_t &= \Ds \frac{d}{d{\lambda}} \left[{
                 \frac{m}{2}
                 {\delta_{ij} \primato{x}^i(\lambda) 
                 \primato{x}^j(\lambda) 
                 \over \primato{t}^2(\lambda)}
                 }\right] \deq 0 \\[4 mm]
   \widehat{\EL}_i &= \Ds -\frac{d}{d{\lambda}} \left[{
                 m {\delta_{ij} \primato{x}^j(\lambda)
                 \over \primato{t}(\lambda)}
                 }\right] \deq 0   ~~~,
\Ea
          }\right.
\nome{ELpar}
\Ee
\Be
\left\{ {
\Ba{rl}
  \hat{E} &= \Ds -\partder{\hat{L}}{\primato{t}}
        = \frac{m}{2} {\delta_{ij} \primato{x}^i(\lambda)
        \primato{x}^j(\lambda) \over \primato{t}^2(\lambda)}
        = \frac{1}{2m} \delta^{ij} \hat{p}_i \hat{p}_j \\[4 mm]
   \hat{p}_i &= \Ds \partder{\hat{L}}{\primato{x}^i}
        = m {\delta_{ij} \primato{x}^j(\lambda)
        \over \primato{t}(\lambda)}
        = m \delta_{ij} \dot{x}^j(t) = p_i  ~~,
\Ea
}\right.
\nome{mompar}
\Ee
respectively, where we have defined the Poisson brackets so that:
\Be
\Ba{rl}
   \{ t(\lambda ) , \hat{E}(\lambda ) \}^\prime &= -1 \\[2 mm]
   \{ x^i(\lambda ),\hat{p}_j(\lambda )\}^\prime &= \delta^i_j ~~.
\Ea
\nome{pbpar}
\Ee
In the enlarged phase-space, coordinatized by
($t,~x^i,~\hat{E},~\hat{p}_i$), we obtain a vanishing canonical
Hamiltonian and the first-class constraint
\Be
\hat{\chi} \equiv \hat{E}-\frac{1}{2m} \delta^{ij} \hat{p}_i \hat{p}_j
           \approx 0 ~~.
\nome{conuno}
\Ee
The constraint $\hat{\chi}$ generates the following transformation
of the configurational variables:
$\hat{\bar{\delta}}_0 t(\lambda) = - \alpha (\lambda ) ~,~~
 \hat{\bar{\delta}}_0 x^i (\lambda) 
= - \alpha (\lambda ) \frac{1}{m} \delta^{ij} \hat{p}_j$.
This is the re-parameterization gauge transformation
$\lambda \rightarrow \lambda - \alpha(\lambda ) / t^\prime$.
The Lagrangian is obviously {\it quasi-invariant} under this 
operation since:
\Be
\hat{{\delta}}_0 \hat{L}_M
    =\frac{d}{d{\lambda}}
                \left[ {-\alpha (\lambda ) \over t^\prime}
                        \hat{L}_M
                \right]  ~.
\Ee

The canonical generators of the extended Galilei algebra are now:
\Be
\hat{\Hc} = \hat{E}~, ~~\hat{\Pc}_i = \hat{p}_i~,
               ~~\hat{\Kc}_i = m~\delta_{ij}~x^j - t~\hat{p}_i ~,
               ~~\hat{\Jc}_i = c^{~~k}_{ij} x^j \hat{p}_k ~,
               ~~\hat{\Mc}   = m ~.
\nome{identpar}
\Ee
and satisfy the Lie-algebra (\ref{liealgebra}) with the primed 
Poisson-brackets (\ref{pbpar}).
Consequently, the generator of the complete phase-space Galilei 
transformation $\hat{\bar{\delta}}_0$~, which is now given by:
\Be
\Ba{rl}
 \hat{\bar{G}} &= \varepsilon \hat{E}  + \varepsilon^i \hat{p}_i
                   + \omega^i \hat{\bar{J}}_i + v^i \hat{\bar{K}}_i \\
         &= \varepsilon \hat{E} + (\eta^i-t v^i) \hat{p}_i
                   + m \delta_{ij}  v^i x^j     ~~,\\
\Ea
\nome{eqNoetherpar}
\Ee
yields the following ``equal-$\lambda$'' infinitesimal transformations:
\Be
\left\{
\Ba{ccl}
\hat{\bar\delta}_0 \lambda &=& 0 \\
\hat{\bar\delta}_0 t   &=& \{ t~ ,\hat{\bar{G}}\}^\prime
                      = - \varepsilon = \hat\delta_0 t = \delta t  \\
\hat{\bar\delta}_0 x^i &=& \{ x^i,\hat{\bar{G}}\}^\prime
                      = \varepsilon^i+c^{~~i}_{jk}\omega^j x^k -t v^i
                      = \hat\delta_0 x^i  = \delta x^i
\;\; ,
\Ea
\right.
\nome{transpar}
\Ee
which coincide with the transformations (\ref{galiglobal}). \newline
We have now
\mbox{
$\left.{ \hat{\bar{\delta}}_0 \hat{p}_i } \right|_{\hat{p} 
    = \partial \hat{L}_M / \partial \primato{x} }
    = \hat\delta_0 
      \left[{m \delta_{ij} \primato{x}^j \over \primato{t} }
                 \right]
$},
without any use of Euler-Lagrange equations. On the other hand,
under the transformations (\ref{transpar}), it follows:
\Be
\hat{\delta}_0 \hat{L}_M  = \frac{d}{d{\lambda}} \left[{
                                  {- m \delta_{ij} v^i x^j } }\right]
 ~~~,
\nome{invarpar} 
\Ee
and
\Be
\hat{\bar{\delta}}_0 \hat{\chi} = 0 ~~, \nome{invarparA} 
\Ee
so that the canonical generators (\ref{identpar}) are 
again constants of the motion.
Furthermore, the first-class constraint is Galilei invariant, and
the {\it quasi-invariance} of the Lagrangian is an effect 
of the {\it central-charge} term alone.

\setcounter{equation}{0}
\section{``Gauging'' the extended Galilei algebra for the free mass-point}

We proceed now to {\it gauging} the Galilei transformations along the
standard line of Utiyama \cite{Uty}\rlap . Since the Newtonian time is
absolute, the most general finite transformation allowed for the time
coordinate is of the form $t\rightarrow \primato{t} = t + f(t)$~.
Therefore a first guess of how to {\it gauge} the group  amounts to
replace the complete infinitesimal generators (\ref{eqNoether}) and
(\ref{eqNoetherpar}) with:
\Be
\Ba{rl}
\bar{G} &= \varepsilon (t)~\bar{H}  + \varepsilon^i ({\BMx},t)~p_i
                     + \omega^i ({\BMx},t)~\bar{J}_i 
                     + v^i ({\BMx},t)~\bar{K}_i \\[1 mm]
        &= \varepsilon (t)~\bar{H}
                     + [\eta^i({\BMx},t)~-t v^i({\BMx},t)~] p_i
                     + m~ \delta_{ij} v^i ({\BMx},t)~x^j ~~,
\Ea
\nome{eqNoetherGau}
\Ee
and
\Be
\Ba{rl}
\hat{\bar{G}} &= \varepsilon (t)~\hat{E}
                     + \varepsilon^i({\BMx},t)~ \hat{p}_i
                     + \omega^i({\BMx},t)~ \hat{\bar{J}}_i
                     + v^i ({\BMx},t)~\hat{\bar{K}}_i \\[1 mm]
                  &= \varepsilon (t)~\hat{E}
                     + [\eta^i({\BMx},t)~-t v^i({\BMx},t)]~\hat{p}_i
                     + m~ \delta_{ij}  v^i ({\BMx},t)~x^j ~~,
\Ea
\nome{eqNoetherparGau}
\Ee
respectively, with $\varepsilon$ independent of $\BMx$, and
\Be
\eta^i({\BMx},t) = \varepsilon^i ({\BMx},t)~+
                    c^{~~i}_{jk}  \omega^j({\BMx},t)~ x^k  ~~.
\nome{defetaGau}
\Ee

  Notice that we have not absorbed the term $tv^i ({\BMx},t)$ into
$\eta^i({\BMx},t)$~, as it would be natural from the point of view of
the configuration space. Actually, this would not be as much natural
in phase-space and, in addition, from the group-theoretical point of
view, it would be confusing: it would mix the role of the {\it central
charge} with that of the three-dimensional Euclidean subalgebra.

We see from Eq.(\ref{eqNoetherGau}) and Eq.(\ref{eqNoetherparGau})
that  {\it gauging} the extended Galilei algebra in the case of the
mass-point realization is equivalent to {\it gauging} the algebra
generated by energy (respectively {\it time-translation}, within the
re-parameterization invariant picture), linear momentum, and the {\it
central-charge}
$\Mc = m$, with parameters  $\varepsilon (t)$~,
$\eta^i({\BMx},t)~-t v^i({\BMx},t)$~,
$\delta_{ij} v^i ({\BMx},t)~x^j$~, respectively.

Corresponding to the complete generator (\ref{eqNoetherGau}) , we
obtain:
\Be
\left\{
\Ba{ll}
\bar{\delta}_0 ~t    &= 0 \\
\bar{\delta}_0 x^i  &= \Ds \varepsilon (t) \frac{\delta^{ij} p_j}{m} 
                              +\eta^i ({\BMx},t) - t~v^i({\BMx},t) ~~,\\
\Ea
\right.
\nome{transfGauA}
\Ee
and
\Be
\left\{
\Ba{rcl}
\bar{\delta}_0 p_i       &=& \Ds p_k\partder{}{x^i}[\eta^k({\BMx},t)
                                - t~v^k({\BMx},t)]
                                + m \partder{}{x^i}
                                  [\delta_{lk}x^l v^k({\BMx},t)]\\[2 mm]
\bar{\delta}_0 \bar{H}   &=& \Ds {1 \over m} 
                              \delta^{ij} p_i \bar{\delta} p_j 
                           \\[2 mm]
\bar{\delta}_0 \bar{J}_i &=& \Ds c^{~~k}_{ij} p_k
                               [\eta^j({\BMx},t)-t~v^j({\BMx},t)]
                               -c^{~~k}_{ij} x^j  p_r   \partder{}{x^k}
                               [\eta^r({\BMx},t)-t~v^r({\BMx},t)]
                          \\[2 mm]
                       & & \Ds -m c^{~~k}_{ij} x^j \partder{}{x^k}
                               [\delta_{lm}x^l v^m({\BMx},t)]
                          \\[2 mm]
\bar{\delta}_0 \bar{K}_i &=& \Ds \varepsilon(t) p_i  
                            + t p_k \partder{}{x^i}
                               [\eta^k({\BMx},t)-t~v^k({\BMx},t)]
                          \\[2 mm]
                    & & \Ds +m \delta_{ij} 
                        [\eta^j({\BMx},t)-t~v^j({\BMx},t)]
                        -m~ t \partder{}{x^i}
                        [\delta_{lm}x^l v^m({\BMx},t)]  ~~.
\Ea
\right.
\nome{transfGauB}
\Ee
Finally, using $p_i = m \delta_{ij} \dot{x}^j$~, we obtain
\Be
\left\{
\Ba{lcl}
{\delta}_0 ~t  &=& 0 \\
{\delta}_0 x^i  &=& \varepsilon (t) \dot{x}^i
                    +\eta^i ({\BMx},t) - t~v^i({\BMx},t)   \\[2 mm]
{\delta}_0 \dot{x}^i  
         &=& \Ds \frac{d}{dt} [ \varepsilon (t) \dot{x}^i ]
                      +\dot{x}^k\partder{}{x^k}
                        [\eta^i({\BMx},t)-t~v^i({\BMx},t)]  \\[2 mm]
         & & \Ds + \partder{}{t}[\eta^i ({\BMx},t) - t~v^i({\BMx},t)] 
~~~.
\Ea
\right.
\nome{transfGauC}
\Ee

  Eqs.(\ref{transfGauC}) reduce to Eqs.(\ref{ETgaliglobal})
in the limit of global flat symmetry and they can be taken
as a {\it definition} of the {\it Galilei gauge transformations} 
in configuration space.
Notice that the {\it quasi-invariance} of the Lagrangian
under the global flat transformations is now broken and that
$\bar{\delta}_0 p_i |_{p_i=m\delta_{ij}\dot{x}^j}$ is not
equal to $\delta_0 [m \delta_{ij}\dot{x}^j]$~.

  On the other hand, within the re-parameterization invariant picture,
corresponding to the complete generator (\ref{eqNoetherparGau}), we get:
\Be
\left\{
\Ba{lcl}
\hat{\bar\delta}_0  ~\lambda    
&=& 0 \\
\hat{\bar\delta}_0 ~t(\lambda )  
&=& -\varepsilon (t(\lambda ))   \\
\hat{\bar\delta}_0 x^i(\lambda )
&=&  \eta^i ({\BMx},t) - t~v^i({\BMx},t) \\[2 mm]
\hat{\bar\delta}_0 \primato{t}(\lambda )  
&=& \Ds - \primato{t} ~\frac{d\varepsilon (t(\lambda ))}{dt} \\[2 mm]
\hat{\bar\delta}_0 \primato{x}^i(\lambda )
&=& \Ds  \primato{x}^k \partder{}{x^k}
           [\eta^i ({\BMx},t) - t~v^i({\BMx},t)]
        +\primato{t} \partder{}{t}
           [\eta^i ({\BMx},t) - t~v^i({\BMx},t)]  ~,
\Ea
\right.
\nome{transfparGauA}
\Ee
and
\Be
\left\{
\Ba{rcl}
\hat{\bar\delta}_0 \hat{p}_i  &=& \Ds -\hat{p}_k\partder{}{x^i}
                             [\eta^k({\BMx},t)-t~v^k({\BMx},t)]
                             -m \partder{}{x^i}
                             [\delta_{lk}x^l v^k({\BMx},t)]\\[2 mm]
\hat{\bar\delta}_0 \hat{E}    &=& \Ds \hat{E} \frac{d\varepsilon(t)}{dt}
                             +\hat{p}_i \partder{}{t}
                             [\eta^i({\BMx},t)-t~v^i({\BMx},t)]
                             + m \partder{}{t}
                             [\delta_{ij}x^i v^j({\BMx},t)]\\[2 mm]
\hat{\bar\delta}_0 \hat{\Jc}_i &=& \Ds c^{~~k}_{ij}  \hat{p}_k
                             [\eta^j({\BMx},t)-t~v^j({\BMx},t)]
                             -c^{~~k}_{ij} x^j
                             \hat{p}_r \partder{}{x^k}
                             [\eta^r({\BMx},t)-t~v^r({\BMx},t)]
                             \\[2 mm]
                             & & \Ds  -m c_{ij}^{~~k} x^j
                             \partder{}{x^k}
                             [\delta_{lm}x^l v^m({\BMx},t)]\\[2 mm]
\hat{\bar\delta}_0 \hat{\Kc}_i &=& \Ds \varepsilon(t) \hat{p}_i
                             +t \hat{p}_k \partder{}{x^i}
                             [\eta^k({\BMx},t)-t~v^k({\BMx},t)]
                             \\[2 mm]
                             & & \Ds +m \delta_{ij} 
                             [\eta^j({\BMx},t)-t~v^j({\BMx},t)]
                             -m t \partder{}{x^i}
                             [\delta_{lm}x^l v^m({\BMx},t)]   ~~.
\Ea
\right.
\nome{transfparGauB}
\Ee

In conclusion, from Eq.(\ref{transfGauC}), we see that ${\BMx}(t)$
undergoes a time-dependent general Euclidean coordinate transformation
(although deformed by an effect from Galilei boosts) plus a
velocity-dependent transformation induced by the condition
$\bar{\delta}_0 t = 0$~. On the other hand, from
Eq.(\ref{transfparGauA}), we see that $t(\lambda )$ undergoes a
transformation which does not involve the Euclidean coordinates and,
simultaneously, ${\BMx}(\lambda )$ undergoes a time-dependent general
Euclidean coordinate transformation (still deformed by an effect from
Galilei boosts). Being a consequence of the absolute nature of the
Newtonian time, these features do not appear in the {\it gauging} of
the Poincar\'e realization corresponding to   the free relativistic
mass-point with Lagrangian $L_M = - m \sqrt{ \eta_{\mu\nu}
\primato{x}^\mu (\lambda ) \primato{x}^\nu (\lambda )}$~ (see, for
instance, \cite{GpP}). Actually, the {\it gauging} of the Poincar\'e 
transformations induces a general coordinate transformation of $x^\mu
(\lambda )$ which is indistinguishable from the simple {\it gauging}
of the space-time translations alone. As well-known, the {\it gauging}
of the Lorentz transformations is made evident by means of the
``soldering procedure'' which amounts to say that the vectors
belonging to the tangent bundle of the curved space-time must
transform as four vectors under ``local'' Lorentz transformation.  The
``soldering'' is done by means of a set of {\it vierbeins}  ${\bf
E}^A_\mu (x)$ so that the flat transformation $\primato{x}^\mu
\rightarrow \primato{x}^\nu + \varepsilon^\mu_{\cdot\nu}
\primato{x}^\mu$ (where $\varepsilon^\mu_{\cdot\nu}$ are the
parameters of the Lorentz transformations) is replaced by ${\bf
E}^A_\mu (x)\primato{x}^\mu \rightarrow {\bf E}^A_\mu (x)
\primato{x}^\mu + \varepsilon^\Sc_{\cdot B} {\bf E}^A_\mu
(x)\primato{x}^\mu$.

We want to discuss now the problem of the invariance (possibly {\it
quasi-invariance}) of the Lagrangian with respect to the local Galilei
transformations just defined. In the Newtonian case, unlike the
relativistic one, the absolute nature of time prevents us from using
the standard ``soldering'' procedure, if not for the Euclidean
subalgebra generated by $\hat{\Pc}_k$ and $\hat{\Jc}_i$. Actually,
while the effect of the space rotations in the term $\eta^i({\BMx},t)$
is indistinguishable from the effect generated  in it by space
translations, the effect of ``local'' space rotations and translations
can be distinguished for three-vectors by introducing the
``soldering'' with {\it dreibeins} ${\bf E}^a_i({\BMx},t)$. Notice,
however, that further complications arise here from the fact that the
general Euclidean transformations are time-dependent and that time
translations and Galilei boosts introduce new terms into the
transformation of the velocity. Accordingly, the  rule for the
transition from  the global flat transformation of three-velocity to
its general transformation is  obtained by defining
$\omega^a({\BMx},t) = {\bf E}^a_i({\BMx},t) \omega^i ({\BMx},t)$, and
by imposing, within the two alternative pictures introduced above, the
following transformation properties
\Be
\left\{
\Ba{rlcl}
   & {\delta}_0 ~\dot{x}^i &=& \Ds {d \over dt} [\varepsilon  \dot{x}^i]
                            +  c^{~~k}_{ij}  \omega^j \dot{x}^k
                            - v^i          \\[3 mm]
\Rightarrow
  & {\delta}_{0[t]}  [{\bf E}^a_i \dot{x}^i]
               &=& \Ds {d \over dt} [\varepsilon {\bf E}^a_i \dot{x}^i]
                  + c^{~~a}_{bc}  \omega^b
                  {\bf E}^c_i \dot{x}^i + [?] 
\Ea \right.
\nome{dreibeinA}
\Ee
\Be
\left\{
\Ba{rlcl}
  &\hat{\delta}_0 ~\primato{x}^i 
      &=& c^{~~i}_{jk}  \omega^j \primato{x}^k
         -v^i \primato{t}  \\[1 mm]
\Rightarrow
  &\hat{\delta}_{0[\lambda]} [{\bf E}^a_i \primato{x}^i]
         &=& c^{~~a}_{bc}  \omega^b
         {\bf E}^c_i \primato{x}^i + [?]  ~~,\\
\Ea \right.
\nome{dreibeinA1}
\Ee
where the question marks in Eqs.(\ref{dreibeinA}) and
(\ref{dreibeinA1})  stand for possible terms connected to time
translation   and Galilei boosts: this point will be settled later
on.

By means of the invertible {\it dreibeins} ${\bf E}^a_i$, a
Euclidean metric in the three-space is naturally introduced in the
form
\Be
\delta_{ij} \Rightarrow g_{ij}
            \equiv    \delta_{ab} {\bf E}^a_i {\bf E}^b_j ~~,
\nome{metricaA}
\Ee
while the inverse metric can be likewise defined in term of the inverse
{\it dreibeins} ${\bf H}_a^i$ (${\bf H}_a^i {\bf E}^b_i = \delta^a_b$, 
${\bf H}_a^i {\bf E}^a_j = \delta^i_j$) as:
\Be
\delta^{ij} \rightarrow g^{ij}
            \equiv    \delta^{ab} {\bf H}_a^i {\bf H}_b^j
~~,~~
 g^{ij}  g_{jk} = \delta^i_k
~~.
\nome{metricaB}
\Ee

Let us now discuss, within the standard picture, the issue of the
invariance of the Lagrangian under the newly introduced local
transformations, by considering first the local time-independent
Euclidean transformations $\delta_0^{\SSs{[e]}}$ ($\varepsilon = 0$,
$v^i = 0$, $\omega^i =\omega^i ({\BMx})$, $\varepsilon^i
=\varepsilon^i ({\BMx})$); then pure time transformations
$\delta_0^{\SSs{[t]}}$ ($\varepsilon = \varepsilon (t)$, $v^i = 0$,
$\omega^i = 0$, $\varepsilon^i = 0$), and the combination of both
$\delta_0^{\SSs{[e]}}$ and $\delta_0^{\SSs{[t]}}$, say
$\delta_0^{\SSs{[et]}}$ ($\varepsilon = \varepsilon (t)$, $v^i = 0$,
$\omega^i =\omega^i ({\BMx})$, $\varepsilon^i =\varepsilon^i
({\BMx})$); finally the most general transformations including local
time-dependent Euclidean transformations and Galilei boosts.
\\[3 mm]
\noindent
{\bf i)}  In the case of local time-independent
Euclidean transformations $\delta_0^{\SSs{[e]}}$,  defined by
$\varepsilon = 0$, $v^i = 0$, $\omega^i =\omega^i ({\BMx})$,
$\varepsilon^i =\varepsilon^i ({\BMx})$, we have:
\Be
\Ba{rcl}
 {\delta}_0^{\SSs{[e]}}
      \dot{x}^i &=&  \Ds \dot{x}^k \partder{}{x^k}
     [\varepsilon^i({\BMx}) 
     + c^{~~i}_{jl} \omega^j({\BMx}) x^l ] 
     \\[2 mm]
     &=& \Ds  \dot{x}^k  \partder{\eta^i_{\SSs{[e]}}({\BMx})}{x^k} 
     ~~,
\Ea
\nome{LocalEuclide}
\Ee
where $\eta^i_{\SSs{[e]}}({\BMx}) 
=\varepsilon^i({\BMx}) + c^{~~i}_{jl} \omega^j({\BMx}) x^l$
is the restriction of $\eta^i({\BMx},t)$ to time independent 
Euclidean transformation.  Then, by imposing
\Be
{\delta}_{0[t]}^{\SSs{[e]}}
    [{\bf E}^a_i \dot{x}^i] = c^{~~a}_{bc} \omega^b({\BMx})
                                             [{\bf E}^c_i \dot{x}^i] ~,
\nome{EuclideInvA}
\Ee
it follows:
\Be
\Ba{lcr}
\delta_{0[t]}^{\SSs{[e]}} {\bf E}^a_i ({\BMx},t)
     &=& \Ds {\delta_0^{\SSs{[e]}}}  {\bf E}^a_i ({\BMx},t)
     +\partder{{\bf E}^a_i ({\BMx},t)}{x^k}
     {\delta}_0^{\SSs{[e]}} x^k  \\[2 mm]
     &=& \Ds  - \partder{\eta^i_{\SSs{[e]}}({\BMx}) }{x^k} {\bf E}^a_i
     + c^{~~a}_{bc} \omega^b({\BMx}) {\bf E}^c_i ~~,
\Ea
\nome{TetradiTAZ}
\Ee
i.e.
\Be
{\delta}_{0[t]}^{\SSs{[e]}}
{\bf E}^a_i = - \partder{\eta^i_{\SSs{[e]}}({\BMx}) }{x^k} {\bf E}^a_i
                + c^{~~a}_{bc} \omega^b({\BMx}) {\bf E}^c_i
                - {\bf E}^a_{i,k} \eta^k_{\SSs{[e]}}({\BMx})
~.
\nome{TetradiTranA}
\Ee
in agreement to the conventions established in Eqs.(\ref{2.12}).
It is then immediately seen that the invariance of the Lagrangian under
time-independent local Euclidean transformation can be recovered
by means of the substitution:
\Be
\dot{x}^i \Rightarrow {\bf E}^a_i \dot{x}^i   ~~,
\Ee
which leads to
\Be
\Ba{rcl}
L_M       =  \frac{1}{2} m \delta_{ij} \dot{x}^i \dot{x}^j
             ~~~\Rightarrow L_M^{g_{[e]}}
          &\equiv& \frac{1}{2} m \delta_{ab} {\bf E}^a_i \dot{x}^i
                                        {\bf E}^b_j \dot{x}^j \\[1 mm]
          &=& \frac{1}{2} m g_{ij} \dot{x}^i \dot{x}^j ~~.\\
\Ea
\nome{LocalEuLag}
\Ee
In fact, since the transformation rules for the three-dimensional
metric field (\ref{metricaA}) are given by:
\Be
{\delta}_{0[t]}^{\SSs{[e]}}
       g_{ij} = {\delta}_0^{\SSs{[e]}} g_{ij}
                  + g_{ij,k} \eta^k_{\SSs{[e]}}({\BMx})  
              = \Ds - \partder{\eta^k_{\SSs{[e]}}({\BMx}) }{x^i} g_{kj}
                  - \partder{\eta^k_{\SSs{[e]}}({\BMx}) }{x^j} g_{ik} 
~~,
\nome{metricaTranA}
\Ee
(which are the correct infinitesimal transformation properties of a
covariant second-rank three-dimensional tensor), it follows:
\Be
{\delta}_{0[t]}^{\SSs{[e]}} L_M^{g_{[e]}} = 0
~~.
\nome{LagInvarA}
\Ee
Therefore, the preliminary guess for the {\it gauging} of
the Galilei transformation (see Eqs. \ref{eqNoetherGau},
\ref{transfGauA}-\ref{transfGauC}, 
or \ref{eqNoetherparGau}, \ref{transfparGauA} 
and \ref{transfparGauB})
is made geometrically consistent by replacing the flat metric 
$\delta_{ij}$ and $\delta^{ij}$ in those formulae by 
$g_{ij}$ and $g^{ij}$, respectively.

  Within the  re-parameterization invariant picture, the invariance is
recovered by the corresponding substitution,
\Be
\primato{x}^i       \Rightarrow  {\bf E}^a_i \primato{x}^i
\Ee
which leads to
\Be
\hat{L}_M (\lambda) \Rightarrow \hat{L}_M^{g\SSs{[e]}} (\lambda)
          = \frac{1}{2} m {g_{ij} \primato{x}^i(\lambda)
          \primato{x}^j(\lambda) \over \primato{t}(\lambda)} ~.
\nome{LocalEuLagpar}
\Ee
In fact, from Eq.(\ref{metricaTranA}) with
${\delta}_{0[t]}^{\SSs{[e]}}$ formally replaced by
$\hat{\delta}_{0[\lambda]}^{\SSs{[e]}}$, it follows:
\Be
\hat{\delta}_{0[\lambda]}^{\SSs{[e]}}  \hat{L}_M^{\SSs{[e]}} = 0
~.
\nome{LagInvarApar}
\Ee

\noindent
{\bf ii)}  Consider now the pure time-translations
$\hat{\delta}_0^{\SSs{[t]}}$. For the sake of simplicity, we shall
confine to the re-parameterization invariant picture. We have:
\Be
\left\{
\Ba{lcl}
\hat{\delta}_0^{\SSs{[t]}} ~t  &=& - \epsilon (t)   \\
\hat{\delta}_0^{\SSs{[t]}} x^i &=& 0 ~~.
\Ea
\right.
\nome{PureTime}
\Ee
It is immediate to see that, by means  of a {\it einbein} 
substitution of the form
\Be
\primato{t} \Rightarrow \Theta (t) \primato{t}
~~,
\nome{PureTimeRule}
\Ee
the Lagrangian:
\Be
\hat{L}_M^{g\SSs{[t]}} (\lambda)
              \equiv \frac{1}{2} m {\delta_{ij} \primato{x}^i(\lambda)
              \primato{x}^j(\lambda) \over
              \Theta (t)   \primato{t}(\lambda)}  ~~,
\nome{Lt}
\Ee
is invariant under the transformations $\hat{\delta}_0^{\SSs{[t]}}$,
provided we impose:
\Be
\hat{\delta}_{0[\lambda]}^{\SSs{[t]}} [\Theta (t) \primato{t} ] = 0 ~~,
\Ee
which, in turn, gives:
\Be
\hat{\delta}_{0[\lambda]}^{\SSs{[t]}}
        \Theta (t) = \hat{\delta}_0^{\SSs{[t]}} \Theta (t)
                     - {\varepsilon}(t) \frac{d\Theta (t)}{dt}
                   = \dot{\varepsilon}(t) \Theta (t)
~~.
\nome{TranTThet}
\Ee

~~\\[3 mm]
\noindent
{\bf iii)} We can summarize the results found up to 
now in the re-parameterization
invariant picture, by saying that the modified matter Lagrangian
\Be
\hat{L}_M^{g\SSs{[et]}} (\lambda)
               \equiv \frac{1}{2} m {g_{ij} \primato{x}^i(\lambda)
                      \primato{x}^j(\lambda) \over
                      \Theta (t)   \primato{t}(\lambda)}  ~~,
\nome{GauLagEt}
\Ee
is strictly invariant under the local transformations:
\Be
\left\{ {
\Ba{ccccl}
\hat{\delta}_0^{\SSs{[et]}}  t
    &=& \hat{\delta}_0^{\SSs{[t]}}  t
    &=& - \epsilon (t)            \\[1 mm]
\hat{\delta}_0^{\SSs{[et]}} {x}^i
    &=&\hat{\delta}_0^{\SSs{[e]}} {x}^i
    &=&\eta^i_{\SSs{[e]}}({\BMx}) \\[1 mm]
\hat{\delta}_0^{\SSs{[et]}} \primato{t}
    &=&\hat{\delta}_0^{\SSs{[t]}} \primato{t}
    &=&- \dot{\varepsilon} (t) \primato{t} \\[1 mm]
\hat{\delta}_0^{\SSs{[et]}} \primato{x}^i
    &=&\hat{\delta}_0^{\SSs{[e]}} \primato{x}^i
    &=&\Ds \primato{x}^k \partder{\eta^i_{\SSs{[e]}}({\BMx})}{x^k} ~~,
\Ea
}\right.
\nome{GauTranEt}
\Ee
if we adopt the following transformation rules for the fields:
\Be
\left\{
\Ba{lclcl}
\hat{\delta}_{0[\lambda]}^{\SSs{[et]}} \Theta (t)
    &=& \hat{\delta}_{0[\lambda]}^{\SSs{[t]}} \Theta (t)
    &=& \dot{\varepsilon}(t) \Theta (t) \\[2 mm]
\hat{\delta}_{0[\lambda]}^{\SSs{[et]}} {\bf E}^a_i
    &=& \hat{\delta}_{0[\lambda]}^{\SSs{[e]}} {\bf E}^a_i
    &=& \Ds - \partder{\eta^i_{\SSs{[e]}}({\BMx}) }{x^k} {\bf E}^a_i
    + c^{~~a}_{bc} \omega^b({\BMx}) {\bf E}^c_i \\[2 mm]
\hat{\delta}_{0[\lambda]}^{\SSs{[et]}} g_{ij}
    &=& \hat{\delta}_{0[\lambda]}^{\SSs{[e]}} g_{ij}
    &=& \Ds  - \partder{\eta^k_{\SSs{[e]}}({\BMx}) }{x^i} g_{kj}
             - \partder{\eta^k_{\SSs{[e]}}({\BMx}) }{x^j} g_{ik} ~~.
\Ea \right.
\nome{DeltaRg}
\Ee

~~\\[3 mm]
\noindent
{\bf iv)} Finally, we have to show that the invariance (possibly {\it
quasi-invariance}) of the matter Lagrangian under the most general
transformations, including local time-dependent Euclidean
transformations and Galilei boosts, can be recovered by introducing
{\it four} additional external fields, say $A_0({\BMx},t)$,
$A_i({\BMx},t)$. Within the re-parameterization invariant picture, our
scope will be achieved if we succeed in defining a new matter
Lagrangian $\hat{L}_M^g$ such that: \newline {\bf a)} under the {\it
gauge} transformations (\ref{transfparGauA})  and the corresponding
transformations induced on all the  additional fields, be {\it
quasi-invariant} in the form:
\Be
\hat{\delta}_{0[\lambda]} \hat{L}_M^{g} = {d \Fc \over d\lambda} ~~,
\nome{LagQuasInv}
\Ee
{\bf b)} its {\it global flat} limit, together with that of its
transformation properties, coincide with the expression
(\ref{actionpar}) and (\ref{invarpar}), respectively.

First of all, in order to reproduce the {\it cocycle} term of
Eq.(\ref{invarpar}) in the global flat limit, $\Fc$ will be chosen
as $\Fc = - m g_{ij} v^i x^j$. Then, the matter Lagrangian will be
defined as
\Be
\hat{L}_M^{g} (\lambda)
          = {1\over\Theta\primato{t}} \frac{m}{2}
             \left[ {  g_{ij} \primato{x}^i  \primato{x}^j
                      + 2 A_i \primato{x}^i  \primato{t}
                      + 2 A_0 \primato{t}    \primato{t} }
             \right] ~~.
\nome{GauLagPar}
\Ee
Second, since we have to preserve the transformation properties
of ${\bf E}^a_i$, $g_{ij}$ and $\Theta$ already established
for the case $v^i({\BMx},t) = 0$ in Eqs.(\ref{DeltaRg}),
we shall assume
\Be
\left\{
\Ba{rl}
\hat{\delta}_{0[\lambda]} \Theta  &= \Ds  \hat{\delta}_0 \Theta
          - {\epsilon}(t) \frac{d\Theta (t)}{dt} \\[2 mm]
          &= \Ds \dot{\epsilon}(t) \Theta (t)  \\[2 mm]
\hat{\delta}_{0[\lambda]} {\bf E}^a_i &= \Ds \hat{\delta}_0  {\bf E}^a_i
          +  {\bf E}^a_{i,k} \tilde\eta^k ({\BMx},t)
          - \partder{ {\bf E}^a_i}{t} \varepsilon (t) \\[2 mm]
          &= \Ds c^{~~a}_{bc} \omega^b ({\BMx},t)  {\bf E}^c_i
          - \partder{\tilde\eta^k({\BMx},t)}{x^i}{\bf E}^a_k \\[2 mm]
\Ea
\right.
\nome{GauTranTotA}
\Ee
$$
\Rightarrow \Ba{lr} \hat{\delta}_{0[\lambda]}  g_{ij} 
            &= \Ds \hat{\delta}_0  g_{ij}
            + g_{ij,k} \tilde\eta^k ({\BMx},t)
            - \partder{g_{ij}}{t} \varepsilon (t) \\[2 mm]
            &= \Ds - \partder{\tilde\eta^k ({\BMx},t) }{x^i} g_{kj}
            - \partder{\tilde\eta^k ({\BMx},t) }{x^i} g_{kj} ~,
            \Ea
$$
where we have introduced the notation 
$\tilde\eta^k ({\BMx},t) \equiv \eta^k ({\BMx},t) - t v^k({\BMx},t)$.
Then, the {\it quasi-invariance} (\ref{LagQuasInv}) of the matter 
Lagrangian (\ref{GauLagPar}) is guaranteed if the additional fields
$A_0({\BMx},t)$ , $A_i({\BMx},t)$
transform as follows:
\Be
\left\{ {
\Ba{rcl}
\hat{\delta}_{0[\lambda]}   A_0     &=& \Ds \hat{\delta}_0 A_0
                         - {\varepsilon} \partder{A_0}{t}
                         + A_{i,j} \tilde\eta^j  \\[2 mm]
                     &=& \Ds 2 \dot{\varepsilon} A_0
                         - A_i \partder{\tilde\eta^i}{t}
                         + \Theta \partder{\Fc}{t} \\[2 mm]
\hat{\delta}_{0[\lambda]}   A_i     &=& \Ds  \hat{\delta}_0 A_i
                        - {\varepsilon} \partder{A_i}{t}
                        + A_{i,j} \tilde\eta^j \\[2 mm]
                     &=& \Ds \dot{\varepsilon} A_i
                        - A_j \partder{\tilde\eta^j}{x^i}
                        - g_{ij}  \partder{\tilde\eta^j}{t}
                        + \Theta \partder{\Fc}{x^i} ~~.
\Ea
}\right.
\nome{GauTranTotB}
\Ee
Notice that the fields ${\bf E}^a_i$, $g_{ij}$, $\Theta$, $A_0$ and
$A_i$ have the global flat Galilean limits $\delta^a_i$,
$\delta_{ij}$, $1$, $0$ and $0$, respectively. Here, the $A_i$'s play
the role of the components ${\bf E}^0_i (x)$ of  the relativistic {\it
vierbeins}, while the two fields $\Theta$ and $A_0$ correspond to a
splitting of the {\it vierbein} component ${\bf E}^0_0 (x)$, as it
will be clear later on.

   In a similar way, it can be seen that, within the standard picture,
the Lagrangian
\Be
{L}_M^{g} (t) = {1 \over \Theta} \frac{m}{2}
                           \left[{   g_{ij} \dot{x}^i  \dot{x}^j
                                    + 2 A_i \dot{x}^i
                                    + 2 A_0
                            } \right]    ~~,
\nome{GauLag}
\Ee
  is {\it quasi-invariant} under the field transformation
rules (\ref{GauTranTotA}), (\ref{GauTranTotB}),
with ${\delta}_0 = {{\hat{\delta}}}_0$, and the mass-point coordinate
transformations given by (\ref{transfGauC}), in the sense that we have
\Be
{\delta}_{0[t]} {L}_M^{g} =   {d \Fc \over dt} 
                       + \varepsilon {d [ {L}_M^{g} ] \over dt}   ~,
\nome{LagQuasInvT}
\Ee
so that
\Be
\Delta {\Sc}_M^{g} = \int_{t_1}^{t_2} dt  \left[ { {d \Fc \over dt} 
                         +{d [ \varepsilon  {L}_M^{g} ] \over dt}
                                  } \right] = 0  ~~.
\Ee
The condition for the invariance of the theory under the  {\it gauged}
Galilei transformations can now be easily formulated also in the
Hamiltonian formalism. From Eq.(\ref{GauLagPar}), we get the following
expressions for energy and linear momentum
\Be
\left\{
\Ba{rcl}
\hat{E} &=& \Ds \partder{\hat{L}^g_M}{\primato{t}} 
           =\frac{1}{\Theta \primato{t}^2} \frac{m}{2}
               g_{ij} \primato{x}^i\primato{x}^j
            - \frac{m}{\Theta} A_0 \\[2 mm]
\hat{p}_i &=& \Ds \partder{\hat{L}^g_M}{\primato{x}^i} 
           = \frac{m}{\Theta \primato{t}} \left({
                 g_{ij} \primato{x}^j + A_i \primato{t}
                 }\right)  ~~, 
\Ea
\right.
\nome{SostHam}
\Ee
while the first-class constraint (\ref{conuno}) becomes
\Be
\hat{\chi}_g =  {1 \over \Theta}    
                \left[\hat{E} + {m \over \Theta} A_0 \right]
              - {g^{ij} \over 2 m}  
                \left[\hat{p}_i - {m \over \Theta} A_i \right]
                \left[\hat{p}_j - {m \over \Theta} A_j \right]
~~.
\nome{conunoGau}
\Ee
Since, under the transformation (\ref{transfparGauB})
(with $\delta_{ij} \Rightarrow g_{ij}$), we have:
\Be
\Ba{rl}
\hat{\bar{\delta}}_{0[\lambda]} \hat\chi 
     &=\Ds {1 \over \Theta} \hat{\bar\delta}_0 \hat{E}
       -{1 \over \Theta^2} \left[{
       \hat{E} + {m \over \Theta} A_0 }\right]
       \hat\delta_{0[\lambda]} \Theta   
       + {1 \over \Theta^2} m  \hat\delta_{0[\lambda]} A_0   \\[1 mm]
       &~\Ds - {1 \over 2 m} \hat\delta_{0[\lambda]} g^{ij}
       \left[\hat{p}_i - {m \over \Theta} A_i \right]
       \left[\hat{p}_j - {m \over \Theta} A_j \right]  
       \\[1 mm]
       &~\Ds - {g^{ij} \over m}
       \left[\hat{p}_i - {m \over \Theta} A_i \right]
       \left[\hat{\bar{\delta}}_0 \hat{p}_j
       - {m \over \Theta}  \hat\delta_{0[\lambda]} A_j
       + {m \over \Theta^2} A_j \hat\delta_{0[\lambda]} \Theta 
       \right] ~~,
\Ea
\nome{varconunoGau}
\Ee
the invariance of the constraint
\Be
\hat{\bar{\delta}}_{0[\lambda]} \hat\chi_g = 0 ~~,
\Ee
which is the Hamiltonian analogue to the invariance
of the Lagrangian, is guaranteed if the fields  transform
according to
\Be
\left\{
\Ba{rl}
\hat{\delta}_{0[\lambda]}~\Theta  &= \dot{\epsilon}(t) \Theta (t)  
\\[2 mm]
\hat{\delta}_{0[\lambda]}  g_{ij} &= \Ds 
                  - \partder{\tilde\eta^k ({\BMx},t) }{x^i} g_{kj}
                  - \partder{\tilde\eta^k ({\BMx},t) }{x^i} g_{kj}
\\[2 mm]
\hat{\delta}_{0[\lambda]} A_0     &= \Ds 2 \dot{\varepsilon} A_0
                  - A_i \partder{\tilde\eta^i}{t}
                  - \Theta \partder{}{t}
                  \left[{g_{ij} v^i x^j}\right] \\[2 mm]
\hat{\delta}_{0[\lambda]} A_i     &= \Ds \dot{\varepsilon} A_i
                  - A_j \partder{\tilde\eta^j}{x^i}
                  - g_{ij}  \partder{\tilde\eta^j}{t}
                  - \Theta \partder{}{x^i}
                  \left[{g_{ij} v^i x^j}\right]  ~~.
\Ea
\right.
\nome{tranCampGau}
\Ee
Note that since
\Be
\delta f(\BMx(t),t) = \hat{\delta}_{0[\lambda]}
f(\BMx(t(\lambda)),t(\lambda))
~~,
\Ee
Eqs.(\ref{tranCampGau}) can be easily adapted to the standard picture.

Let us remark in addition that were it not for the
presence of the {\it einbein} $\Theta$, the modified
Hamiltonian constraints (\ref{conunoGau}) could have
been made invariant only in the {\it week} sense
$\hat{\bar{\delta}}_{0[\lambda]} \hat\chi_g =
\varepsilon
\hat\chi_g \simeq 0$.

Finally, the {\it cocycle} term which appeared in a
generic form in the transformations (\ref{GauTranTotB})
of the fields $A_0$ and $A_i$, is now {\it explicitly
determined} (up to a constant) in a form which
reproduces the standard expression in the global flat
Galilean limit.  This is due to the fact that the
Hamiltonian formalism of the reparemetrization
invariant scheme requires a first class constraint
which, in absence of external fields, says that the
Galilei Casimir invariant representing the internal
energy ($E-\frac{1}{2m}\delta^{ij}p_i p_j$) is equal to
zero.

In presence of external fields, the constraints given
by Eqs.(\ref{conunoGau}), says again that the Casimir
invariant vanishes.  This is clearly a consequence of
the fact that time is absolute and of the fact that
preservation of the constraints (i.e. of the vanishing
of the Casimir invariant) is possible only provided the
projective realization is taken into account through
the {\it cocycle} terms in Eq.(\ref{tranCampGau}). The
analogue of this phenomenon in the standard Lagrangian
picture is expressed by the fact that the
identification of the total energy in different
coordinate systems connected by general Galilean
coordinate transformation requires explicitly the
appearance of the {\it cocycle} term within the
transformation rules of the relevant quantities. On the
other hand it is obvious that the {\it cocycle} term is
not associated to actual forces since it appears in the
variation of the Lagrangian as a total derivative
irrelevant for the equations of motion.

\setcounter{equation}{0}
\section{Dynamics of a mass-point in the external "gauge" fields}

We want to discuss now the equations of motion of the mass-point acted
upon, as a {\it test particle}, by the newly introduced compensating
external fields. For the sake of simplicity, we will use the
Lagrangian in the standard picture:
\Be
{L}_M^{g} (t) = {m \over \Theta } \left[ {
                      \frac{1}{2} g_{ij} \dot{x}^i  \dot{x}^j
                      +  A_i \dot{x}^i
                      +  A_0 } \right] ~~.
\nome{lagrangianaMat}
\Ee
Upon variation, we get the Euler-Lagrange equations:
\Be
\Ba{rcl}
  \ddot{x}^i + \Gamma^i_{kl} \dot{x}^k \dot{x}^l
           &=& \Ds - { \dot{\Theta} \over \Theta } 
                      \left[ { \dot{x}^i + g^{ij} A_j }\right]
            - g^{ij} \partder{ g_{jl} }{t} \dot{x}^l   \\[1 mm]
           &~& \Ds +  g^{ij} 
            \left[ { \partder{A_0}{x^j} - \partder{A_j}{t} } \right]
             +  g^{ij} 
            \left[ { \partder{A_l}{x^j} - \partder{A_j}{x^l} } \right] 
             \dot{x}^l ~~,
\Ea
\nome{ElLagEq}
\Ee
where:
\Be
\Gamma^l_{ij} = \left\{ \matrix{ l \cr i~j \cr} \right\}
     = \frac{1}{2} {\bf H}_a^l [ {\bf E}^a_{i,j} + {\bf E}^a_{j,i} ]
~~,
\nome{GammaTetrad}
\Ee
is the three-dimensional metric affinity of $g_{ij}$. \newline Let
us stress that the function $\Theta (t)$ which appears in the
Lagrangian (\ref{lagrangianaMat}) and in the equations of motion
(\ref{ElLagEq}) has no real dynamical content. In fact, by
redefining the evolution parameter $t$ and the field $A_0$, $A_i$
according to
\Be
\left\{ {
\Ba{rcl}
T (t)       &\equiv& \Ds \int_0^t d\tau \Theta (\tau ) \\[2 mm]
\tilde{A}_0 &\equiv& \Ds {A_0 \over \Theta^2} \\[2 mm]
\tilde{A}_i &\equiv& \Ds {A_i \over \Theta}   ~~,
\Ea
}\right.
\nome{Tempo}
\Ee
it follows:
\Be
\Ba{rcl}
\Ds \frac{d^2{x}^i}{dT^2} 
+ \Gamma^i_{kl} \frac{d{x}^k}{dT} \frac{d{x}^l}{dT}
           &=& \Ds - g^{ij} \partder{ g_{jl} }{T} \frac{d{x}^l}{dT}  
                   + g^{ij} \left[ { \partder{\tilde{A_l}}{x^j}
                     -\partder{\tilde{A_i}}{x^l}  } \right]
                     \frac{d{x}^l}{dT}  \\[2 mm]
           & & \Ds +  g^{ij} \left[ { \partder{\tilde{A_0}}{x^j}
                     -\partder{\tilde{A_i}}{T}    } \right] ~. 
\Ea
\nome{ElLagTEq}
\Ee

In order to gain a better physical insight, we will discuss the
standard Newtonian gravitational problem in general Galilean
coordinates. Before doing that, let us reproduce in our notations the
Kucha\v{r}'s  classification of the relevant reference frames obtained
by passive coordinate transformations:    \newline
1)~~ {\it non rotating observers}: $\Theta=1$, $A_i=0$; \newline
2)~~ {\it rigid observers}, $\Theta=1$, $g_{ij}=\delta_{ij}$
     (an example belonging to this class is provided by 
      eqs.(\ref{rigidB}) below; Wheeler' definition of {\it Galilean 
     observer} is a subcase of {\it rigid observers}, see eq.(4.22); 
\newline
3)~~ {\it freely-falling observers}: $\Theta=1$, $A_0=0$ \newline
4)~~ {\it Gaussian {\rm (freely falling, non rotating)} observer}:
       $\Theta=1$, $A_i=0$, $A_0=0$ (these are the analogues
       of the {\it Fermi-Walker observers} in general relativity); 
\newline
5)~~ {\it Galilean observers} (rigid, non rotating): 
       $\Theta=1$, $g_{ij}=\delta_{ij}$, $A_i=0$, $A_0=-\varphi$: 
       they are called {\it absolute Galilean observers} by
       Wheeler \cite{Wheel} and {\it inertial} by 
       Levi-Civita \cite{LC}; \newline
6)~~ {\it inertial observers} (rigid, freely falling, non rotating):
       $\Theta=1$, $g_{ij}=\delta_{ij}$, $A_i=0$, $A_0=0$; \newline

  Let $[y^a,t]$ be the coordinates used by a {\it Galilean observer}
(rigid and non-rotating).
Newton's equations can be written:
\Be
m \frac{d^2 y^a}{dt^2} = - m \delta^{ab} \partder{\varphi}{y^b}  ~,
\nome{NewtonE}
\Ee
where $\varphi ({\BMy},t)$ is the gravitational potential
satisfying the Poisson equation 
($\Delta = \delta^{ij}\partial_i\partial_j$)
\Be
\Delta \varphi ({\BMz},t) = 4 \pi G m \delta [{\BMz} - {\BMy}(t) ] ~~.
\nome{PoissonStan}
\Ee
This situation corresponds to:
\Be
 \Theta=1, \qquad g_{ij}=\delta_{ij}, 
\qquad A_0 = - \varphi, \qquad A_i=0 ~. 
\nome{PosA}
\Ee
Eqs.(\ref{NewtonE}) and (\ref{PoissonStan}) are form-invariant under
the group of Galilei transformations (\ref{galiglobal}) that connect
{\it Galilean observers}. In particular, the gravitational potential
is a {\it scalar} under these transformations (see for example
Wheeler \cite{Wheel} \S12.17).  \\[.2 cm]

Let us perform now a passive transformation to an arbitrary coordinate
system $(T,\xi^a)$:
\Be
\left\{
\Ba{rl}
t   &\rightarrow T(t)   \\
y^a &\rightarrow \xi^a ({\BMx}(t),t) ~~.\\
\Ea
\right.
\nome{Passiva}
\Ee
We have:
\Be
\Ba{rcl}
\Ds \frac{d^2 \xi^a}{dT^2} 
        &=& \Ds {1 \over \dot{T}^2 } \left[ {
                \partder{\xi^a}{x^i} \ddot{x}^i } \right] 
                +{\ddot{T} \over \dot{T}^t } \left[ {
                \partder{\xi^a}{x^i} \dot{x}^i
                + \partder{\xi^a}{t}  } \right]  \\[1 mm]
        & & \Ds +{1 \over \dot{T}^2 } \left[ {
                \partder{^2\xi^a}{x^i \partial x^j} \dot{x}^i\dot{x}^j
                +2\partder{^2\xi^a}{x^i \partial t} \dot{x}^i
                + \partder{^2\xi^a}{t \partial t}  } \right]  ~~, 
\Ea
\nome{NewtonTransform}
\Ee
and, after some algebraic manipulations, using Eq.(\ref{GammaTetrad}),
\Be
\Ba{rcl}
 \Ds \ddot{x}^i + \partder{x^i}{y^a} \partder{^2\xi^a}{x^k \partial x^l}
     \dot{x}^i\dot{x}^j  
     &=& \Ds  - \partder{x^i}{y^a} \left[{
               2\partder{^2\xi^a}{x^i \partial t} \dot{x}^i
               + \partder{^2\xi^a}{t \partial t}  } \right]  \\[1 mm]
     & & \Ds - \partder{x^i}{y^a} \left[{
               {\ddot{T} \over \dot{T} } \left[ {
               \partder{\xi^a}{x^i} \dot{x}^i
               +\partder{\xi^a}{t}  } \right] }\right] \\[1 mm]
     & & \Ds - \partder{x^i}{y^a} \dot{T}^2
               \delta^{ab} \partder{\varphi}{y^b}  ~~,
\Ea
\nome{NonInerziale}
\Ee
where $\dot{x}$ means derivation respect the evolution parameter $t$.
The identification  of the equations (\ref{NonInerziale}) and
(\ref{ElLagEq}) expresses the non-relativistic {\it equivalence
principle} which is implicit in the application of the {\it gauge}
technique to a space-time symmetry group.
We have:
\Be
\left\{
\Ba{rl}
    \Theta &= \dot{T} \\[1 mm]
    g_{ij} &= \Ds \delta_{ab} \partder{\xi^a}{x^i}\partder{\xi^b}{x^j}
                  ~~;~{\bf E}^a_i= \partder{\xi^a}{x^i} \\[2 mm]
    g^{ij} &= \Ds \delta^{ab} \partder{x^i}{\xi^a}\partder{x^j}{\xi^b}
                 ~~;~{\bf H}_a^i= \partder{x^i}{y^a} \\[2 mm]
    A_0    &= \Ds  -  \varphi \dot{T}^2
                   + \frac{1}{2} \delta_{ab} \partder{\xi^a}{t} 
                                             \partder{\xi^b}{t} 
\\[2 mm]
    A_i    &= \Ds \delta_{ab} \partder{\xi^a}{x^i} \partder{\xi^b}{t} 
~~.
\Ea
\right.
\nome{NoninerzialeB}
\Ee
Note that
\Be
   \tilde{A} = \frac{1}{\Theta^2} ( A_0 - g^{ij} A_i A_j ) = - \varphi
\nome{defA}
\Ee
Then this last combination of the fields ${A}_0$ and $A_i$ has 
to be identified with Newton's potential, while the $A_i$'s
are inertial fields.

      Let us remark that, while the Poisson equation (\ref{PoissonStan})
      can be rewritten in the new coordinates
      $z^b \rightarrow {\xi}^b (\primato{{\BMz}},t)$
      in the form
      \Be
      \frac{1}{\sqrt{g}} \partder{}{\primato{z}^i}
      \left[ \sqrt{g} g^{ij} \partder{\varphi}{\primato{z}^j}
      \right] =   4 \pi G m \delta [\primato{\BMz} - {\BMx}(t) ]
      ~~,
      \nome{PoissonGen}
      \Ee
      this equation does not determine the gravitational potential 
      since this latter is not a geometrical object and its functional
      form must depend on other fields. 
      Therefore, the transformed Poisson equation
      (\ref{PoissonGen}) cannot be considered the equation 
      for the field $A_0$ 
      in an arbitrary reference frame and its integration cannot 
      give rise to a term like
      $\frac{1}{2} \delta_{ab} \partder{\xi^a}{t} \partder{\xi^b}{t}$,
      as it should be according to Eq.(\ref{NoninerzialeB}). Instead,
it is the quantity $\tilde{A}$ of Eqs.(\ref{defA}) that plays
now the role of the Newton's potential $\varphi$. Then
Eqs.(\ref{PoissonGen}) can be rewritten in the form
      \Be
      \frac{1}{\sqrt{g}} \partder{}{\primato{z}^i}
      \left[ \sqrt{g} g^{ij} \partder{\tilde{A}}{\primato{z}^j}
      \right] =   - 4 \pi G m \delta [\primato{\BMz} - {\BMx}(t) ]
      ~~.
      \nome{PoissonGenA}
      \Ee

The existence of the {\it cocycle} terms in the transformation rules
of the potentials (see Eqs.(\ref{tranCampGau})) does not invalidate
the {\it non-relativistic equivalence principle} since, as shown at
the end of the previous section, it affects only the definition of the
total energy in different coordinate systems. The dynamics of a
mass-point in the above external fields  can be made
Galilean-generally-covariant only if the second Galilean Casimir 
invariant ($E-P^2/2m$) vanishes, as in the flat case.

Let us consider now various cases of passive coordinate
transformations. For instance the kinematical group of passive
coordinate transformations for the {\it rigid observers} (i.e., an
arbitrary rigid but rotating frame). First, there are the {\it
Galilean coordinate transformations} (according to Wheeler's
terminology), which preserve the rigid and non-rotating character of
the coordinates:
\Be
\left\{
\Ba{rclcrcl}
   t &\rightarrow& t - \varepsilon  &,& \varepsilon &=& \mbox{cost.} \\
 y^a &\rightarrow& \xi^a ({\BMx},t)
               = {\rm R}^a_i  x^i + \varepsilon^a (t)
                 ~~~ &,& {\rm R}^a_i  &=& \mbox{cost.} ~~.
\Ea
\right.
\nome{GalRigid}
\Ee
We have now  $\Theta = 1$, $g_{ij}=\delta_{ij}$, $A_0 = - \varphi +
\half \delta_{ab} \dot\varepsilon^a\dot\varepsilon^b$, $A_i =
\delta_{ab} {\rm R}^a_i \dot\varepsilon^b$, and Eq.(\ref{NewtonE}) is
correspondingly modified.

These results can be obtained as a particular case ($ \dot{\rm R}^a_i
= 0$) of the passive coordinate transformation corresponding to an
{\it arbitrary rigid} and {\it rotating} motion
\Be
\left\{
\Ba{rl}
   t &\rightarrow t - \varepsilon \\
 y^a &\rightarrow \xi^a ({\BMx},t)
               = {\rm R}^a_i (t) x^i + \varepsilon^a (t) ~~.\\
\Ea
\right.
\nome{rigid}
\Ee
In this case we have:
\Be
\left\{
\Ba{rl}
\Ds \partder{ \xi^a }{ x^j }&\equiv {\bf E}^a_j ({\BMx},t)
                            = {\rm R}^a_j (t)  \\[2 mm]
\Ds \partder{ \xi^a }{ t } &= \dot{\rm R}^a_j (t) x^j
                             + \dot{\varepsilon}^a (t) \\[2 mm]
\Ds \partder{^2 \xi^a }{ x^j \partial t }
                           &= \dot{\rm R}^a_j (t)  \\[2 mm]
\Ds \partder{^2 \xi^a }{ t^2 }
                           &= \ddot{\rm R}^a_j (t)  x^j
                             + \ddot{\varepsilon}^a (t) ~~,
\Ea
\right.
\nome{rigidA}
\Ee
and Eqs.(\ref{NoninerzialeB}) become:
\Be
\left\{
\Ba{rcl}
    \Theta &=& 1 \\[1 mm]
    g_{ij} &=& \Ds \delta_{ij}  ~;~ \partder{ g_{ij} }{t} = 0
                         ~;~ \Gamma^k_{ij} = 0  \\[2 mm]
    A_0    &=& \Ds -  \varphi + \frac{1}{2} \delta_{ab}
                 [ \dot{{\rm R}}_j^a x^j + \dot{\varepsilon}^a ]
                 [ \dot{{\rm R}}_j^b x^j + \dot{\varepsilon}^b ] 
\\[2 mm]
    A_i    &=& \Ds \delta_{ab} {\rm R}_i^a 
                 [ \dot{{\rm R}}_j^b x^j + \dot{\varepsilon}^b ] ~~.
\Ea
\right.
\nome{rigidB}
\Ee
Then, the equations of motions take the form:
\Be
\Ba{rcl}
 \ddot{x}^l  
   &=& \Ds \delta^{lk} \left\{{
           \partder{ A_k }{t} - \partder{ A_0 }{x^k}
           +  \left[ { \partder{ A_k }{x^j} - \partder{ A_j }{x^k} }
                     \right]  \dot{x}^j  }\right\}  
\\[2.4 mm]
   &=& \Ds \delta^{lk} \left\{{
     - \partder{ \phi }{x^k}
     - \delta_{ab} {\rm R}^a_k \ddot{\varepsilon}^b
     - 2 \delta_{ab} {\rm R}^a_k \dot{\rm R}^b_j \dot{x}^j}\right. 
\\[2 mm]
   & & \Ds ~~~~~~\left. {
     -\frac{1}{2} \delta_{ab}\left[{ {\rm R}^a_k \ddot{\rm R}^b_j
                             + \ddot{\rm R}^a_k {\rm R}^b_j  }
                             \right]  x^j
     -\frac{1}{2} \delta_{ab}\left[{ {\rm R}^a_k \ddot{\rm R}^a_j
                             -\ddot{\rm R}^a_k {\rm R}^a_j  }
                             \right]  x^j
     } \right\} ~~.      
\Ea
\nome{EqMotRig}
\Ee
Since the angular velocity vector can be expressed as
\Be
\omega^k (t) = \frac{1}{2} \epsilon^{krs}
                \delta_{ab} {\rm R}^a_r (t) {\rm \dot{R}}^b_s (t)
~~,
\nome{DefOmega}
\Ee
the physical meaning of the various terms can be immediately
identified as follows:
\Be
\Ba{rcl}
\Ds m [ \partder{A_k}{x^j} - \partder{A_j}{x^k} ] \dot{x}^j
   &=& 2 m {\rm R}^a_j (t) \dot{\rm R}^a_k (t) \dot{x}^j
      = 2 m [ \vec{\omega} \wedge \dot{\vec{x}} ]_k
        ~~\hfill\mbox{(Coriolis~force)} \\[1.5 mm]
\Ds m [ \partder{A_k}{t} - \partder{A_0}{x^k} ]
   &=& \Ds - m \partder{\varphi}{x^k}     
             ~~\hfill\mbox{(gravitational~force)} \\[1.5 mm]
   & & \Ds - m \delta_{ab} {\rm R}^a_k \ddot{\varepsilon}^b
             ~~\hfill\mbox{(translational~inertial~force)} \\[1.5 mm]
   & & \Ds + m [ \vec{\omega} \wedge ( \vec{\omega} \wedge \vec{x} ) ]_k
             ~~\hfill\mbox{(centrifugal~force)} \\[1.5 mm]
   &~& \Ds + m [ \dot{\vec{\omega}} \wedge \vec{x}  ]_k ~~.
             ~~\hfill\mbox{(Jacobi~force)} 
\Ea
\Ee
Finally, the equations of motion, which are the transformed of Newton's
equations (\ref{NewtonE}) in this generalized
reference frame, are:
\Be
m \ddot{x^k} = - m \delta^{kl} \partder{ \varphi}{x^l}
          - m \delta^{kl} {\rm R}^a_l \ddot{\varepsilon}^a
          + m [ \vec{\omega} \wedge ( \vec{\omega} \wedge \vec{x} ) ]^k
          + 2 m [ \vec{\omega} \wedge \dot{\vec{x}} ]^k
          + m [ \dot{\vec{\omega}} \wedge \vec{x}  ]^k
~~.
\nome{NonInertiaNew}
\Ee
On the other hand the Kinematical group of the subclass of {\it rigid
observers}, called {\it Galilean} by Wheeler, is defined by
Eqs.(\ref{GalRigid}). Finally, the Kinematical group of {\it Galilean
observers} (Wheeler's {\it absolute Galilean}) is defined by
$R_i^a={\rm const}$ and $\varepsilon^a={\rm const}$.

When $\dot{\rm R}^a_i = 0$, the modification of Eq.(\ref{NewtonE}) is
due only to the {\it translational-inertial force}. Of course,
corresponding to a transformation of the form (\ref{Passiva}), metric,
affinity, and {\it dreibeins} are equivalent to the global flat ones
and therefore do not represent true  additional dynamical variables.
It will be interesting to see whether it is  possible to build up a
Galilean theory in which the metric and the fields $A_i$ assume an
intrinsic dynamical content. This question will be dealt with in the
following Sections.

\setcounter{equation}{0}
\section{Galilean limit of the relativistic mass-point theory}

In order to understand the generality of the results so far obtained,
it is profitable to make recourse to the axiomatic formulation of the
so-called {\it Newtonian space-time structures} and to  reconsider our
formulation as a suitable non-relativistic limit of a Poincar\'e
invariant theory. Axiomatic formulations of the possible geometries of
Newtonian space-times has been introduced by
Havas \cite{Havas}, Trautman \cite{TrautA}\rlap,  
K\"{u}nzle \cite{Kunz}\rlap, and Kucha\v{r} \cite{Kuch}\rlap.

Following K\"{u}nzle, we define:

\begin{itemize}
\item   A {\bf Galilei structure} over a four-dimensional manifold $\Vc$
        is a pair $(h^{\mu\nu},t_\nu)$, where $h^{\mu\nu}$ is a
        symmetric covariant tensor of rank 3 and $t_\mu$ is a 1-form
        having the property that is the generator of 
        the kernel of $h^{\mu\nu}$,
        $\forall x\in \Vc$. \newline
        The triple $(\Vc ;h^{\mu\nu},t_\nu)$  is called a 
        {\bf Galilei Manifold}.
        A vector $u^\mu$ is called a {\it unit time-like vector} if
        $u^\mu t_\mu =1$, and a contravariant tensor is called
        {\it space-like} if it vanish when contracted with $t_\mu$
        on any index.

\item   A linear symmetric connection is called {\bf a Galilei 
        connection} $\nabla$ if it is defined on a Galilei manifold  
        $(\Vc ;h^{\mu\nu},t_\nu)$ and satisfies
        \Be
        \nabla_\rho h^{\mu\nu} = \nabla_\rho t_\mu = 0  ~~.
        \Ee
        It can be shown that such a connection exist if and only 
        if $t_\mu$ is a closed 1-form and it is uniquely defined 
        up to an arbitrary two-form $\chi$ according to:
        \Be
          \Gamma^\alpha_{\beta\gamma} = 
                 \stackrel{u~~~}{\Gamma^\alpha_{\beta\gamma}}
                 +t_\beta h^{\alpha\rho} \chi_{\rho\gamma}
                 +t_\gamma h^{\alpha\rho} \chi_{\rho\beta} ~~,
        \nome{ConnesGal}
        \Ee
        where $\Gamma^\alpha_{\beta\gamma}$ are the connection 
        coefficients and
        \Be
          \stackrel{u~~~}{\Gamma^\alpha_{\beta\gamma}} \equiv
             h^{\alpha\rho} \left[
                 \half h_{\beta\mu}h_{\gamma\nu} h^{\mu\nu}_{,\rho}
                 +u^\sigma_{,\sigma} h_{\rho (\beta} t_{\gamma )}
                 \right]
            - h_{\rho (\beta} h_{,\gamma )}^{\alpha\rho}
            - t_{\beta} u_{,\gamma )}^{\alpha} ~;
        \Ee
        here $u^\alpha$ ($u^\alpha t_\alpha =1$) is an arbitrary 
        given time-like unit vector field, and $\gamma_{\alpha\beta}$
        is the associated covariant space metric defined by
        \Be
        \gamma_{\alpha\beta} u^\beta = 0
        \qquad
        \mbox{and}
        \qquad
        \gamma_{\alpha\rho} \gamma^{\rho\beta} 
            = \delta_{\alpha}^{\beta} - t_{\alpha} u^{\beta} 
        ~~.
        \Ee
\item  A symmetric Galilei connection $\nabla$ is called 
       {\bf Newtonian}, and the quadruple 
       $(\Vc ;h^{\mu\nu},t_\nu ;\Gamma)$ is 
       called correspondingly a {\bf Newtonian Manifold},
       if the two-form $\chi$ in Eq.(\ref{ConnesGal}) is closed.

\item  Newtonian manifolds in which $t_\mu$ is also {\it exact} will be
       called {\it Special Newtonian Manifolds}: the hypothesis 
       that $t_\mu$ be an {\it exact} one form 
       implies the existence of a {\it global absolute time}.
       The standard Newtonian connection is indeed obtained by choosing 
       $\chi_{\alpha\beta}=\varphi_{,\alpha}t_\beta 
                           -\varphi_{,\beta} t_\alpha $
       where $\varphi$ is the Newton's potential.

\item  Finally, the covariant {\it space metric} can be introduced by
       means of the relations:
       \Be
       \Ba{rl}
       h_{\mu\nu} h^{\nu\rho} &= \delta^\rho_\mu - t_{\mu} u^\rho \\
       h_{\mu\nu} u^\nu       &= 0 ~~~.\\
       \Ea
       \nome{MetricaBassa}
       \Ee
\end{itemize}

It is clear that the contravariant metric allows to assign lengths to
{\it space-like} vectors but no lengths whatsoever to {\it time-like} vectors.
These being the premises, we have:
\begin{itemize}
\item[1)] the field equations, i.e. Newton's law of gravitation
          $\Delta\varphi = 4\pi G \rho$ ($\rho$ mass density), can be
          rewritten as 
          $R_{\alpha\beta}=4\pi G\rho t_{\alpha}t_{\beta}$;
\item[2)] the matter Lagrangian for a mass-point, in the
          given Newtonian space-time, is (see Ref. \cite{Kuch}):
\Be
\Ba{rcl}
A_M &=& \Ds \int d\lambda L_M(\lambda )  ~~, \\[1 mm]
L_M &=& \Ds \frac{m}{2}  \left( {1 \over t_{\rho} \primato{x}^\rho }
                         \right)
                     h_{\mu\nu} \primato{x}^\mu \primato{x}^\nu
       - m \varphi \left( { t_{\rho} \primato{x}^\rho } \right)   ~~. 
\Ea
\nome{KucharLag}
\Ee
Note that only within {\it Special Newtonian Manifolds}
the first of Eqs.(\ref{KucharLag})
can be rewritten in the form $A_M=\int dT \tilde{L}_M(T)$.
\end{itemize}

At this point, if our fields $g_{ij}$, $\Theta$, $A_i$, $A_0$
are supposed to be resident within a {\it Special Newtonian} 
space-time, it is easy to make all the identifications 
which are relevant to our
formulation.
In particular, in a convenient coordinate chart $x^\mu=[t;x^k]$
of the {\it Special Newtonian} space time, 
we have:
\Be
\left\{ {
\Ba{rclcrcl}
     \varphi &=&\Ds  - {1 \over \Theta^2} 
             \left({A_0 - \frac{1}{2} g^{ij} A_i A_j
             }\right)
& &          & &  \\[2 mm]
     t_{\mu} &=& [ \Theta (t) ; \vec{0} ] 
             =[\dot{T} ; \vec{0} ] = t_{,\mu}
& &  u^\mu   &=&\Ds  {1 \over \Theta} [ 1 ; - g^{ij} A_j ] \\
     h^{\mu\nu} &=&\Ds  \left| \matrix{ 0 & 0 \\ 0 & g^{ij} \\}
             \right|
& &  h_{\mu\nu} &=& \left| \matrix{ g^{rs} A_r A_s & A_j \\
                                       A_i & g_{ij} \\}
             \right| \\[2 mm]
\left[ {\nabla_0} \right]_0^i
             &=&\Ds  g^{ij} \left[ \partder{A_j}{t} 
                    - \frac{1}{\Theta} \partder{\Theta}{t} A_j
                    -\partder{A_0}{x^j} \right]
& &\left[ {\nabla_0} \right]_0^0
                &=&\Ds  {1 \over \Theta} {d\Theta \over dt} \\[2 mm]
\left[ {\nabla_0} \right]_k^i
                &=&\Ds  \half g^{ij} \left[ A_{j,k} - A_{k,j}
                    + \partder{g_{jk}}{t} \right]
& &\left[ {\nabla_0} \right]_k^0
                &=&  0  \\[2 mm]
\left[ {\nabla_j} \right]_k^i
                &=&  \Gamma^i_{jk}
& &\left[ {\nabla_j} \right]_k^0
                &=&  0  ~~~. \\
\Ea
}\right.
\nome{KucharIdent}
\Ee
so that the Lagrangian (\ref{KucharLag}) becomes:
\Be
\Ba{rcl}
{L}_M^g &=& \Ds \frac{1}{t_{\rho} \primato{x}^\rho} 
                \frac{m}{2} h_{\mu\nu}
                \primato{x}^\mu \primato{x}^\nu
                - m \varphi t_{\rho} \primato{x}^\rho \\[1 mm]
      &=& \Ds {m \over \Theta \primato{t} }
              \left[ {
               \frac{1}{2} g_{ij} \primato{x}^i  \primato{x}^j
                          +  A_i  \primato{x}^i  \primato{t}
                          +  A_0  \primato{t}    \primato{t}
                         } \right] ~~, 
\Ea
\nome{lagrangianaMatCon}
\Ee
which is precisely our Lagrangian (\ref{GauLagPar})
(see also Eqs.(\ref{NoninerzialeB})).

  We are now in a position to perform (see for example
\cite{GpP}) in a general way the non-relativistic limit of the
mass-point Lagrangian that is invariant with respect to the
Poincar\'e group {\it gauged} {\it \'a la Utiyama}. We have: \Be
L^{Rg}_M = - m c \sqrt{ - g_{\mu\nu} \primato{x}^\mu \primato{x}^\nu }
~~,
\nome{LagrangianaR}
\Ee
where we we have assumed the $(-1,+1,+1,+1)$ convention for
the signature of the metric $g_{\mu\nu}$.

In order to perform the limit, we must write the explicit dependence
of the metric tensor on $c^2$ ($c$ velocity of light). As shown by 
Dautcourt \cite{Daut}\rlap, Ehlers \cite{EHL} and
K\"unzle \cite{Kunz1}\rlap ,~ the correct parameterization to start with
is:
\Be
 g_{\mu\nu} = - c^2  t_{\mu} t_{\nu} + \stackrel{\circ}{g}_{\mu\nu} ~,
\nome{gbasso}
\Ee
and, for the inverse metric,
\Be
g^{\mu\nu} = h^{\mu\nu} - \frac{1}{c^2} \kappa^{\mu\nu}
~~.
\Ee
From the relation
\Be
 g^{\mu\nu}g_{\nu\rho} = \delta^\mu_\rho ~~,
\Ee
we see that, at first order in $1/c^2$, the following identities are
fulfilled:
\Be
\left\{
\Ba{lcl}
 h^{\mu\nu}t_{\nu} &=& 0 \\
 h^{\mu\nu}\stackrel{\circ}{g}_{\nu\rho} &=& \delta^\mu_\rho
                 -t_{\rho} \kappa^{\mu\nu} t_{\nu} ~~~.\\
\Ea
\right.
\nome{RelSvilM}
\Ee

Therefore, in the limit $c\rightarrow +\infty$, we must identify
$h^{\mu\nu}$ with the Newtonian {\it space metric} and
$t_{\mu}t_{\nu}$  with the Newtonian {\it time metric}.

All the objects of the standard ({\it Special Newtonian Manifold} 
$t_\mu = t_{,\mu}$) Newtonian theory can then be reconstructed  
by the relations:
\Be
\left\{
\Ba{rcl}
  u^\mu t_{,\mu} &=& 1 \\
  u^\nu\stackrel{0}{g}_{\nu\rho} &=&  2~\varphi~t_{,\rho} \\
  h^{\mu\nu}\stackrel{0}{g}_{\nu\rho} &=&  \delta^\mu_\rho
                   -t_{,\rho}~ \kappa^{\mu\nu}~ t_{,\nu} \\
  h_{\mu\nu} &=& \stackrel{\circ}{g}_{\mu\nu} 
                 - 2 \varphi~ t_{,\mu} t_{,\nu} 
\Ea
\right.
\nome{pseudometr}
\Ee
that give the expressions $\varphi,u^\mu,h^{\mu\nu},h_{\mu\nu}$ 
as functions of $t_{,\mu}$ and $\stackrel{\circ}{g}_{\mu\nu}$. 
The fields so defined automatically fulfill all the 
conditions required for the underlying space-time
to be a {\bf Newtonian manifold}. Within the coordinate chart 
used in Eqs.(\ref{KucharIdent}), it results
\Be
\stackrel{\circ}{g}_{\mu\nu} 
  =  h_{\mu\nu} + 
    \left|
    \matrix{ 2 \varphi \Theta^2 & 0 \cr 0 & 0 \cr }
    \right|
~~.
\Ee

By inserting the metric tensor (\ref{gbasso}) into the relativistic
Lagrangian (\ref{LagrangianaR}) and taking into account
Eqs.(\ref{pseudometr}), the expansion in terms of $c^2$ becomes:
\Be
L = c^2  \left[ { -  m t_{,\nu}~ \primato{x}^\nu } \right]
    + \left[ {
         \frac{1}{t_{,\rho} \primato{x}^\rho} \frac{m}{2} h_{\mu\nu}
                           \primato{x}^\mu \primato{x}^\nu
    - m ~\varphi~ t_{,\rho} \primato{x}^\rho
       } \right]
    + O(1/c^2)  ~~.
\nome{SvilLagG}
\Ee
We see that the zero$^{\mbox{\underline{\rm th}}}$  order part in
$c^2$, reproduces the matter Lagrangian (\ref{KucharLag}), while the
coefficient of $c^2$ (which in the adapted frame (\ref{KucharIdent})
becomes $\left[
-m\frac{dT}{dt}\frac{dt}{d\lambda}=-m\Theta\frac{dt}{d\lambda}
\right]$) is, within the standard picture, the {\it central charge}
$-m\Theta$ of the extended Galilei Group ($-m$ if one chooses $t=T$).

\setcounter{equation}{0}
\section{Searching for field equations}

The expressions of the fields appearing in
Eq.(\ref{NoninerzialeB}) are the generic-frame
expressions of external fields resident in a spatially
flat Newtonian space. On the other hand, having in mind
the relations among the three-dimensional {\it gauge}
fields that we have introduced and the four-dimensional
metric, as contained in
Eqs.(\ref{KucharIdent},\ref{gbasso},\ref{pseudometr}),
it seems natural to look for dynamical field equations
in three dimensions by exploiting some limiting
procedure over a four-dimensional theory. Now, let us
observe that the contraction $c\rightarrow
\infty$ from the Poincar\'e algebra to the extended Galilei
algebra is a well defined procedure in the case of the
single mass-point that we have studied in the previous
section, due to the fact that the action is a
Poincar\'e invariant: it amounts indeed to a uniquely
defined contraction on the scalar representation. On
the other hand, the field equations of motion do not
transform like a scalar representation and it is well
known that contracting a non-trivial representation is
a delicate matter: a-priori, different contractions of
the same equations could result. Therefore taking into
account the fact that the Galilean matter Lagrangian
(\ref{KucharLag}) is nothing but the zero$^{\rm th}$
order term of the $1/c^2$ expansion (\ref{SvilLagG}) of
the general relativistic mass-point Lagrangian, we will
build up the wanted Galilean variational problem by
means of the zero$^{\rm th}$-order term of the $1/c^2$
expansion of the full four-dimensional Einstein-Hilbert
action for the gravitational field plus a single
mass-point \cite{Landau}: 
\Be
\Sc = \Sc_F + \Sc_M 
    = \frac{c^3}{16\pi G} \int d^4\!z \sqrt{-{^4\! g}} \;\;{^4\! R} 
      - m c \int d\lambda 
            \sqrt{ - g_{\mu\nu} \primato{x}^\mu\primato{x}^\nu } 
\nome{EinsHilb} 
\Ee

First of all, we shall restrict ourselves to globally
hyperbolic space-time manifolds for which a {\it
global} 3+1 splitting exists.  The associated action
principle will be formulated by re-expressing the
Einstein-Hilbert Action (\ref{EinsHilb}) in the form
given by Arnowitt-Deser-Misner \cite{ADM} and
De-Witt \cite{DeWitt67}\rlap .~ This is tantamount to
assume the existence of an absolute-time foliation of
space-time $(t=t(z^\mu ))$ and of a global coordinate
system in which $t_{,\mu}=(\Theta (t),\vec{0})$ as in
Eq.(\ref{KucharLag}). In such a system, $g_{\mu\nu}$
must be of the form of Eq.(\ref{gbasso}) with $t_\mu =
t_{,\mu}$ and the integration measure $d^4\!z$ can be
rewritten as $dtd^3\!z$.

Then, owing to
Eqs.(\ref{KucharIdent},\ref{gbasso},\ref{pseudometr}),
we can put
\Be
\Ba{rl}
g_{\mu\nu} &= - c^2 t_{,\mu} t_{,\nu} + \stackrel{\circ}{g}_{\mu\nu} 
\\[2 mm]
           &= - c^2 \left| \matrix{ \Theta^2 &  0 \cr
                                       0     &  0 \cr } \right|
              +\left| \matrix{ 2 A_0 & A_j    \cr
                                A_i  & g_{ij} \cr } \right|
              +{1 \over c^2}
                \left| \matrix{ 2 \alpha_0 & \alpha_j    \cr
                                  \alpha_i  & \gamma_{ij} \cr } \right|
              +{1 \over c^4}
                \left| \matrix{  ...  & ...    \cr
                                 ...  & \beta_{ij} \cr } \right|
              +O({1 \over c^6}) \\
\Ea
\nome{IdenKuchB}
\Ee
and
\Be
\Ba{rcl}
g^{\mu\nu}
    = \left| \matrix{ 0 &  0      \cr
                             0 &  g^{ij} \cr } \right|
    -{1 \over c^2 \Theta^2}
     \left| \matrix{  1  & - g^{jk} A_k    \cr
        -g^{ik} A_k & g^{ik}g^{jl} ( \gamma_{kl} + A_k A_l) \cr } 
     \right|
    -{1 \over c^4 \Theta^4}
         \left| \matrix{  A  & ... \cr
                         ... & ...  \cr } \right|
              +O({1 \over c^6}) ~. \\
\nome{IdenKuchA}
\Ea
\Ee
While some terms of order $c^{-4}$ do contribute to the
zero$^{\rm th}$ order term we are interested in, {\it
no term of order $c^{-6}$ can survive the contraction},
so that they will be ignored from now on. Note that the
parameterization (\ref{IdenKuchB}) of $g_{\mu\nu}$ is
the most general one which can be {\it locally}
connected to the flat Minkowski metric $\eta_{\mu\nu}$
by means of a general coordinate transformation. By
relaxing this last restriction, i.e. allowing for
$\Theta = \Theta (t,{\BMz})$, one obtains a
parameterization of $g_{\mu\nu}$, which needs also a
Weyl transformation to be {\it locally} connected with
$\eta_{\mu\nu}$. This corresponds in some way to allow
for a classical analogue of the {\it dilaton} degree of
freedom dynamically interacting with the other fields
by paying the price of abandoning the Galilean
interpretation of $\Theta(t)$ given in the previous
section. Yet, this additional liberty allows for interesting
results and will be exploited explicitly
in the second paper of the present series.

Following Arnowitt, Deser and Misner \cite{ADM}\rlap ,\ we
write the covariant and the contravariant
four-dimensional metric in the form:
\Be
\Ba{rl}
^4\! g_{\mu\nu} &=  \left|
              \matrix{ -N^2+ \gTRE^{ij} N_i N_j  &  N_j   \cr
                           N_i              & \gTRE_{ij} \cr }
               \right| \\[5 mm]
^4\! g^{\mu\nu} &=  \left|
          \matrix{ - \frac{1}{N^2}   & \frac{\gTRE^{jk} N_k}{N^2}  \cr
                \frac{\gTRE^{il}N^l}{N^2} &\gTRE^{ij}
                - \frac{\gTRE^{il}N_l \gTRE^{jk}N_k}{N^2}  \cr }
          \right| ~~, 
\Ea
\nome{eqADM}
\Ee
where $\gTRE_{ij} \gTRE^{jk} = \delta^k_i$. Then, neglecting 
surface terms, the action (\ref{EinsHilb}) can be written:
\Be
\Ba{rcl}
\Sc   &=&  \Ds \frac{c^3}{16\pi G} \int dtd^3\! z \sqrt{\gTRE} N
                \left[ { {^3\!}R
                        + \gTRE^{ik}~\gTRE^{jl} 
                          (K_{ij}K_{kl} - K_{ik}K_{jl})
                        } \right]   \\[4 mm]
      & & \Ds + mc \int d\lambda 
              \sqrt{ (N^2- {^3}g^{ij}N_i N_j) \primato{t}\primato{t} 
                      -N_i \primato{x}^i\primato{t} 
                      -{^3}g_{ij} \primato{x}^i \primato{x}^j }
  ~~, 
\Ea
\nome{AzADM}
\Ee
where:
\Be
\left\{
\Ba{ccl}
  \gTRE    &=& \det{\gTRE_{ij}} \\[2 mm]
 {^3\!}\Gamma^k_{ij}
           &=& \Ds \gTRE^{kl} \frac{1}{2} ( \gTRE_{il,j}+ \gTRE_{jl,i}
                               -\gTRE_{ij,l} ) \\[2 mm]
  {^3\!}R_{ij} &=& \Ds {^3\!}\Gamma^k_{ik,j} - {^3\!}\Gamma^k_{ij,k}
           +{^3\!}\Gamma^k_{il} {^3\!}\Gamma^l_{jk}
           -{^3\!}\Gamma^k_{ij} {^3\!}\Gamma^l_{lk}  \\[2 mm]
  {^3\!}R  &=& \Ds \gTRE^{ij} ~{^3\!}R_{ij}   \\[1 mm]
  K_{ij}   &=& \Ds \frac{1}{2N} \left( {
           {^3\!}\nabla_i N_j
           +{^3\!}\nabla_j N_i
           -\partder{\gTRE_{ij}}{t}
           }\right)
~~\mbox{(\it extrinsic curvature)}~,
\Ea
\right.
\nome{definitions}
\Ee
having denoted by ${^3\!}\nabla$ the covariant three-space derivative
with respect to the Christoffel connection of $\gTRE_{ij}$
In terms of the notation we have previously introduced for
the expansion of the covariant four-dimensional metric, we have
\Be
\left\{.
\Ba{rl}
  \gTRE_{ij} &\Ds \equiv g_{ij} + \frac{1}{c^2} \gamma_{ij}
                            + \frac{1}{c^4} \beta_{ij} 
                            +O({1 \over c^6})  \\[3 mm] 
   {^3\!}R   &\Ds \equiv R + \frac{1}{c^2} R_1(g_{ij},\gamma_{ij})
               +\frac{1}{c^4} R_2(g_{ij},\gamma_{ij},\beta_{ij}) 
               +O({1 \over c^6}) \\[3 mm] 
   N_i       &\Ds \equiv A_i  + \frac{1}{c^2} \alpha_i
                          + \frac{1}{c^4} \beta_i 
                          +O({1 \over c^6}) \\[3 mm] 
   N^2       &\Ds \equiv c^2 \Theta^2 - 2 A
                + \frac{2}{c^2} \left[{
                          \alpha_0 - g^{ij} \alpha_i A_j
                        - \frac{1}{2} \gamma_{rs} g^{ri} g^{sj} A_i A_j
                          }\right]
                +O({1 \over c^4})     \\[3 mm] 
   N K_{ij}  &\Ds \equiv {^3\!}B_{ij} 
               = B_{ij} + \frac{1}{c^2} B_{ij}^{\SSs{(1)}}  
                        + O({1 \over c^4})   ~~,
\Ea
\right.
\nome{IdentADMK}
\Ee
where we have defined:
\Be
\Ba{rcl}
 A      &=&  A_0 -\frac{1}{2}g^{ij}A_iA_j   \\
 B_{ij} &=& \frac{1}{2} [{^3\!}\nabla_i A_{j}+{^3\!}\nabla_j A_i 
                          -\partder{g_{ij}}{t} ] \\
 B_{ij}^{\SSs{(1)}}
   &=& \frac{1}{2} [{^3\!}\nabla_i \alpha_{j}+{^3\!}\nabla_j \alpha_i
                        -\partder{\gamma_{ij}}{t}
                        - A_k g^{kl} ( {^3\!}\nabla_j \gamma_{il}
                                      +{^3\!}\nabla_i \gamma_{lj}
                                      -{^3\!}\nabla_l \gamma_{ij} ) ] ~.
\Ea
\nome{6.8}
\Ee
\noindent
Then, inserting eqs.(\ref{definitions}-\ref{6.8}) into
eq.(\ref{AzADM}), it follows:
\Be
\Ba{rcl}
\Sc   &=&\Ds \Sc_F + \Sc_M =  \\[3 mm]
     &=&\Ds ~~c^4 \left\{ {
            \frac{1}{16\pi G} {\Ds \int} dtd^3z \sqrt{g} \Theta  R
                     }\right\} 
\\[4.5 mm]
     & &\Ds +c^2 \left\{ {
         \frac{1}{16\pi G} {\Ds \int} dtd^3z \sqrt{g}
         \left[ { \Theta R_1 + \frac{\Theta}{2} g^{ij} \gamma_{ij} R
         - \frac{A}{\Theta} {R}
         + \Theta g^{ik} g^{jl} (B_{ij}B_{kl} - B_{ik}B_{jl})
         } \right]  }\right. \\
     & &\Ds  ~~~~~~\left.{ - m  {\Ds \int} d\lambda \Theta \primato{t}
                   }\right\} 
\\[4.5 mm]
     & &\Ds +~~~\left\{ {
            \frac{1}{16\pi G} {\Ds \int} dtd^3z \sqrt{g}
                \left[{ \Theta R_2
                        + \frac{1}{2} g^{ij} \gamma_{ij} \Theta  R_1
                        + \frac{1}{2} g^{ij} \beta_{ij} \Theta R
                }\right.}\right.   \\[3 mm]
     & &\Ds  ~~~~~~~~~~~~~~
         - \frac{1}{2} \Theta
         \left( g^{ik} g^{jl} - \frac{1}{2} g^{ij} g^{kl} \right)
         \gamma_{ij} \gamma_{kl} R
                 - { A \over \Theta } R_1  \\[3 mm]
     & &\Ds  ~~~~~~~~~~~~~
                        - { 1\over \Theta }
                          \left({ \alpha_0 - g^{ij} A_i \alpha_j
                         +\frac{1}{2} A g^{ij} \gamma_{ij}
                         +\frac{1}{2} \gamma_{ij} g^{il} g^{jm} A_l A_m
                          }\right) R
                        - \frac{A^2}{2\Theta^3} {R}   \\[3 mm]
     & &\Ds  ~~~~~~~~~~~~~
                        + \frac{2}{\Theta}
                          g^{ik} g^{jl} ( B_{ij}B_{kl}^{\SSs{(1)}}
                                         -B_{ik}B_{jl}^{\SSs{(1)}})
                        - \frac{2}{\Theta}
                          g^{ik} g^{jr} \gamma_{rs} g^{sl}
                          (B_{ij}B_{kl} - B_{ik}B_{jl})   \\[3 mm]
     & &\Ds  ~~~~~~~~~~~~~  \left.{\left.{
                        + \frac{A}{\Theta^3}
                          g^{ik} g^{jl} (B_{ij}B_{kl} - B_{ik}B_{jl})
                       } \right] }\right.   \\[3 mm]
     & &\Ds ~~~~~~\left.{\Ds  +m  {\Ds \int} 
                  d\lambda {m \over \Theta \primato{t}}
                  \left[ \frac{1}{2}g_{ij} 
                         (\primato{x}^i + g^{ik} A_k \primato{t} )
                         (\primato{x}^i + g^{ik} A_k \primato{t} )
                             +A_0 \primato{t} \primato{t}   
                       \right]
                   }\right\} 
\\[4.5 mm]
     & &+ O(1/c^2)  ~~~,\\
\Ea
\nome{developT}
\Ee
so that the zero$^{\rm th}$-order term, identified as
the {\it total Action} $\tilde{\Sc}$, is:
\Be
\Ba{rcl}
\tilde{\Sc} &=& \Ds \frac{1}{16\pi G} \int dtd^3z \sqrt{g}
         \left[{ \Theta R_2 + \frac{1}{2} g^{ij} \gamma_{ij} \Theta  R_1
         + \frac{1}{2} g^{ij} \beta_{ij} \Theta R
         }\right.  \\[3 mm]
     & & \Ds ~~~~~~~~~~~~~~ - \frac{1}{2} \Theta
         \left( g^{ik} g^{jl} - \frac{1}{2} g^{ij} g^{kl} \right)
         \gamma_{ij} \gamma_{kl} R
         - { A \over \Theta } R_1  \\[3 mm]
     & & \Ds ~~~~~~~~~~~~~
         - { 1\over \Theta }
         \left({ \alpha_0 - g^{ij} A_i \alpha_j
         +\frac{1}{2} A g^{ij} \gamma_{ij}
         +\frac{1}{2} \gamma_{ij} g^{il} g^{jm} A_l A_m
         }\right) R
         - \frac{A^2}{2\Theta^3} {R}   \\[3 mm]
     & & \Ds ~~~~~~~~~~~~~
                      + \frac{2}{\Theta}
                      g^{ik} g^{jl} ( B_{ij}B_{kl}^{\SSs{(1)}}
                                     -B_{ik}B_{jl}^{\SSs{(1)}})
                      - \frac{2}{\Theta}
                      g^{ik} g^{jr} \gamma_{rs} g^{sl}
                      (B_{ij}B_{kl} - B_{ik}B_{jl})   \\[3 mm]
     & & \Ds ~~~~~~~~~~~~~  \left.{\left.{
                      + \frac{A}{\Theta^3}
                      g^{ik} g^{jl} (B_{ij}B_{kl} - B_{ik}B_{jl})
                      } \right] }\right.   \\[3 mm]
     & & \Ds ~~~~~~ +m  {\Ds \int} d\lambda {m \over \Theta \primato{t}}
                       \left[ \frac{1}{2}g_{ij} 
                       (\primato{x}^i + g^{ik} A_k \primato{t} )
                       (\primato{x}^i + g^{ik} A_k \primato{t} )
                             +A_0 \primato{t} \primato{t}   
                       \right] ~~.
\Ea
\nome{6.10}
\Ee
We see that {\it 27 fields survive the contraction}, namely $\Theta$,
$A$, $A_i$, $g_{ij}$, $\alpha_0$, $\alpha_i$, $\gamma_{ij}$,
$\beta_{ij}$, where $A$ has been used as independent variable instead
of $A_0$; on the other hand $\beta_0$ and $\beta_i$ disappear at this
order.

Let us now look for the invariance of the Action. In section 3
we have shown that, provided we assume
the transformation properties (\ref{tranCampGau}, 
the matter part of the Action (\ref{6.10}) is invariant
under the {\it gauged} Galilei transformation). 
Now, in order to formulate the {\it local
Galilei invariance} for the action (\ref{6.10}),
appropriate {\it gauge} transformations for the fields
$\gamma_{ij}$, $\beta_{ij}$, $\alpha_i$ and $\alpha_0$ must be postulated.
The correct choice is 
\Be
\Ba{rcl}
{\delta} \gamma_{ij} &=& \Ds 
                  - \partder{\tilde\eta^k ({\BMx},t) }{x^i} \gamma_{kj}
                  - \partder{\tilde\eta^k ({\BMx},t) }{x^i} \gamma_{kj}
\\[2 mm]
{\delta} \alpha_0     &=& \Ds 2 \dot{\varepsilon} \alpha_0
                  - \alpha_i \partder{\tilde\eta^i}{t}
\\[2 mm]
{\delta} \alpha_i  &=& \Ds \dot{\varepsilon} \alpha_i
                  - \alpha_j \partder{\tilde\eta^j}{x^i}
                  - \gamma_{ij}  \partder{\tilde\eta^j}{t} ~~,
\Ea
\nome{INV27a}
\Ee
and, indeed, by direct calculation it turns out the the total variation
of the Action (\ref{6.10}) under the transformation $\delta$ defined by
(\ref{transfparGauA}), (\ref{tranCampGau}) and (\ref{INV27a}) is
given by:
\Be
\Ba{rcl}
{\delta} \tilde{\Sc} 
    &=&\Ds  \int dt d^3\! z \left\{ { \dot\varepsilon \tilde{\Lc} 
                     +  \Theta \EL_{A} 
               \left( \partder{\Fc}{t} - A_r g^{rs} \partder{\Fc}{z^s} 
                    \right) 
              +\Theta \EL_{A_i} \partder{\Fc}{z^i}   }\right. \\[2 mm]
           & &\Ds ~~ \left. {
              +\frac{1}{8\pi G} \partder{}{z^i}
                        \left( \frac{\sqrt{g} A}{\Theta^2} 
                               [B^{ij} - ({\rm Tr}B) g^{ij} ]
                               \partder{\Fc}{z^j} 
                               \right)  } \right\} ~~.
\Ea
\nome{INV27}
\Ee
As a matter of fact, this result means that 
a {\it quasi-invariance} of the Galilean total Action (\ref{6.10}holds, 
neglecting surface terms,
{\it modulo} the Euler-Lagrange equations
for the fields $A_i$ ($\EL_{A_i}$) and
$A$ ($\EL_{A}$) that are given by:
\Be
\Ba{rcl}
 \EL_A     &=& \Ds \frac{\delta \tilde{\Sc}}{\delta A} \\
 \EL_{A_i} &=& \Ds\frac{\delta \tilde{\Sc}}{\delta A_i}
              - \frac{\partial}{\partial z^k}
                \frac{\delta \tilde{\Sc}}{\delta \partial_k A_i} \\
\Ea
\Ee
\par
\noindent
Let us remark that this peculiarity is precisely what it should be expected
in the case of a variational principle corresponding to a {\it singular}
Lagrangian.

By analogy to the free mass-point case, the terms $c^4 ... + c^2 ...$
of eqs.(\ref{developT}) could be rewritten as  $c^2(\Mc + c^2 \Nc)$
where  $\Mc+c^2\Nc$ ought to be interpreted as the central charge  of
the {\it asymptotic} Galilei group. To avoid an infinite
central-charge, the theory should, in some sense,  provide the
condition $\Nc=0$ automatically.  The discussion of the {\it
asymptotic} Galilei group will be dealt with in a separate paper. This
analysis will require taking into account the $1/c^2$ expansion of the
neglected surface terms, as they are needed in the evaluation of the
asymptotic Poincar\'e group in the case of asymptotically-flat 
space-times (see
for example ref.\cite{RT}).

\setcounter{equation}{0}

\section[Galilean Covariant Formulations of Newtonian Gravity]
{Galilean Covariant Formulations of Newtonian \\ Gravity}

\subsection{The Newtonian Theory in Arbitrary Reference Frames}

In Section 4 it was shown that the field $\Theta( t)$ has no real
dynamical content since its effect amounts only to a redefinition of
the evolution parameter t in the expression $T(t)=\int_0^t d\tau
\Theta (\tau )$ . It is easy to show that this fact is still true for
its role within the total Action (\ref{6.10}). Indeed, if we
redefine the fields $A_0$, $A_i$, $\alpha_0$ and $\alpha_{i}$ as
follows (see eq.(\ref{Tempo}); from now on $^3\!\nabla$
will be replace by $\nabla$),
\Be
\left\{ {
\Ba{rl}
\tilde{A}_0  &\equiv \Ds {A_0 \over \Theta^2}
                             ~~;\qquad  \tilde{A} 
              \equiv {A   \over \Theta^2}                \\[3 mm]
\tilde{A}_i       &\equiv \Ds {A_i \over \Theta}         \\[3 mm]
\tilde{\alpha}_0  &\equiv \Ds {\alpha_0 \over \Theta^2}
              ~~;\qquad  
\tilde{\alpha}_i  \equiv {\alpha_i \over \Theta}  \\[3 mm]
\tilde{B}_{ij} &\equiv \Ds  {{B}_{ij}  \over \Theta}
                       =  \frac{1}{2} \left[ 
                               \nabla_i \tilde{A}_{j}
                             + \nabla_j \tilde{A}_i
                             - \partder{g_{ij}}{T} 
                             \right]  \\
\tilde{B}_{ij}^{\SSs{(1)}} 
       &\equiv \Ds  {{B}_{ij}^{\SSs{(1)}}  \over \Theta}
                       =  \frac{1}{2} \left[ 
                               \nabla_i \tilde{\alpha}_j
                             + \nabla_j \tilde{\alpha}_i
                             - \partder{\gamma_{ij}}{T} 
                             - \tilde{A}_k g^{kl} 
                               ( \nabla_i\gamma_{jl}
                                +\nabla_j\gamma_{il}
                                -\nabla_l\gamma_{ij} ) 
                             \right] ~~, \\
\Ea
}\right.
\nome{TempoN27}
\Ee
the Action (\ref{6.10}) becomes:
\Be
\Ba{rcl}
\tilde{\Sc} &=& \int dTd^3z \tilde\Lc \\
     &=& \Ds \frac{1}{16\pi G} \int dTd^3z 
                \bigg[  \tilde{R}_2  
                        -  \tilde{A}  \tilde{R}_1 \\[3 mm]
     & & \Ds ~~~~~~~~~~~~~
                  -  \sqrt{g} R 
                  \left({- \frac{\tilde{A}^2}{2}
                  + \tilde{\alpha}_0 
                  - g^{ij} \tilde{A}_i \tilde{\alpha}_j
                  +\frac{1}{2} \tilde{A} g^{ij} \gamma_{ij}
                  +\frac{1}{2} \gamma_{ij} g^{il} g^{jm} 
                  \tilde{A}_l \tilde{A}_m
                  }\right)  \\[3 mm]
     & & \Ds ~~~~~~~~~~~~~
                        + {2} g^{ik} g^{jl} 
                            ( \tilde{B}_{ij}\tilde{B}_{kl}^{\SSs{(1)}}
                             -\tilde{B}_{ik}\tilde{B}_{jl}^{\SSs{(1)}})
                        - {2} g^{ik} g^{jr} \gamma_{rs} g^{sl}
                          (\tilde{B}_{ij}\tilde{B}_{kl} 
                          -\tilde{B}_{ik}\tilde{B}_{jl})   \\[3 mm]
     & & \Ds ~~~~~~~~~~~~~  \left.{\left.{
                        + {\tilde{A}}
                          g^{ik} g^{jl} (\tilde{B}_{ij}\tilde{B}_{kl} 
                                        -\tilde{B}_{ik}\tilde{B}_{jl})
                       } \right] }\right.   \\[3 mm]
     & & \Ds ~~~~~~ +m  {\Ds \int} dTd\BMz~ 
                       \left[ \frac{1}{2}g_{ij} 
                              (\frac{d{x}^i}{dT} + g^{ik} \tilde{A}_k )
                              (\frac{d{x}^i}{dT} + g^{ik} \tilde{A}_k )
                             +\tilde{A} 
                       \right] \delta^3[\BMz-\BMx(T)]~~,
\Ea
\nome{A27}
\Ee
where, for future convenience, we have introduced the notations
\Be
\left\{
\Ba{rcl}
\tilde{R}_1 (g,\gamma) 
    &=&\Ds  \sqrt{g} {R}_1 (g,\gamma) 
        +\frac{1}{2} g^{ij} \gamma_{ij} \sqrt{g} R \\[2 mm] 
\tilde{R}_2 (g,\gamma,\beta) 
    &=&\Ds  \sqrt{g} {R}_2 (g,\gamma,\beta) 
        +\frac{1}{2} g^{ij} \gamma_{ij} R_1 \\[2 mm] 
    & &\Ds  +\frac{1}{2}  \sqrt{g} R \left[ g^{ij} \beta_{ij} 
                           -(g^{ik}g^{jl} - \frac{1}{2} g^{ij}g^{kl})
                            \gamma_{ij}\gamma_{kl} 
                     \right]~~.
\Ea
\right.
\Ee
It is then seen that the Action (\ref{A27}) is independent of $\Theta(t)$.

For future reference we give here the explicit expressions of the 
quantities $\tilde{R}_1$
and $\tilde{R}_2$. They are:
\Be
\left\{
\Ba{rcl}
\tilde{R}_1 (g,\gamma) 
    &=&\Ds \sqrt{g} \left[{ -\left( 
        R^{ij} -\frac{1}{2} g^{ij}  R 
        \right) \gamma_{ij}
         + \left( g^{ik}g^{jl}-g^{ij}g^{kl} \right)
         \nabla_k\nabla_l \gamma_{ij}
        }\right] \\[2 mm] 
\tilde{R}_2 (g,\gamma,\beta) 
    &=&\Ds \sqrt{g} \left[{ -\left( 
        R^{ij} -\frac{1}{2} g^{ij}  R 
        \right) \beta_{ij} 
         + \left( g^{ik}g^{jl}-g^{ij}g^{kl} \right)
         \nabla_k\nabla_l \beta_{ij}
        }\right] \\[2 mm] 
    & &+\Ds \frac{\sqrt{g}}{2} g^{ab}\gamma_{ab}
        \left[{ -\left( 
        R^{ij} -\frac{1}{2} g^{ij}  R 
        \right) \gamma_{ij}
        + \left( g^{ik}g^{jl}-g^{ij}g^{kl} \right)
         \nabla_k\nabla_l \gamma_{ij}
        }\right] \\[2 mm] 
    & &+\Ds \sqrt{g} g^{ab}\gamma_{ib}\gamma_{jb}
          \left( 
        R^{ij} -\frac{1}{2} g^{ij}  R 
        \right)  \\[2 mm] 
    & &+\Ds \sqrt{g} g^{ab}g^{ri}g^{js} \gamma_{rs}
          \left[ \nabla_a\nabla_b\gamma_{ij}+\nabla_i\nabla_i\gamma_{ab}
                -\nabla_a\nabla_i\gamma_{jb}-\nabla_i\nabla_a\gamma_{bj} 
          \right]  \\[2 mm] 
    & &+\Ds \sqrt{g} g^{ab}g^{ij}g^{rs} 
          \left[ \nabla_r\gamma_{ij} \nabla_a\gamma_{sb}
                -\frac{1}{4} \nabla_r\gamma_{ij} \nabla_s\gamma_{ab}
                +\frac{3}{4} \nabla_r\gamma_{ia} \nabla_s\gamma_{jb}
          \right. 
\\[2 mm] & & ~~~~~~~~~~~~~~~~~~\Ds
          \left.   
                -\frac{1}{2} \nabla_r\gamma_{ia} \nabla_j\gamma_{sb}
                -\nabla_r\gamma_{is} \nabla_a\gamma_{jb}
          \right]  ~~;
\Ea
\right.
\Ee

We shall deal now with the problem of investigating the true dynamical
degrees of freedom of the theory by means of a constraint analysis
within the Hamiltonian formalism. The canonical momenta
[$\dot{f}=\partder{f}{T}$] are defined by:
\Be
\left\{
\Ba{rcl}
\pi^{ij} &=&\Ds \frac{\delta \tilde\Sc}{\delta \dot{g}_{ij}}
\\       &=&\Ds  \frac{-\sqrt{g}}{16\pi G} \bigg[
             (g^{ik}g^{jl}-g^{ij}g^{kl})
            - (g^{im} \gamma_{mn} g^{nk} g^{jl}
              +g^{ik} g^{jm} \gamma_{mn} g^{nl}
\\ & &\Ds  ~~~~~~~-g^{im} \gamma_{mn} g^{nj} g^{kl}
              -g^{ij} g^{km} \gamma_{mn} g^{nl})
             \bigg] \tilde{B}_{kl}^{\SSs{(1)}}
\\       & &\Ds +\frac{-\sqrt{g}\tilde{A} }{16\pi G} 
              (g^{ik}g^{jl}-g^{ij}g^{kl})
              \tilde{B}_{kl}
\\[4 mm]
\pi^{ij}_\gamma &=&\Ds  \frac{\delta\tilde\Sc}{\delta\dot{\gamma}_{ij}}
         = \Ds  \frac{-\sqrt{g}}{16\pi G} (g^{ik}g^{jl}-g^{ij}g^{kl})
              \tilde{B}_{kl}   
\\[3 mm]
p_{i} &=&\Ds  \frac{\delta\tilde\Sc}{\delta\dot{x}^{i}} 
       = g_{ij} (\frac{d{x}^i}{dT} + g^{ik} \tilde{A}_k )
         \delta^3[\BMz-\BMx(T)]  ~~,
\Ea
\right.
\Ee
and 
\Be
\left\{
\Ba{rclcrclcrcl}
\Ds \pi_A        &=&\Ds \frac{\delta\tilde\Sc}{\delta\dot{\tilde{A}}}        
= 0 
&\qquad&
\Ds \pi_{\alpha_0}&=&\Ds
 \frac{\delta\tilde\Sc}{\delta\dot{\tilde{\alpha}}_0} 
= 0
&\qquad&
\Ds \pi^{ij}_\beta&=&\Ds 
\frac{\delta\tilde\Sc}{\delta\dot{\beta}_{ij}}       
= 0 \\[5 mm]
\Ds \pi^i        &=&\Ds 
\frac{\delta\tilde\Sc}{\delta\dot{\tilde{A}}_i}      
= 0 
&\qquad&
\Ds \pi^i_\alpha  &=&\Ds
 \frac{\delta\tilde\Sc}{\delta\dot{\tilde{\alpha}}_i} 
= 0 ~~~.
& & && 
\Ea
\right.
\Ee
Since the Lagrangian $\tilde\Lc$ is independent of the corresponding
{\it velocities}, the latter momenta define in fact 
14 {\it primary} constraints.

The Dirac Hamiltonian density is given by:
\Be
\Ba{rcl}
\tilde{H}_d  &=&\Ds  \frac{1}{16\pi G} \left\{ \tilde{A}\tilde{R}_1 
              - \tilde{R}_2
       + \sqrt{g}R \left( \frac{\tilde{A}^2}{2} + \tilde{\alpha}_0
                         -g^{ij}\tilde{A}_i\tilde{\alpha}_j
                         +\frac{1}{2}\gamma_{ij}g^{ir}g^{js}
                          \tilde{A}_r \tilde{A}_s \right)\right\}
\\ & &+\Ds  \frac{32\pi G}{\sqrt{g}}
       \left( g_{ik}g_{jl}-\frac{1}{2} g_{ij}g_{kl} \right)
       \pi^{ij}_\gamma\pi^{kl}
      -\frac{16\pi G \tilde{A}}{\sqrt{g}} 
       \left( g_{ik}g_{jl}-\frac{1}{2} g_{ij}g_{kl} \right)
       \pi^{ij}_\gamma\pi^{kl}_\gamma
\\ & &\Ds +\frac{32\pi G}{\sqrt{g}} 
       \left( g_{ik}\gamma_{jl}-\frac{1}{2} g_{ij}\gamma_{kl} \right)
       \pi^{ij}_\gamma\pi^{kl}_\gamma
\\ & &\Ds +\left( \frac{1}{2m} g^{ij} p_i p_j - m \tilde{A} \right)
       \delta^3[\BMz-x(T)]
\\ & &\Ds - [ \tilde{\alpha}_k 
             -\tilde{A}_i g^{ij} \gamma_{jk} ] \tilde{\Phi}^k 
          -\tilde{A}_i g^{ij} \tilde{\phi}_k 
\\ & &\Ds +\lambda^A \pi_A +\lambda^{\alpha_0} \pi_{\alpha_0} 
      +\lambda_i \pi^i +\lambda^\alpha_i \pi_\alpha^i 
      +\lambda^\beta_{ij} \pi_\beta^{ij} ~~,
\Ea
\nome{H27}
\Ee
where we have introduced {\it ad hoc} notations for
the following important quantities
\Be
\left\{
\Ba{rcl}
  \tilde{\phi}_k &=& 2 g_{ij} \nabla_k \pi^{kl} 
            + p_k \delta^3[\BMz-x(T)]
            + 2 \nabla_r [ \pi_\gamma^{rs} \gamma_{sk} ]
            -  \pi_\gamma^{rs} \nabla_k\gamma_{rs}
\\[2 mm]
  \tilde{\Phi}^k &=& \nabla_l \pi_\gamma^{kl}  ~~.
\Ea
\right.
\nome{7.8}
\Ee
We will apply now the Dirac-Bergmann procedure.
By imposing time-conservation of the {\it primary} constraints, we 
obtain the 14 {\it secondary} (not all independent) ones:
\Be
\left\{
\Ba{rcl}
\dot{\pi}_{\alpha_0}   &=&\Ds  
                 - \frac{1}{16\pi G} \sqrt{g}R \simeq 0 \\[2 mm]
\dot{\pi}^{ij}_\beta   &=&\Ds  
                 = -\frac{1}{16\pi G} \left[
                     R^{ij} - \frac{1}{2} g^{ij} R
                    \right] \simeq 0 ~~.
\\[2 mm]
\dot{\pi}_A    &=&\Ds -\frac{\tilde{R}_1 }{16\pi G} 
                   + m \delta^3[\BMz-x(T)]
                   -\frac{\sqrt{g}\tilde{A}}{16\pi G}  R
                   +\frac{16\pi G}{\sqrt{g}} 
                    (g_{ik}g_{jl} -\frac{1}{2} g_{ij} g_{kl})
                    \pi^{ij}_\gamma\pi^{rs}_\gamma
                   \simeq 0 \\[2 mm]
\dot{\pi}^i     &=&\Ds  
                  g^{ik} \tilde{\phi}_k
                   + \frac{\sqrt{g}}{16\pi G} R 
                   \left( g^{ik} \tilde{\alpha}_k 
                         -g^{ik} \gamma_{kl} g^{lm} \tilde{A}_m
                    \right) \simeq 0\\[2 mm]
\dot{\pi}^i_\alpha        &=&\Ds  
                  \Ds \tilde{\Phi}^k + \frac{\sqrt{g}}{16\pi G}  R 
                            g^{kl} \tilde{A}_l \simeq 0 ~~.
\Ea
\right.
\nome{7.9}
\Ee
An equivalent, more expressive, set of 10 {\it
secondary} constraints is:
\Be
\left\{
\Ba{rcl}
\chi_{\alpha_0} &\equiv& \sqrt{g} R \simeq 0 
\\[2 mm]  
\chi_R^{ij} &\equiv& \sqrt{g} R^{ij} - \frac{1}{3} \sqrt{g} R \simeq 0 
\\[2 mm]  
\chi_1      &\equiv&\Ds   -\frac{1}{16\pi G} \tilde{R}_1 
                      + m \delta^3[\BMz-x(T)]
                      +\frac{16\pi G}{\sqrt{g}} 
                      (g_{ik}g_{jl} -\frac{1}{2} g_{ij} g_{kl})
                     \pi^{ij}_\gamma\pi^{rs}_\gamma
                     \simeq 0 \\
\tilde\phi_k      &\simeq&\Ds 0   \\
\tilde\Phi^k      &\simeq&\Ds 0  ~~. \\
\Ea
\right.
\nome{27SEC}
\Ee
By imposing time-conservation of the {\it secondary} constraints,
we obtain the {\it tertiary} constraints in the form :
\Be
\left\{
\Ba{rcl}
\dot\chi_{\alpha_0} &\simeq&  0 \\  
\dot{\chi}_R^{ij} &\simeq&  \Ds 
              -16 \pi G \sqrt{g} \left[ 
              \frac{1}{2} \nabla^i\nabla^j 
              \left(\frac{\pi_\gamma}{\sqrt{g}}\right)
              + \Delta\left(\frac{\pi_\gamma^{ij}}{\sqrt{g}}\right)
              - \frac{1}{2} g^{ij} 
              \Delta\left(\frac{\pi_\gamma}{\sqrt{g}}\right)
              \right]
             \simeq 0 \\[4 mm]  
\dot{\chi}_1  &\simeq& 
             \Ds \sqrt{g} g^{kl}g^{rs} \nabla_k\nabla_l \gamma^{rs}
             - \frac{(16\pi G)^2}{2 \sqrt{g}} g_{rs} \pi_\gamma^{rs}
               (g_{ik}g_{jl} - \frac{1}{2} g_{ij}g_{kl})
              \pi_\gamma^{ij}\pi_\gamma^{kl}
  \simeq 0 \\[2 mm]
\dot{\tilde\phi}_k      &\simeq&\Ds 0   \\
\dot{\tilde\Phi}^k      &\simeq&\Ds 0 ~~,  \\
\Ea
\right.
\nome{27TERa}
\Ee
where $\Delta$ is the three-dimensional Laplace operator ($\Delta
= \nabla^k\nabla_k$). In this way we get seven 
(not independent) {\it tertiary}
constraints. Note in this connection that, because of
the constraint $R^{ij}\simeq 0$, the covariant derivatives
commute on the constraint's surface. Also, recall finally that all the
fields' momenta are tensor densities of weight $+1$. 
\par
At this
point, it is expedient to introduced the well-known {\it
transverse-traceless-de\-compo\-si\-tion} of symmetric tensors, which,
due to the {\it secondary} constraint $\chi_R^{ij} \simeq 0$, 
can now be referred to the
globally flat metric (on the constraints surface) $g_{ij}$.
Specifically:
\Be
\left\{
\Ba{rcl}
   \pi^{ij} &=& \pi_{TT}^{ij} 
                + \frac{1}{2} \left[ g^{ij} \pi_{T} 
                         -\Delta^{-1} \nabla^i\nabla^j \pi_{T}
                              \right]
                + \nabla^i \pi_{L}^{j} 
                + \nabla^j \pi_{L}^{i} 
\\[3 mm]
   \gamma_{ij} &=& \gamma^{TT}_{ij} 
                + \frac{1}{2} \left[ g_{ij} \gamma^T 
                         -\Delta^{-1}\nabla_i\nabla_j\gamma^T 
                              \right]
                + \nabla_i\gamma^{L}_{j} + \nabla_j\gamma^{L}_{i}
\\[3 mm]
   \pi_\gamma^{ij} &=& \pi_{\gamma TT}^{ij} 
                + \frac{1}{2} \left[ g^{ij} \pi_{\gamma T} 
                         -\Delta^{-1} \nabla^i\nabla^j \pi_{\gamma T}
                              \right]
                + \nabla^i \pi_{\gamma L}^{j} 
                + \nabla^j \pi_{\gamma L}^{i} 
\\[3 mm]
   \beta_{ij} &=& \beta^{TT}_{ij} 
                + \frac{1}{2} \left[ g_{ij} \beta^T 
                         -\Delta^{-1}\nabla_i\nabla_j\beta^T 
                              \right]
                + \nabla_i\beta^{L}_{j} + \nabla_j\beta^{L}_{i}
\\[3 mm]
   \pi_\beta^{ij} &=& \pi_{\beta TT}^{ij} 
                + \frac{1}{2} \left[ g^{ij} \pi_{\beta T} 
                         -\Delta^{-1}\nabla^i\nabla^j\pi_{\beta T} 
                              \right]
               + \nabla^i\pi_{\beta L}^{j} + \nabla^j \pi_{\beta L}^{i}
\\[3 mm]
   \lambda^{\beta}_{ij} &=& \lambda^{\beta TT}_{ij} 
                + \frac{1}{2} \left[ g_{ij} \lambda^{\beta T}
                         -\Delta^{-1}\nabla_i\nabla_j\lambda^{\beta T} 
                              \right]
                + \nabla_i\lambda^{\beta L}_{j} 
                + \nabla_j\lambda^{\beta L}_{i}
~~, 
\Ea
\right.
\nome{27TTdecomposition}
\Ee
where $g^{ij} \gamma^{TT}_{ij} = \nabla^i \gamma^{TT}_{ai} =0$, 
$g^{ij} \beta^{TT}_{ij} = \nabla^i \beta^{TT}_{ai} =0$  and
$g_{ij}\pi_{TT}^{ij}=\nabla_{i}\pi_{TT}^{ai}=0$,
$g_{ij}\pi_{\gamma TT}^{ij}=\nabla_{i}\pi_{\gamma TT}^{ai}=0$,
$g_{ij}\pi_{\beta TT}^{ij}=\nabla_{i}\pi_{\beta TT}^{ai}=0$. In
terms of these quantities, it can be seen that the chain that
starts from the {\it primary} constraint $\pi_\beta^{ij}\simeq 0$
gets contributions only from the TT part, $\pi_{\beta TT}^{ij}$.
Consistently, the {\it longitudinal} and {\it trace} parts,
$\pi_{\beta L}^{i}$ and $\pi_{\beta T}$, do not generate any
chain.

At this stage of the procedure, the constraints
$\tilde{\Phi}^k\simeq 0$, $\dot\chi_R^{ij}\simeq 0$ can be
rewritten as:
\Be
\left\{
\Ba{rcl}
 \tilde{\Phi}^k  &=& \nabla^k\nabla_i \pi_{\gamma L}^i 
                     + \Delta \pi_{\gamma L}^k
\\[2 mm]
 \dot\chi_R^{ij} &=& 16 \pi G \left[ 
                        \Delta \pi_{\gamma TT}^{ij} 
                       +\Delta \cdot (\nabla^i\pi_{\gamma L}^j
                               +\nabla^j\pi_{\gamma L}^i)
                       + g^{ij} \Delta\cdot \nabla_k\pi_{\gamma L}^k 
                       + \nabla^i\nabla^j \nabla_k\pi_{\gamma L}^k
                       \right] ~~.
\Ea
\right.
\nome{7.13}
\Ee
It is then apparent from these expressions that, provided that the asymptotic
boundary conditions are such as to allow the inference $\Delta f
= 0 ~
\Longrightarrow ~ f=0$, a condition that is also necessary
to the inversion of the {\it transverse-traceless
decomposition}, the constraints $\dot\chi_R^{ij}$ and
$\tilde\Phi^k$ are equivalent to:
\Be
\left\{
\Ba{rcl}
 \pi_{\gamma TT}^{ij} &\simeq& 0  ~~, \\
  \pi_{\gamma L}^i    &\simeq& 0  ~~.
\Ea
\right.
\nome{7.14}
\Ee
Therefore we have in fact only three independent {\it tertiary}
constraints, namely $\dot{\chi}_1 \simeq 0$ and the
two independent components of the first line of eqs.(\ref{7.14}). 

Using these conditions for reexpressing the constraints
$\chi_1\simeq 0$ and $\dot{\chi}_1\simeq 0$ in terms of the {\it
transverse traceless variables}, we obtain:
\Be
\left\{
\Ba{rcl}
 \chi_1 &=& \Ds \frac{\sqrt{g}}{16\pi G} \Delta \gamma^T 
           + m \delta^3 [\BMz - \BMx(t)]
\\[3 mm] 
    & & \Ds +\frac{16 \pi G}{\sqrt{g}}\frac{1}{4}
             \Bigg[ (\pi_{\gamma T})^2 
              -     (\Delta^{(-1)}\nabla^i\nabla^j\pi_{\gamma T}) 
               \cdot(\Delta^{(-1)}\nabla_i\nabla_j\pi_{\gamma T}) \Bigg]
           \simeq 0 ~~~,
\\[4 mm]  
 \dot\chi_1 &=&  - \pi_{\gamma T} \cdot ( \Delta \cdot\nabla^k\gamma^L_k)
                + \left[ \frac{1}{2} \nabla_i\nabla_j \gamma^T
                        -(\nabla_i\nabla_j \nabla^k\gamma^L_k )
                \right]
                  \cdot (\Delta^{(-1)}\nabla^i\nabla^j\pi_{\gamma T})
\\[3 mm]
  & & \Ds -\frac{16 \pi G}{\sqrt{g}}\frac{\pi_{\gamma T}}{8}
             \Bigg[ (\pi_{\gamma T})^2 
              -     (\Delta^{(-1)}\nabla^i\nabla^j\pi_{\gamma T})
               \cdot(\Delta^{(-1)}\nabla_i\nabla_j\pi_{\gamma T}) \Bigg]
           \simeq 0   ~~~.
\Ea
\right.
\nome{27TERb}
\Ee
At this stage we have the following situation: {\it i})
$\pi_{\beta T} \simeq 0$, $\pi^i_{\beta L}\simeq 0$ do not
generate {\it secondary} constraints; {\it ii})
$\pi_{\alpha_0}\simeq 0$ generate the {\it secondary} 
$\chi_{\alpha_0} \simeq 0$ and no {\it tertiary}; {\it iii})
$\pi^i_\alpha \simeq 0$ generate the {\it secondaries} 
$\tilde{\Phi}^i \simeq 0$, i.e. $\pi^i_{\gamma L} \simeq 0$ and
no {\it tertiary}; {\it iv}) $\pi^i\simeq 0$ generate the {\it
secondaries} $\tilde{\phi}_i\simeq 0$ and no {\it tertiary}; {\it
v}) $\pi_A\simeq 0$ generates the {\it secondary} $\chi_1\simeq
0$ and then the {\it tertiary} $\dot\chi_1\simeq 0$; {\it vi})
$\pi^{ij}_{\beta TT} \simeq 0$ (only two independent constraints)
generate $\chi^{ij}_R \simeq 0$ (only two independent constraints
due to the Bianchi identities) and then the two {\it tertiaries}
$\dot\chi^{ij}_R \simeq 0$.

We have only to find the {\it quaternary} constraints generated
by the time derivatives of the {\it tertiary}  constraints
$\dot\chi_1 \simeq 0$ and $\dot\chi^{ij}_R \simeq 0$. While 
$\ddot\chi_1 \simeq 0$ is given in appendix A, we have: 
\Be
\Ba{rcl}
  \ddot\chi_R^{ij} &=& \Ds \frac{\sqrt{g}}{2\cdot16 \pi G}
                       ~\Delta\Delta \gamma^{TT}_{ij} + 
        \frac{16\pi G}{2\sqrt{g}} 
        \left[ \nabla^r\pi_{\gamma T} \nabla^s \pi_{\gamma T}  \right.
\\[2 mm]
&&~~~~~~~~~-\nabla^r\nabla^s\nabla^i\nabla^j\Delta^{(-1)}\pi_{\gamma T}
               \nabla_i\nabla_j \Delta^{(-1)}\pi_{\gamma T}
\\[2 mm]
  & &~~~~~~~~~+ \frac{1}{2} \nabla^i\nabla^r\Delta^{(-1)}\pi_{\gamma T}
                \nabla_i\nabla^s \pi_{\gamma T}
              + \frac{1}{2} \nabla^i\nabla^s\Delta^{(-1)}\pi_{\gamma T}
                \nabla_i\nabla^r \pi_{\gamma T}
\\[2 mm]
  & &~~~~~~~~~\left.
              - \frac{1}{2} \nabla^r\nabla^s\Delta^{(-1)}\pi_{\gamma T}
                \Delta \pi_{\gamma T}
              + \frac{1}{2} \nabla^r\nabla^s \pi_{\gamma T}
         \right]\simeq 0 ~~~.
\Ea
\nome{7.16}
\Ee
Before ending the discussion of these chains of constraints, let us 
remark that the relevant sector of solutions of eqs.(\ref{27TERb})
is 
\Be
\left\{
\Ba{rcl}
     \pi_{\gamma_T}      &\simeq& 0 \\[2 mm]
 \sqrt{g}\Delta\gamma^T &\simeq& - 16\pi G m \delta^3 [\BMz - \BMx(t)]
~~~.
\Ea
\right.
\nome{7.17}
\Ee
Using eqs.(\ref{7.17}) inside eqs.(\ref{7.16}), we get 
$  \ddot\chi_R^{ij} \simeq \frac{\sqrt{g}}{2\cdot16 \pi G}
~g^{ir} g^{js} \Delta\Delta \gamma^{TT}_{rs} \simeq 0 $,
which implies $\gamma^{TT}_{rs} \simeq 0$. By using
eqs.(\ref{7.17}) and $\gamma^{TT}_{rs}\simeq 0$ in 
$\ddot{\chi}_1$ (see Appendix A),
we obtain 
\Be
\Ba{rcl}
\ddot\chi_1 &\simeq&\Ds \frac{\sqrt{g}}{16 \pi G}
             \left[ \left( \Delta \gamma^T 
                          - 2 \Delta \cdot \nabla^k \gamma^L_k
                    \right) g^{ij}
                    + 2 \nabla^i\nabla^j\nabla^k \gamma^L_k
             \right] \cdot
\\[4 mm] & & ~~\cdot \nabla_i\nabla_j 
             \left(\tilde{A} - \frac{1}{4}\gamma^T \right)~,
\Ea
\nome{EQa27}
\Ee
which implies $\tilde{A}-\frac{1}{4} \gamma^T \simeq 0$ as a 
relevant solution. Therefore we get in the end 
\Be
 \Delta \tilde{A} \simeq - 4 \pi G m \delta^3 [\BMz - \BMx(t)] ~~~,
\nome{EQp27}
\Ee 
i.e. the Poisson equation in a {\it three-dimensional
general covariant} form. This means that $\ddot\chi_1 \simeq
0$ is the equation which replaces the Poisson equation
in an {\it arbitrary-absolute time respecting} frame; 
the important result just
obtained is that, provided that the Newton potential $A_0 =
-\varphi$, seen by the {\it Galilean} observers, is replaced by the
effective potential $\tilde{A} = \frac{A_0}{\Theta^2} -
\frac{1}{2} \frac{ g^{ij} A_i A_j}{\Theta^2}$, then we get the 
Poisson equation for $\tilde{A}$
as the most relevant solution (see Eq.
\ref{PoissonGenA}) in every {\it allowed} reference frame.

Some words should be spent about the ``invariance'' of
Eq.(\ref{EQp27}). Since we have shown in
Eq.(\ref{INV27}) that the Action is {\it
quasi-invariant} modulo the equation of motion, one
could expect that Eq.(\ref{EQp27}) be invariant under
{\it all} the {\it local} Galilei transformations, just as all other
equations are. Yet, this is not true because Eq.(\ref{INV27}) is
not invariant under {\it local} Galilei boosts because
it gets contributions from the {\it cocycle} term. This
does not invalidated the invariance of the theory,
however.  Indeed, since the Action is {\it
quasi-invariant modulo} equations of motion, there is
anyway a {\it conserved charge} associated to the
boosts \cite{Lusanna}.  Therefore, the full invariance
of Poisson equation should be accounted for by the
transformations generated by these {\it conserved
charge}.

Finally, time conservation of the {\it quaternary}
constraints gives the {\it quinquenary} constraints.
One of these latter, precisely that following from the
$\chi_1$ chain, fixes the {\it multiplier} $\lambda_A$.
On the other hand, the chain originated by
$\chi_R^{rs}$ continues along three more time
derivations. To avoid cumbersome expression, we give
the simplified forms of the leading terms for the
previous relevant sector; using all the constraints
already worked out, it follows:
\Be
\left\{
\Ba{rcl}
\Ds {\buildrel{(3)}\over{\chi}}_1 
= \Ds \frac{d^3}{dT^3} \chi_1 &\simeq&\Ds \frac{\sqrt{g}}{16 \pi G}
         \left[ \left( \Delta \gamma^T 
                      - 2 \Delta \cdot \nabla^k \gamma^L_k
                \right) g^{ij}
                + 2 \nabla^i\nabla^j\nabla^k \gamma^L_k
         \right] \cdot \nabla_i\nabla_j \lambda_A + .... \simeq 0 
\\[4 mm] 
\Ds {\buildrel{\Sss (3)}\over{\chi}_R}{}^{rs} 
= \Ds \frac{d^3}{dT^3} \chi^{rs}_R 
&\simeq& 
\Delta\Delta \pi_{TT}^{rs} \simeq 0 \\[4 mm] 
\Ds {\buildrel{\Sss (4)}\over{\chi}_R}{}^{rs} 
= \Ds \frac{d^4}{dT^4} \chi^{rs}_R &\simeq& 
\Ds \frac{\sqrt{g}g^{ri}g^{sj}}{16 \pi G}
\Delta\Delta\Delta \beta^{TT}_{ij}  + .... \simeq 0 \\[4 mm] 
\Ds {\buildrel{\Sss (5)}\over{\chi}_R}{}^{rs} 
= \Ds \frac{d^5}{dT^5} \chi^{rs}_R &\simeq& 
\Ds \frac{\sqrt{g}g^{ri}g^{sj}}{16 \pi G}
\Delta\Delta\Delta \lambda^{\beta{TT}}_{ij}  + .... \simeq 0 ~~, 
\\[2 mm] 
\Ea
\right.
\nome{EndC27}
\Ee

The last one ends the chain and fixes the {\it
transverse-traceless} part $\lambda^{\beta{TT}}_{ij}$ of the {\it
multipliers} $\lambda_\beta^{ij}$.

Therefore, since
$\lambda_A$ is determined from eq.(\ref{EndC27},
the chain of $\pi_A\simeq 0$ contains two
pairs of {\it second class} constraints
($\pi_A$,$\ddot\chi_1$), ($\chi_1$,$\dot\chi_1$) ).
On the other hand, each of the two independents chains of $\pi^{ij}_{\beta
TT}\simeq 0$ contains three pairs of {\it
second class} constraints ($\pi_{\beta
TT}^{ij}$,${\buildrel{\Sss (4)}\over{\chi}_R}{}^{rs}$),
($\chi_{R}^{rs}$,${\buildrel{\Sss
(3)}\over{\chi}_R}{}^{rs}$),
($\dot\chi_{R}^{rs}$,$\ddot\chi_{R}^{rs}$), since the
sixth time derivative of these {\it primary}
constraints determine the two independent Dirac
multipliers $\lambda^{\beta TT}_{ij}$.

In conclusion there are 18 {\it first-class}
constraints and 16 {\it second-class} constraints.
While the variables
$\tilde{A}$,$\gamma^T$,$\beta_{ij}^{TT}$,
$\gamma_{ij}^{TT}$ and two components of $g_{ij}$ are
determined by half of the {\it second-class}
constraints (the other half determines their canonical
momenta), the variables $\tilde{A}_i$, 3 of the
$g_{ij}$, $\tilde{\alpha}_i$, $\gamma_i^{L}$,
$\tilde{\alpha_0}$, $\pi_T$ (conjugated to one of the
$g_{ij}$), $\beta_i^L$, $\beta^T$ are {\it gauge}
variables (their conjugated variables are determined by
the {\it first-class} constraints); correspondingly, the
eleven Dirac multipliers $\lambda_i,
\lambda^{\alpha_0},
\lambda^{\alpha}_i, \lambda^{\beta T}, \lambda^{\beta L}_i$
remain arbitrary.  
\par
Thus, apart from the {\it particle}
degrees of freedom, no physical {\it field} degrees of
freedom survive, as indeed it should be, and the role
of the Newton potential is taken by $\tilde{A}$, which
satisfies a Poisson equation in the most relevant
sector of solutions. It would be interesting to see
whether unconventional sectors are allowed corresponding 
to more general solutions for the gravitational 
potential\footnote{ This could possibly be of some interest in
connection with the debate about the so-called fifth
force}\rlap .

The logical connections of the various constraints
involved is described in Fig.1, which summarizes what
is being fixed by each chains.

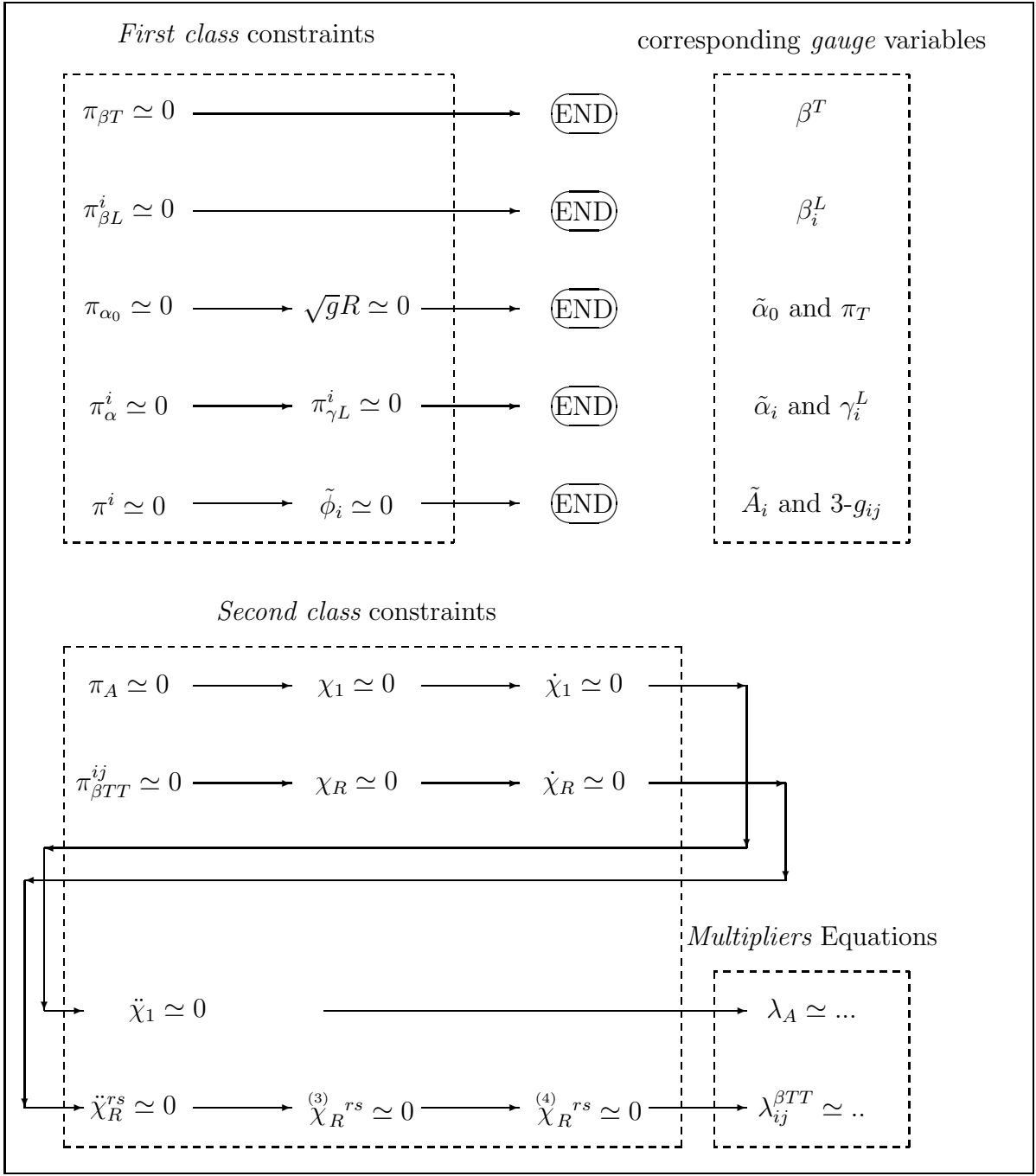
\begin{figure} 
\noindent 
\unitlength=1.00mm
\linethickness{0.4pt}
\begin{picture}(159.00,187.00)
\put(125.00 ,183.00){\makebox(0,0)[ct]{corresponding {\it gauge} variables}}
\put(38.00  ,181.00){\makebox(0,0)[cb]{{\it First class} constraints}}
\put(20.00  ,170.00){\makebox(0,0)[cc]{$\pi_{\beta T} \simeq 0$}}
\put(90.00  ,170.00){\makebox(0,0)[cc]{END}}
\put(30.00  ,170.00){\vector(1,0){50.00}}
\put(90.00  ,170.00){\oval(10.00,6.00)[]}
\put(125.00 ,170.00){\makebox(0,0)[cc]{$\beta^T$}}
\put(20.00  ,155.00){\makebox(0,0)[cc]{$\pi_{\beta L}^i \simeq 0$}}
\put(90.00  ,155.00){\makebox(0,0)[cc]{END}}
\put(30.00  ,155.00){\vector(1,0){50.00}}
\put(90.00  ,155.00){\oval(10.00,6.00)[]}
\put(125.00 ,155.00){\makebox(0,0)[cc]{$\beta^L_i$}}
\put(20.00  ,140.00){\makebox(0,0)[cc]{$\pi_{\alpha_0} \simeq 0$}}
\put(55.00  ,140.00){\makebox(0,0)[cc]{$\sqrt{g} R \simeq 0$}}
\put(90.00  ,140.00){\makebox(0,0)[cc]{END}}
\put(30.00  ,140.00){\vector(1,0){15.00}}
\put(65.00  ,140.00){\vector(1,0){15.00}}
\put(90.00  ,140.00){\oval(10.00,6.00)[]}
\put(125.00 ,140.00){\makebox(0,0)[cc]{$\tilde{\alpha}_0$ and $\pi_T$}} 
\put(20.00  ,125.00){\makebox(0,0)[cc]{$\pi^i_\alpha \simeq 0$}}
\put(55.00  ,125.00){\makebox(0,0)[cc]{$\pi_{\gamma L}^i \simeq 0$}}
\put(90.00  ,125.00){\makebox(0,0)[cc]{END}}
\put(30.00  ,125.00){\vector(1,0){15.00}}
\put(65.00  ,125.00){\vector(1,0){15.00}}
\put(90.00  ,125.00){\oval(10.00,6.00)[]}
\put(125.00 ,125.00){\makebox(0,0)[cc]{$\tilde{\alpha}_i$ and $\gamma^L_i$}} 
\put(20.00  ,110.00){\makebox(0,0)[cc]{$\pi^i \simeq 0$}}
\put(55.00  ,110.00){\makebox(0,0)[cc]{$\tilde\phi_i \simeq 0$}}
\put(90.00  ,110.00){\makebox(0,0)[cc]{END}}
\put(30.00  ,110.00){\vector(1,0){15.00}}
\put(65.00  ,110.00){\vector(1,0){15.00}}
\put(90.00  ,110.00){\oval(10.00,6.00)[]}
\put(125.00 ,110.00){\makebox(0,0)[cc]{$\tilde{A}_i$ and 3-$g_{ij}$}} 
\put(10.00  ,104.00){\dashbox{1.00}(60.00,72.00)[cc]{}}
\put(110.00 ,104.00){\dashbox{1.00}(30.00,72.00)[cc]{}}
\put(55.00  , 95.00){\makebox(0,0)[ct]{{\it Second class} constraints}}
\put(20.00  , 82.00){\makebox(0,0)[cc]{$\pi_A \simeq 0$}}
\put(55.00  , 82.00){\makebox(0,0)[cc]{$\chi_1\simeq 0$}}
\put(90.00  , 82.00){\makebox(0,0)[cc]{$\dot\chi_1 \simeq 0$}}
\put(30.00  , 82.00){\vector(1,0){15.00}}
\put(65.00  , 82.00){\vector(1,0){15.00}}
\put(100.00 , 82.00){\vector(1,0){15.00}}
\put(115.00 , 82.00){\vector(0,-1){25.00}}
\put(20.00  , 67.00){\makebox(0,0)[cc]{$\pi_{\beta TT}^{ij} \simeq 0$}}
\put(55.00  , 67.00){\makebox(0,0)[cc]{$\chi_R \simeq 0$}}
\put(90.00  , 67.00){\makebox(0,0)[cc]{$\dot\chi_R \simeq 0$}}
\put(30.00  , 67.00){\vector(1,0){15.00}}
\put(65.00  , 67.00){\vector(1,0){15.00}}
\put(100.00 , 67.00){\vector(1,0){21.00}}
\put(121.00 , 67.00){\vector(0,-1){15.00}}
\put(115.00 , 57.00){\vector(-1,0){108.00}}
\put(7.00   , 57.00){\vector(0,-1){25.00}}
\put(121.00 , 52.00){\vector(-1,0){117.00}}
\put(4.00   , 52.00){\vector(0,-1){35.00}}
\put(125.00 , 45.00){\makebox(0,0)[ct]{{\it Multipliers} Equations}}
\put(7.00   , 32.00){\vector(1,0){6.00}}
\put(20.00  , 32.00){\makebox(0,0)[lc]{$\ddot\chi_1 \simeq 0$}}
\put(50.00  , 32.00){\vector(1,0){65.00}}
\put(125.00 , 32.00){\makebox(0,0)[cc]{$\lambda_A \simeq ...$}}
\put(4.00   , 17.00){\vector(1,0){9.00}}
\put(20.00  , 17.00){\makebox(0,0)[cc]{
${\ddot{\chi}}_R^{rs} \simeq 0$}}
\put(55.00  , 17.00){\makebox(0,0)[cc]{
${\buildrel{\Sss (3)}\over{\chi}_R}{}^{rs} \simeq 0$}}
\put(90.00  , 17.00){\makebox(0,0)[cc]{
${\buildrel{\Sss (4)}\over{\chi}_R}{}^{rs} \simeq 0$}}
\put(125.00 , 17.00){\makebox(0,0)[cc]{$\lambda^{\beta TT}_{ij}\simeq ..$}}
\put(30.00  , 17.00){\vector(1,0){15.00}}
\put(65.00  , 17.00){\vector(1,0){15.00}}
\put(100.00 , 17.00){\vector(1,0){15.00}}
\put(110.00 , 11.00){\dashbox{1.00}(30.00,27.00)[ct]{}}
\put(10.00  , 11.00){\dashbox{1.00}(95.00,77.00)[cc]{}}
\put(1.00   ,  7.00){\framebox(158.00,180.00)[cc]{}}
\end{picture}
\hfill\hfill\break 
\caption{What is being fixed by the constraints' 
chains for the {\it 27-fields}
theory.} 
\end{figure}

We can conclude this section by noting that, in force of
eqs.(\ref{27SEC}) and (\ref{27TERa}), the condition for the
finiteness of the central-charge term is indeed satisfied, and
that $\Nc =0$ holds. It remains an open task that of performing
the $1/c^2$ expansion of the neglected surface term. It is likely
that clarifying this issue will be relevant also to the
understanding of the role of the {\it cocycle} contribution to
the {\it local} Galilei transformations of the Poisson equation.

\subsection{The Newtonian Theory in Galilean Reference Frames}

Starting from the general scheme of the {\it 27-fields} theory it
is now interesting to see that, by confining to a post-Newtonian
like \cite{Weeler1} parameterization for the four-dimensional
covariant metric tensor, defined by  $\Theta = 1$, $g_{ij} =
\delta_{ij}$, $A_i=0$ and $A=-\varphi$ (i.e. the fields as seen by
the {\it Galilean}
observers: see Eq.(\ref{PosA})), one obtains the maximum of
similarity to Newton's theory, i.e. a non-generally covariant
formulation which is valid only in {\it Galilean} reference frame
connected by {\it global} Galilei transformations. It should be clear,
however, that in this way we are dealing in fact with a {\it
different} variational problem with respect to the previous one.
Putting
\Be
{^4}\! g_{\mu\nu} = \left| \matrix{- c^2 - 2 \varphi
                 +\frac{2 \alpha_0}{c^2}& \frac{\alpha_i}{c^2} \cr
                 \frac{\alpha_i}{c^2}  &
                 \delta_{ij} + {\gamma_{ij} \over c^2}
                 + {\beta_{ij} \over c^4} \cr}
                 \right|
~~,
\Ee
the explicit expressions of the quantities $R_1$ and $R_2$ defined
in Eq.(\ref{IdentADMK}) become:
\Be
\left\{ {
\Ba{rcl}
   R_1 &=&\delta^{ij}\delta^{rs} [ \gamma_{ir,sj} - \gamma_{ij,rs} ] \\
   R_2 &=&\gamma_{lm} \delta^{lr} \delta^{ms} \delta^{ij}
            [ \gamma_{ij,rs} + \gamma_{rs,ij} - 2 \gamma_{ir,sj} ] \\
       & &+ \delta^{lm} \delta^{rs} \delta^{ij}
            [ \gamma_{rs,l} \gamma_{mi,j}
             -\frac{1}{4}  \gamma_{ij,l} \gamma_{rs,m}
             - \gamma_{li,j} \gamma_{mr,s}  \\
       & & ~~~~~~~~~~~~ + \frac{3}{4}  \gamma_{ir,l} \gamma_{js,m}
          - \frac{1}{2}  \gamma_{ir,l} \gamma_{jm,s}  ] \\
       & &+\delta^{ij}\delta^{rs} [ \beta_{ir,sj} - \beta_{ij,rs} ] ~~~,
\Ea  
} \right.
\nome{EspressioniR}
\Ee
and the total action $\tilde{\Sc}$ (\ref{6.10}) results:
\Be
\Ba{rcl}
\tilde{\Sc} &\equiv& \frac{1}{16\pi G} {\Ds \int} dtd^3z  \Lc_f
   + m {\Ds \int} dtd^3\! z \Lc_m \delta^3[{\BMz}-{\BMx}(t)] \\[2 mm]
      &=&~\frac{1}{16\pi G} {\Ds \int} dtd^3z
            \left[ {  
                  (\varphi+\frac{1}{2}\delta^{ij}\gamma_{ij}) R_1
                  + R_2
                  } \right]   \\[2 mm]
      & &+ m  {\Ds \int} dtd^3\! z
                  \left[ \frac{1}{2}\delta_{ij} \dot{x}^i \dot{x}^j
                  -\varphi  \right] \delta^3[{\BMz}-{\BMx}(t)]
~~~.
\Ea
\nome{ExpansionLCa}
\Ee
It is seen that the matter Lagrangian $\Lc_m$ has precisely the form 
which is to be expected for a {\it Galilean observer} if Eq.(\ref{PosA})
are inserted in Eq.(\ref{lagrangianaMat}). Therefore $(t,\BMz)$ define
a system of coordinates for a {\it Galilean} reference frame.  \newline
Note that:  \newline
(1) $\alpha_0$ and $\alpha_i$ do not appear 
    in the lagrangian $\Lc_f$; \newline
(2) $\Lc_f$ depends on $\beta_{ij}$ in a pure additive way 
    through the term
    $\delta^{ij}\delta^{rs} [ \beta_{ir,sj} - \beta_{ij,rs} ]$
    (see Eqs.(\ref{EspressioniR})),
     which is again a {\it surface} term; 
    moreover $\beta_{ij}$  
    is not coupled to the other fields. \newline

We can put accordingly $\alpha_0 = \alpha_i = \beta_{ij} = 0$
without altering the dynamics of this theory
which indeed depends now on $\varphi$ and $\gamma_{ij}$ only.

Let us note, moreover, that the $c^4$-order term in the expansion
(\ref{developT}) is automatically zero in this case,
while the $c^2$-order term becomes:
\Be
     \frac{1}{16\pi G} {\Ds \int} dtd^3z \left[ {
     R_1 - 16\pi G m \delta^3\! [{\BMz}-{\BMx}(t)] }\right]
~.
\nome{7.5}
\Ee
The resulting Lagrangian is:
\Be
{\Lc} = \frac{1}{16\pi G} \left[ {  
                         (\varphi+\frac{1}{2}\delta^{ij}\gamma_{ij}) R_1
                         + R_2^\prime  } \right] 
       + m  \left[ \frac{1}{2}\delta_{ij} \dot{x}^i \dot{x}^j
                   -\varphi  \right] \delta^3[{\BMz}-{\BMx}(t)]
~~~,
\nome{ResLlc}
\Ee
where $R_2^\prime = R_2 |_{\beta_{ij}=0}$, does 
not depend on the velocities
$\dot{\varphi}$ and $\dot{\gamma}_{ij}$,
so that all the field-momenta play the role of
{\it primary} constraints. 

The Hamiltonian formulation is defined by:
\Be
\left\{
\Ba{rcl}
 \pi_\varphi     &=& 0 \\
 \pi^{rs}_\gamma &=& 0 \\
 p_k  &=& m \delta_{kl} \dot{x}^l \\
 H_c  &=& \frac{1}{16\pi G} {\Ds \int} d^3z
                \left[ {  - ( \varphi + \frac{1}{2} \delta^{ij} 
                            \gamma_{ij}) R_1
                          - R_2^\prime
                        } \right] \\
           & & + {\Ds \int} d^3\! z
                 \left[ \frac{1}{2m}\delta^{ij} p_i p_j
                 + m \varphi  \right] \delta^3[{\BMz}-{\BMx}(t)] ~~,\\
\Ea
\right.
\nome{HamiltonianaLC}
\Ee
so that we have the 7 {\it primary} constraints:
\Be
\left\{
\Ba{rcl}
    \pi_\varphi           &\simeq& 0 \\
    \pi^{rs}_\gamma &\simeq& 0 ~~. \\
\Ea
\right.
\nome{primLC}
\Ee
The Dirac Hamiltonian is:
\Be
 H_d    = H_c  + \int d^3\! z \left[{
               \lambda^\varphi ({\BMz},t) \pi_\varphi
               +\lambda_{rs}^\gamma ({\BMz},t) \pi^{rs}_\gamma
               }\right]  ~~, 
\nome{DiracHlc}
\Ee
where the $\lambda$'s are the Dirac multipliers.
Time-conservation of these constraints generate
the 7 {\it secondary} constraints:
\Be
\Ba{rcl}
 \chi_{\varphi} ({\BMz},t) &\equiv&
 \dot{\pi}_{\varphi} ({\BMz},t) 
               =\Ds \{ {\pi}_\varphi ({\BMz},t) , H_d \} 
               =   \frac{1}{16\pi G} R_1
               - m\delta^3\! [{\BMz}-{\BMx}(t)] \simeq 0    \\[2 mm]
 \chi^{ij}_\gamma ({\BMz},t) &\equiv&
 \dot{\pi}^{ij}_\gamma ({\BMz},t)
               = \{ {\pi}^{ij}_\gamma ({\BMz},t) , H_d \} \\[2 mm]
               &=&\Ds \frac{1}{16\pi G} \left\{ { { 1 \over 2}
                    [\delta^{ir}\delta^{js}
                    -\frac{1}{2} \delta^{ij}\delta^{rs}]
                    \delta^{ab} 
                    [ \gamma_{ab,rs} + \gamma_{rs,ab}
                    -\gamma_{ar,sb} - \gamma_{as,rb} ]  }\right. 
\\[2 mm]
               & &\Ds  \left. ~~~~~~~ + 
                    [\delta^{ir}\delta^{js}-\delta^{ij}\delta^{rs}]
                    \partial_r \partial_s  \varphi
                    \right\} \simeq 0  ~~,
\Ea
\nome{seconLC}
\Ee
while their time conservation gives the following condition
on $\lambda^{\gamma}_{ij}$ and $\lambda^\varphi$:
\Be
\Ba{rcl}
 \psi_{\varphi} ({\BMz},t) &\equiv&
 \dot{\chi}_{\varphi} ({\BMz},t) 
  =\Ds \{ {\chi}_\varphi ({\BMz},t) , H_d \} 
\\[2 mm]
  &=&     \frac{1}{16\pi G} \delta^{ij}\delta^{rs} 
          [ \lambda^\gamma_{ir,sj} - \lambda^\gamma_{ij,rs} ] 
        + p_k \delta^{kl} \partial_l \delta^3\! [{\BMz}-{\BMx}(t)] 
     \simeq 0 
\\[2 mm]
 \psi^{ij}_\gamma ({\BMz},t) &\equiv&
 \dot{\chi}^{ij}_\gamma ({\BMz},t)
             = \{ {\chi}^{ij}_\gamma ({\BMz},t) , H_d \} \\[2 mm]
             &=&\Ds \frac{1}{16\pi G} \left\{ { { 1 \over 2}
             [\delta^{ir}\delta^{js}
             -\frac{1}{2} \delta^{ij}\delta^{rs}]
             \delta^{ab} 
             [ \lambda^\gamma_{ab,rs} + \lambda^\gamma_{rs,ab}
             -\lambda^\gamma_{ar,sb} - \lambda^\gamma_{as,rb} ]
             }\right. \\[2 mm]
             & &\Ds  \left. ~~~~~~~ + 
             [\delta^{ir}\delta^{js}-\delta^{ij}\delta^{rs}]
             \partial_r \partial_s  \lambda^\varphi
             \right\} \simeq 0  ~~.
\Ea                      
\nome{terLC}
\Ee
The {\it secondary} constraints are just the 
Euler-Lagrange field equations.
Let us remark that the constraint $\chi_\varphi \simeq 0$ is just
the expected condition for the vanishing of the 
$c^2$ term (\ref{7.5}) (in the Lagrangian description
this term vanishes because of the Euler-Lagrange equation
for $\varphi$).
 On the other hand, the mass-point equations 
are 
\Be
\dot{p}_k = \{ p_k , H_d[T] \} = - m \partial_k\varphi
~~,
\nome{7.26}
\Ee
i.e. the standard Newton's equations with potential $\varphi$.
Finally, evaluating $R_1$ from the contraction $\delta_{ij}
\chi_\gamma^{ij} \simeq 0$ and by substituting it in the constraint
$\chi_\varphi \simeq 0$, we obtain the classical Poisson equation for
the potential $\varphi ({\BMz},t)$, i.e.:
\Be
 \delta^{ij} \partial_i \partial_j \varphi ({\BMz},t)
      = 4 \pi G m \delta^3\! [{\BMz} - {\BMx}(t)]
~.
\nome{7.27}
\Ee
Let us remark that, if we put an inhomogeneous solution of 
eq.(\ref{7.27}) into eq.(\ref{7.26}), one should get the motion of
the particle under the usual field reaction; this gives rise to
problems of self-energy similar to those of, e.g., the special
relativistic electromagnetic case.

In spite of what could appear from 
Eqs.(\ref{terLC}), not all the 7 {\it secondary} constraints are
independent: as a matter of fact only four of them are
independent and, correspondingly, only four {\it multipliers} are
determined by eqs.(\ref{terLC}). In order to see this explicitly,
let us complete the constraint analysis of the theory. First of
all, we note that three combination of the primary constraints,
given by
\Be
  \Pi^k = \partial_l \pi_\gamma^{kl}~,
\Ee
are {\it first class}. This can be easily checked thanks to the fact
that the following six relations
\Be
\Ba{rcl}
 \partial_l \chi_\gamma^{kl} &\equiv& 0 \\
 \partial_l \psi_\gamma^{kl} &\equiv& 0 
~~.
\Ea
\Ee 
hold identically. Consequently, one has to expect that three of
the $\gamma_{ij}$  and three of the $\lambda_{ij}^\gamma$ are
free quantities. In order to evidentiate explicitly the
multipliers and the fields that are determined by the
constraints, it is profitable again to parameterize  $\gamma_{ij}$
and $\lambda^\gamma_{ij}$ in terms of  the {\it
transverse-traceless decomposition} of  symmetric tensors, as:
\Be
\Ba{rcl}
   \gamma_{ij} &=& \zeta^{TT}_{ij} 
                + \frac{1}{2} \left[ \delta_{ij} \zeta^T 
                                    -\Delta^{-1} \zeta^T_{,ij} 
                              \right]
                + \zeta_{i,j} + \zeta_{j,i}
\\[3 mm]
   \lambda^\gamma_{ij} &=& \lambda^{\gamma TT}_{ij} 
                + \frac{1}{2} \left[ \delta_{ij} \lambda^{\gamma T} 
                              -\Delta^{-1} \lambda^{\gamma T}_{,ij} 
                              \right]
                + \lambda^\gamma_{i,j} + \lambda^\gamma_{j,i}
~~, 
\Ea
\nome{TTdecomposition}
\Ee
where $\delta^{ij} \zeta^{TT}_{ij} = \delta^{ij} \zeta^{TT}_{ai,j} =
0$ and $\delta^{ij} \lambda^{\gamma TT}_{ij} =  \delta^{ij}
\lambda^{\gamma TT}_{ai,j} = 0$. In terms of these quantities, the
{\it secondary} and {\it tertiary} constraints (Eqs.(\ref{seconLC})
and (\ref{terLC})) become:
\Be
\left\{
\Ba{rclcl}
 \chi_{\varphi} &\simeq & 0 
                &\Rightarrow& \Delta \zeta^T 
                             + 16 \pi G m \delta^3\! [{\BMz}-{\BMx}(t)] 
                 \simeq 0 \\[1.5 mm]    
 \chi^{ij}_\gamma &\simeq & 0 
      &\Rightarrow& [\delta^{ir}\delta^{js}-\delta^{ij}\delta^{rs}]
                              \partial_r\partial_s 
                              [\varphi - \frac{1}{4}  \zeta^T ] 
                             +\delta^{ir}\delta^{js}\delta^{lm}
                              \zeta^{TT}_{ij,lm} \simeq 0 ~~,  
\Ea 
\right.
\Ee
\Be
\left\{
\Ba{rclcl}
 \psi_{\varphi} &\simeq & 0 
                &\Rightarrow& 
\Delta \lambda^{\gamma T} 
 - 16 \pi G p_k \delta^{kl} \partial_l \delta^3\! [{\BMz}-{\BMx}(t)] 
                 \simeq 0 \\[1.5 mm]    
 \psi^{ij}_\gamma &\simeq & 0 
                &\Rightarrow& 
          [\delta^{ir}\delta^{js}-\delta^{ij}\delta^{rs}]
          \partial_r\partial_s 
          [\lambda_\varphi -\frac{1}{4}  \lambda^{\gamma T} ] 
                             +\delta^{ir}\delta^{js}\delta^{lm}
                              \lambda^{\gamma TT}_{ij,lm} \simeq 0   ~,
\Ea
\right.
\Ee
respectively. The {\it transverse-traceless} decomposition 
shows that the equations
$\chi^{ij}_\gamma\simeq 0$ cannot be solved for the fields
$\zeta_i$ and the multipliers $\lambda^\gamma_i$.
In particular, as to the multipliers, we can solve only
for:
\Be
\Ba{rcl}
\lambda^\varphi          &=& \lambda^\varphi [z;p_k,x^k]          
\\[1 mm]
\lambda^{\gamma T}       &=& \lambda^{\gamma T} [z;p_k,x^k]        
\\[1 mm]
\lambda^{\gamma TT}_{ij} &=& \lambda^{\gamma TT}_{ij} [z;p_k,x^k]  
~~,
\Ea
\Ee
where asymptotic boundary conditions for the $\lambda$'s allowing
for
the inference $\Delta f =0 \Longrightarrow f=0$ have been assumed.
Substituting these expressions for the multipliers,
the Dirac Hamiltonian becomes:
\Be
\Ba{rcl}
  H_d &=&\Ds  H_c + \int d^3\! z \bigg[
         \lambda^\varphi [z;p_k,x^k] \pi_\varphi
        +\lambda^{\gamma TT}_{ij} [z;p_k,x^k] \pi_\gamma^{ij}
\\[2 mm]
& &\Ds \qquad\qquad\quad
        +\frac{1}{2} \left( \delta_{ij} \lambda^{\gamma T} [z;p_k,x^k] 
             -\frac{1}{\nabla^2} \lambda^{\gamma T}_{,ij} [z;p_k,x^k] 
              \right)
         \pi_\gamma^{ij}
\\[2 mm]
& &\Ds \qquad\qquad\quad
        -2\lambda^\gamma_i \partial_j \pi_\gamma^{ij}
        \bigg]  ~~,
\Ea
\Ee 
an expression which shows that the undetermined  multipliers
$\lambda^\gamma_i$ are associated  to the first class constraints
$\Pi^k$, as it must be. 

As a consequence, the variational problem must be independent of the
quantities $\zeta_{i}$ of Eqs.(\ref{TTdecomposition}). In fact, in
terms of the {\it transverse-traceless} quantities, we have:
\Be
\left\{ {
\Ba{rcl}
   R_1 &=&\Ds  - \Delta \zeta^{T} \\[2 mm]
   R_2^\prime &=&\Ds   \frac{3}{8} \zeta^{T} \Delta \zeta^{T} 
        -\frac{1}{4}  \delta^{ij}\delta^{rs} \delta^{lm} 
        \zeta^{TT}_{ir,l} \zeta^{TT}_{js,m}
        + \delta^{ij} \zeta_{i,j} \Delta \zeta^T
        + \partder{F^k [\zeta^{TT}_{ij}, \zeta^{T}, \zeta_{i}]}{z^k} 
        \\[4 mm]
   F^k &=& \frac{3}{8}  \delta^{ki} \zeta^T \zeta^T_{,i}
          -\frac{5}{16} \delta^{kl} \delta^{ij} \zeta^T_{,i} 
                                    (\Delta^{-1}\zeta^T)_{,jl}
          +\frac{1}{16} \delta^{kl} \delta^{ij} \delta^{rs}
                                    (\Delta^{-1}\zeta^T)_{,ir}
                                    (\Delta^{-1}\zeta^T)_{,jsl}
\\[2 mm]  & & 
          -\frac{1}{2}  \delta^{kl} \delta^{ij} \delta^{rs}
                        \zeta^{TT}_{ir} \zeta^{TT}_{jl,s}
          +  \delta^{kl} \delta^{ij} \delta^{rs}
                        \zeta^{TT}_{ir} \zeta^{TT}_{js,l}
          -\frac{1}{4}  \delta^{kl} \delta^{ij} \delta^{rs}
                       (\Delta^{-1}\zeta^T)_{,ir} \zeta^{TT}_{jl,s}
\\[2 mm]  & & 
          -\frac{1}{2}  \delta^{kl} \delta^{ij} \delta^{rs}
                       (\Delta^{-1}\zeta^T)_{,i} \zeta^{TT}_{jl,rs}
          +\frac{1}{2}  \delta^{kl} \delta^{ij} 
                       \zeta^T_{,i} \zeta^{TT}_{jl}
          + \delta^{kl} \delta^{ij}  \zeta^T_{,l} \zeta_{i,j}  
\\[2 mm]  & & 
      +\frac{1}{2} \delta^{kl} \delta^{ij}  \zeta^T_{,i} \zeta_{l,j}  
      +\frac{1}{2} \delta^{kl} \delta^{ij}  \zeta^T_{,i} \zeta_{j,l}
          -            \delta^{kl} \delta^{ij} \delta^{rs}
                       \zeta_{i,r} \zeta^{TT}_{jl,s}
\\[2 mm]  & & 
          -            \delta^{kl} \delta^{ij} \delta^{rs}
                       \zeta_{r,i} \zeta^{TT}_{jl,s}
          +            \delta^{kl} \delta^{ij} \delta^{rs}
                       \zeta_{i,r} \zeta^{TT}_{js,l}
          +            \delta^{kl} \delta^{ij} \delta^{rs}
                       \zeta_{r,i} \zeta^{TT}_{js,l}
\\[2 mm]  & & 
          +            \delta^{kl} \delta^{ij} \delta^{rs}
                       \zeta^{TT}_{il,r} \zeta_{j,s}
          +            \delta^{kl} \delta^{ij} \delta^{rs}
                       \zeta_{i,r} \zeta_{j,sl}
          -            \delta^{kl} \delta^{ij} \delta^{rs}
                       \zeta_{i,rs} \zeta_{j,l}
\\[2 mm]  & & 
          -\frac{1}{2} \delta^{kl} \delta^{ij} \delta^{rs}
                       (\Delta^{-1}\zeta^T)_{,ir} \zeta_{l,js}
          +\frac{1}{2} \delta^{kl} \delta^{ij} \delta^{rs}
                       (\Delta^{-1}\zeta^T)_{,il} \zeta_{r,js}  
~~.
\Ea  
} \right.
\nome{EspressioniRTT}
\Ee
Thus, neglecting the total divergence $F^k$, and
thanks to suitable cancellations, the variational
problem for a {\it Galilean observer}, 
can be reformulated, as an effective theory, only in
terms of $\zeta^T$ and $\zeta^{TT}_{ij}$, in the form:
\Be
\Ba{rcl}
\Sc   &=&~\frac{1}{16\pi G} {\Ds \int} dtd^3z
           \left[ { - \varphi \Delta \zeta^{T} 
                  - \frac{1}{8} \zeta^{T} \Delta \zeta^{T}  
                  - \frac{1}{4} \delta^{ij}\delta^{rs} \delta^{lm} 
                  \zeta^{TT}_{ir,l} \zeta^{TT}_{js,m}
                  } \right]   \\[2.5 mm]
      & &+ m  {\Ds \int} dtd^3\! z
                  \left[ \frac{1}{2}\delta_{ij} \dot{x}^i \dot{x}^j
                  -\varphi  \right] \delta^3[{\BMz}-{\BMx}(t)]
~~.
\Ea
\nome{ExpansionLCaTT}
\Ee
Let us remark that in eq.(\ref{ExpansionLCaTT}) the
term depending on $\zeta^{TT}_{ij}$ is decoupled from
the other degrees of freedom.  Therefore, in order to
get a variational principle for the Poisson equation,
only the {\it auxiliary}, non propagating, variable $\zeta^T$
is needed.

This theory turns out to be {\it quasi-invariant} under
the global infinitesimal transformations, which
constitute the kinematical group of the Galilean
reference frames (\ref{galiglobal}), as it should be,
provided that $\varphi$ is a scalar field and
$\gamma_{ij}$ is a covariant space 2-tensor, i.e.
\Be
\left\{
\Ba{rcl}
\delta \varphi     &=& 0 \\   
\delta \gamma_{ij} &=& - \omega^l [ c_{li}^{~~k} \gamma_{kj} 
                                   +c_{lj}^{~~k} \gamma_{ik} ] ~~.
\Ea
\right.
\Ee
As a consequence of these transformation properties, the 
{\it transverse traceless} components transform according
to
\Be
\left\{
\Ba{rcl}
\delta \zeta^T     &=& 0 \\[2 mm]   
\delta \zeta_{i}   &=& - \omega^l c_{li}^{~~k} 
                   [ \zeta_{k} -\frac{1}{2}(\Delta^{-1} \zeta^T)_{,k} ] 
\\[2 mm] 
\delta \zeta^{TT}_{ij} &=& - \omega^l [ c_{li}^{~~k} \zeta^{TT}_{kj} 
                                   +c_{lj}^{~~k} \zeta^{TT}_{ik} ] 
~~,
\Ea
\right.
\Ee
so that, finally,
\Be
 \delta \Sc = m \int dt \frac{d}{dt}\left[ \delta_{ij}v^i x^j \right]
~~.
\Ee

\section{Acknowledgments}

Roberto De Pietri wishes to thank C. Rovelli, Al Janis and E.T. Newman for
the hospitality kindly offered to him at the Department of Physics 
and Astronomy. Massimo Pauri would like to express his deep 
appreciation and thanks to
the {\it Center for Philosophy of Science}, for the warm and 
stimulating intellectual
atmosphere experienced there, and the generous partial support obtained
during the completion
of the present work at the University of Pittsburgh.


\renewcommand{\theequation}{A.\arabic{equation}}\setcounter{equation}{0}
\section*{Appendix A: Explicit expression for 
the constraints $\ddot{\chi}_1$ of the 27-fields theory.}

\def\mysavedown#1{\edef\mysubs{\mysubs#1}}
\def\mysaveup#1{\edef\mysups{\mysups#1}}
\def\mydown#1{{\mytensor}_{\vphantom{\mysubs}#1}}
\def\myup#1{{\mytensor}^{\vphantom{\mysups}#1}}
\def\tensor#1#2{
  #1
  \def\mytensor{\vphantom{#1}}
  \def\mysubs{\relax}
  \def\mysups{\relax}
  \let\down=\mysavedown
  \let\up=\mysaveup
  #2
  \let\down=\mydown
  \let\up=\myup
  #2
  }

\def\Alt{\mathop{\hbox{Alt}}\nolimits}
\def\Sym{\mathop{\hbox{Sym}}\nolimits}
\def\Del{\mathop\nabla\nolimits}
\def\grad{\mathop{\hbox{grad}}\nolimits}
\def\div{\mathop{\hbox{div}}\nolimits}
\def\extd{\mathop{d}\nolimits}
\def\Lie{\cal L} 

\def\kappa{\mbox{$(K_\eta)$}}
\def\Ao{\mbox{$\tilde{A}$}}

\def\LgT{\mbox{$\Delta^{-1} \gamma^T$}}
\def\gT{\mbox{$\gamma^T$}}
\def\gTT{\mbox{$\gamma^{TT}$}}
\def\gV{\mbox{$\gamma^L$}}

\def\pT{\mbox{$\pi_{\gamma T}$}}
\def\LpT{\mbox{$\Delta^{-1} \pi_{\gamma T}$}}

Using the notations: $f_{;i} = \nabla_i f$,
$f^{;i} = \nabla^i f$ and $\kappa =(16\pi G/\sqrt{g})$,
the explicit expression of the constraints $\ddot{\chi}_1$ 
takes the form:

\begin{eqnarray*}
\ddot{\chi}_1
&\simeq&
  -{1\over {\kappa }} \tensor{{\it \Ao}}{\down{;}\down{i}\up{i}} 
    \tensor{{\it \gT}}{\down{;}\down{j}\up{j}} + 
  {1\over {4\,\kappa }} \tensor{{\it \gT}}{\down{;}\down{i}\up{i}} 
    \tensor{{\it \gT}}{\down{;}\down{j}\up{j}} + 
\\[1 mm] & & 
  {1\over {\kappa }} \tensor{{\it \Ao}}{\down{;}\down{i}\down{j}} 
    \tensor{{\it \gT}}{\down{;}\up{i}\up{j}} - 
  {1\over {4\,\kappa }} \tensor{{\it \gT}}{\down{;}\down{i}\down{j}} 
    \tensor{{\it \gT}}{\down{;}\up{i}\up{j}} - 
\\[1 mm] & & 
  {1\over {2\,\kappa }} \tensor{{\it \gT}}{\down{;}\down{i}\down{j}} 
    \tensor{{\it \gTT}}{\up{i}\up{j}\down{;}\down{k}\up{k}} - 
  {2\over {\kappa }} \tensor{{\it \Ao}}{\down{;}\down{i}\up{i}} 
    \tensor{{\it \gV}}{\down{j}\down{;}\up{j}\down{k}\up{k}} + 
\\[1 mm] & & 
  {1\over {2\,\kappa }} \tensor{{\it \gT}}{\down{;}\down{i}\up{i}} 
    \tensor{{\it \gV}}{\down{j}\down{;}\up{j}\down{k}\up{k}} + 
  {2\over {\kappa }} \tensor{{\it \Ao}}{\down{;}\down{i}\down{j}} 
    \tensor{{\it \gV}}{\down{k}\down{;}\up{i}\up{j}\up{k}} - 
\\[1 mm] & & 
  {1\over {2\,\kappa }} \tensor{{\it \gT}}{\down{;}\down{i}\down{j}} 
    \tensor{{\it \gV}}{\down{k}\down{;}\up{i}\up{j}\up{k}} - 
  {1\over {\kappa }} \tensor{{\it \gTT}}{
     \down{i}\down{j}\down{;}\down{k}\up{k}} 
    \tensor{{\it \gV}}{\down{l}\down{;}\up{i}\up{j}\up{l}} - 
\\[1 mm] & & 
{2\over {\kappa }}\tensor{{\it \gV}}{\down{i}\down{;}\down{j}\down{k}\up{k}
     } \tensor{{\it \gV}}{\down{l}\down{;}\up{i}\up{j}\up{l}} + 
{2\over {\kappa }}\tensor{{\it \gV}}{\down{i}\down{;}\down{j}\up{j}\down{k}
     } \tensor{{\it \gV}}{\down{l}\down{;}\up{i}\up{k}\up{l}} + 
\\[1 mm] & & 
  {1\over {2\,\kappa }} \tensor{{\it \gT}}{\down{;}\down{i}\down{j}} 
    \tensor{{\it \gV}}{\up{i}\down{;}\up{j}\down{k}\up{k}} - 
  {1\over {2\,\kappa }} \tensor{{\it \gT}}{\down{;}\down{i}\down{j}} 
    \tensor{{\it \gV}}{\up{j}\down{;}\up{i}\down{k}\up{k}} + 
\\[1 mm] & & 
  {{\kappa }\over 4} \tensor{{\it \Ao}}{\down{;}\down{i}\up{i}} 
    \tensor{{\it \LpT}}{\down{;}\down{j}\down{k}} 
     \tensor{{\it \LpT}}{\down{;}\up{j}\up{k}} - 
  {{\kappa }\over {16}} \tensor{{\it \gT}}{\down{;}\down{i}\up{i}} 
    \tensor{{\it \LpT}}{\down{;}\down{j}\down{k}} 
     \tensor{{\it \LpT}}{\down{;}\up{j}\up{k}} + 
\\[1 mm] & & 
  {{\kappa }\over 2} \tensor{{\it \gT}}{\down{;}\down{i}\down{j}} 
    \tensor{{\it \LpT}}{\down{;}\up{i}\down{k}} 
     \tensor{{\it \LpT}}{\down{;}\up{j}\up{k}} + 
  {{\kappa }\over 4} \tensor{{\it \gTT}}{
     \down{i}\down{j}\down{;}\down{k}\down{l}} 
    \tensor{{\it \LpT}}{\down{;}\up{i}\up{j}} 
     \tensor{{\it \LpT}}{\down{;}\up{k}\up{l}} + 
\\[1 mm] & & 
  {{\kappa }\over 2} \tensor{{\it \gV}}{
     \down{i}\down{;}\down{j}\down{k}\down{l}} 
    \tensor{{\it \LpT}}{\down{;}\up{i}\up{j}} 
     \tensor{{\it \LpT}}{\down{;}\up{k}\up{l}} - 
  {{\kappa }\over 8} \tensor{{\it \LgT}}{
     \down{;}\down{i}\down{j}\down{k}\down{l}} 
    \tensor{{\it \LpT}}{\down{;}\up{i}\up{j}} 
     \tensor{{\it \LpT}}{\down{;}\up{k}\up{l}} + 
\\[1 mm] & & 
  \kappa  \tensor{{\it \gV}}{\down{i}\down{;}\up{i}\down{j}\down{k}} 
    \tensor{{\it \LpT}}{\down{;}\up{j}\down{l}} 
     \tensor{{\it \LpT}}{\down{;}\up{k}\up{l}} + 
  {{\kappa }\over 4} \tensor{{\it \gT}}{\down{;}\down{i}} 
    \tensor{{\it \LpT}}{\down{;}\down{j}\down{k}} 
     \tensor{{\it \LpT}}{\down{;}\up{i}\up{j}\up{k}} + 
\\[1 mm] & & 
  {{\kappa }\over 2} \tensor{{\it \gTT}}{\down{i}\down{j}\down{;}\down{k}} 
    \tensor{{\it \LpT}}{\down{;}\up{k}\down{l}} 
     \tensor{{\it \LpT}}{\down{;}\up{i}\up{j}\up{l}} + 
  \kappa  \tensor{{\it \gV}}{\down{i}\down{;}\down{j}\down{k}} 
    \tensor{{\it \LpT}}{\down{;}\up{k}\down{l}} 
     \tensor{{\it \LpT}}{\down{;}\up{i}\up{j}\up{l}} - 
\\[1 mm] & & 
  {{\kappa }\over 4} \tensor{{\it \LgT}}{\down{;}\down{i}\down{j}\down{k}} 
    \tensor{{\it \LpT}}{\down{;}\up{k}\down{l}} 
     \tensor{{\it \LpT}}{\down{;}\up{i}\up{j}\up{l}} + 
  {{\kappa }\over 2} \tensor{{\it \gV}}{\down{i}\down{;}\up{i}\down{j}} 
    \tensor{{\it \LpT}}{\down{;}\down{k}\down{l}} 
     \tensor{{\it \LpT}}{\down{;}\up{j}\up{k}\up{l}} + 
\\[1 mm] & & 
  {{\kappa }\over 4} \tensor{{\it \gTT}}{\down{i}\down{j}} 
    \tensor{{\it \LpT}}{\down{;}\down{k}\down{l}} 
     \tensor{{\it \LpT}}{\down{;}\up{i}\up{j}\up{k}\up{l}} + 
  {{\kappa }\over 2} \tensor{{\it \gV}}{\down{i}\down{;}\down{j}} 
    \tensor{{\it \LpT}}{\down{;}\down{k}\down{l}} 
     \tensor{{\it \LpT}}{\down{;}\up{i}\up{j}\up{k}\up{l}} - 
\\[1 mm] & & 
  {{\kappa }\over 8} \tensor{{\it \LgT}}{\down{;}\down{i}\down{j}} 
    \tensor{{\it \LpT}}{\down{;}\down{k}\down{l}} 
     \tensor{{\it \LpT}}{\down{;}\up{i}\up{j}\up{k}\up{l}} - 
  {{\kappa }\over 2} \tensor{{\it \LpT}}{\down{;}\down{i}\down{j}} 
    \tensor{{\it \pi}}{\down{k}\up{k}\down{;}\up{i}\up{j}} + 
\\[1 mm] & & 
  \kappa  \tensor{{\it \Ao}}{\down{;}\down{i}\down{j}} 
    \tensor{{\it \LpT}}{\down{;}\up{i}\up{j}} {\it \pT} - 
  {{5\,\kappa }\over 8} \tensor{{\it \gT}}{\down{;}\down{i}\down{j}} 
    \tensor{{\it \LpT}}{\down{;}\up{i}\up{j}} {\it \pT} - 
\\[1 mm] & & 
  {{\kappa }\over 2} \tensor{{\it \gTT}}{
     \down{i}\down{j}\down{;}\down{k}\up{k}} 
    \tensor{{\it \LpT}}{\down{;}\up{i}\up{j}} {\it \pT} - 
  \kappa  \tensor{{\it \gV}}{\down{i}\down{;}\down{j}\down{k}\up{k}} 
    \tensor{{\it \LpT}}{\down{;}\up{i}\up{j}} {\it \pT} + 
\\[1 mm] & & 
  {{\kappa }\over 2} \tensor{{\it \gV}}{\down{i}\down{;}\down{j}\up{j}\down{k}
     } \tensor{{\it \LpT}}{\down{;}\up{i}\up{k}} {\it \pT} - 
  {{3\,\kappa }\over 2} \tensor{{\it \gV}}{
     \down{i}\down{;}\up{i}\down{j}\down{k}} 
    \tensor{{\it \LpT}}{\down{;}\up{j}\up{k}} {\it \pT} - 
\\[1 mm] & & 
  {{\kappa }\over 2} \tensor{{\it \gTT}}{\down{i}\down{j}\down{;}\down{k}} 
    \tensor{{\it \LpT}}{\down{;}\up{i}\up{j}\up{k}} {\it \pT} - 
  \kappa  \tensor{{\it \gV}}{\down{i}\down{;}\down{j}\down{k}} 
    \tensor{{\it \LpT}}{\down{;}\up{i}\up{j}\up{k}} {\it \pT} + 
\\[1 mm] & & 
  {{\kappa }\over 4} \tensor{{\it \LgT}}{\down{;}\down{i}\down{j}\down{k}} 
    \tensor{{\it \LpT}}{\down{;}\up{i}\up{j}\up{k}} {\it \pT} + 
  {{\kappa }\over 2} \tensor{{\it \pi}}{\down{i}\up{i}\down{;}\down{j}\up{j}} 
    {\it \pT} - {{5\,\kappa }\over 4} 
   \tensor{{\it \Ao}}{\down{;}\down{i}\up{i}} {\it \pT}^{2} + 
\\[1 mm] & & 
  {{5\,\kappa }\over {16}} \tensor{{\it \gT}}{\down{;}\down{i}\up{i}} 
    {\it \pT}^{2} + {{\kappa }\over 2} 
   \tensor{{\it \gV}}{\down{i}\down{;}\up{i}\down{j}\up{j}} {\it \pT}^{2} + 
  {{{{\kappa }^3}}\over 8} \tensor{{\it \LpT}}{\down{;}\down{i}\down{j}} 
    \tensor{{\it \LpT}}{\down{;}\up{i}\up{j}} {\it \pT}^{2} - 
\\[1 mm] & & 
  {{{{\kappa }^3}}\over 8} {\it \pT}^{4} - 
  \kappa  \tensor{{\it \Ao}}{\down{;}\down{i}} 
    {\it \pT} \tensor{{\it \pT}}{\down{;}\up{i}} + 
  \kappa  \tensor{{\it \Ao}}{\down{;}\down{i}} 
    \tensor{{\it \LpT}}{\down{;}\up{i}\down{j}} 
     \tensor{{\it \pT}}{\down{;}\up{j}} - 
\\[1 mm] & & 
  {{\kappa }\over 4} \tensor{{\it \gT}}{\down{;}\down{i}} 
    \tensor{{\it \LpT}}{\down{;}\up{i}\down{j}} 
     \tensor{{\it \pT}}{\down{;}\up{j}} + 
  {{\kappa }\over 2} \tensor{{\it \gV}}{\down{i}\down{;}\up{i}\down{j}} 
    {\it \pT} \tensor{{\it \pT}}{\down{;}\up{j}} - 
\\[1 mm] & & 
  \kappa  \tensor{{\it \gV}}{\down{i}\down{;}\up{i}\down{j}} 
    \tensor{{\it \LpT}}{\down{;}\up{j}\down{k}} 
     \tensor{{\it \pT}}{\down{;}\up{k}} - 
  {{\kappa }\over 2} {\it \Ao} {\it \pT} 
     \tensor{{\it \pT}}{\down{;}\down{i}\up{i}} + 
\\[1 mm] & & 
  {{\kappa }\over 8} {\it \gT} {\it \pT} 
     \tensor{{\it \pT}}{\down{;}\down{i}\up{i}} + 
  {{\kappa }\over 2} \tensor{{\it \gV}}{\down{i}\down{;}\up{i}} 
    {\it \pT} \tensor{{\it \pT}}{\down{;}\down{j}\up{j}} + 
\\[1 mm] & & 
{{\kappa }\over 2} {\it \Ao} \tensor{{\it \LpT}}{\down{;}\down{i}\down{j}} 
     \tensor{{\it \pT}}{\down{;}\up{i}\up{j}} - 
{{\kappa }\over 8} {\it \gT} \tensor{{\it \LpT}}{\down{;}\down{i}\down{j}} 
     \tensor{{\it \pT}}{\down{;}\up{i}\up{j}} - 
\\[1 mm] & & 
  {{\kappa }\over 4} \tensor{{\it \gTT}}{\down{i}\down{j}} 
    {\it \pT} \tensor{{\it \pT}}{\down{;}\up{i}\up{j}} - 
  {{\kappa }\over 2} \tensor{{\it \gV}}{\down{i}\down{;}\down{j}} 
    {\it \pT} \tensor{{\it \pT}}{\down{;}\up{i}\up{j}} + 
\\[1 mm] & & 
  {{\kappa }\over 8} \tensor{{\it \LgT}}{\down{;}\down{i}\down{j}} 
    {\it \pT} \tensor{{\it \pT}}{\down{;}\up{i}\up{j}} - 
  {{\kappa }\over 2} \tensor{{\it \gV}}{\down{i}\down{;}\up{i}} 
    \tensor{{\it \LpT}}{\down{;}\down{j}\down{k}} 
     \tensor{{\it \pT}}{\down{;}\up{j}\up{k}}
\end{eqnarray*}



\def\vol#1{{{\bf #1}}}


\end{document}